\documentclass[12pt,a4paper]{article}
\input epsf
\usepackage{amsmath,amsfonts,epsfig,color,latexsym}
\usepackage{amssymb}
\usepackage[english, USenglish]{babel}
\usepackage{epsfig}
\usepackage{a4wide}
\usepackage{rotating}
\usepackage{slashed}
\renewcommand{\theequation}{\thesection.\arabic{equation}}

\newcommand{\N}{\mathcal{N}}

\csname @addtoreset\endcsname{equation}{section}

\newcommand{\ftopmod}[4]{\mathcal{H}^{#1,A_1\ldots A_{#2}}_{\bar{I}_1\ldots\bar{I}_{#3}],\bar{J}_1\ldots\bar{J}_{#4}}}

\newcommand{\gtmin}{(\mu\cdot G^-_{T^2})}
\newcommand{\gkmpin}{(\mu\cdot G^-_{K3,+})}

\newcommand{\da}{{\dot\alpha}}
\newcommand{\db}{{\dot\beta}}

\newcommand{\be}{\begin{equation}}
\newcommand{\ee}{\end{equation}}
\newcommand{\bea}{\begin{eqnarray}}
\newcommand{\eaa}{\end{eqnarray}}

\newcommand{\nn}{\nonumber}

\renewcommand{\a}{\alpha}

\newcommand{\pa}{\partial}

\renewcommand{\b}{\beta}

\newcommand{\la}{\lambda}

\newcommand{\q}{\theta }
\newcommand{\bq}{\bar\theta }
\newcommand{\bu}{\bar u}

\newcommand{\ep}{\epsilon}
\newcommand{\m}{\mu}

\newcommand{\tz}{{\Upsilon}}

\newcommand{\cN}{{\cal N}}

\newcommand{\p}[1]{(\ref{#1})}
\newcommand{\bt}[1]{{\bar t}}

\newcommand{\gameas}{\int [d\zeta]}

\newcommand{\grass}{\vartheta}

\renewcommand{\theequation}{\thesection.\arabic{equation}}

\csname @addtoreset\endcsname{equation}{section}
\author{\\[-0.3cm]\Large I.~Antoniadis${^{\text{a}}}$\footnote{\tt ignatios.antoniadis@cern.ch}
\footnote{On leave from CPHT (UMR CNRS 7644) Ecole Polytechnique, F-91128 Palaiseau}~, S.~Hohenegger${^{\text{b}}}$\footnote{{\tt shoheneg@mppmu.mpg.de}}~, K.S.~Narain${^{\text{c}}}$\footnote{{\tt narain@ictp.trieste.it}}~, E.~Sokatchev${^{\text{ade}}}$\footnote{{\tt emeri.sokatchev@cern.ch}}
}
\title{\begin{flushright}{\vspace{-0.8cm}\small CERN-PH-TH/2011-111\\ \small LAPTH-021/11\\[-0.4cm]\small MPP-2011-75}\end{flushright}
\vspace{0.8cm}
\bf{Generalized $\N=2$ Topological Amplitudes and Holomorphic Anomaly Equation}}
\dateUSenglish
\date{}
\graphicspath{{figures/}}

\begin{document}

\begin{titlepage}

\maketitle
\begin{center}
\renewcommand{\thefootnote}{\fnsymbol{footnote}}\vspace{-0.5cm}
${}^{\text{a}}$ Department of Physics, CERN - Theory Division, CH-1211 Geneva 23, Switzerland\\[0.2cm]
${}^{\text{b}}$ Max--Planck--Institut f\"ur
Physik (Werner--Heisenberg--Institut)\\
F\"ohringer Ring 6, 80805 M\"unchen, Germany\\[0.2cm]
${}^{\text{c}}$ High Energy Section, The Abdus Salam International Center for Theoretical Physics,
\\Strada Costiera, 11-34014 Trieste, Italy\\[0.2cm]
${}^{\text{d}}$ Institut Universitaire de France, 103 bd Saint-Michel,
F-75005 Paris, France \\[0.2cm] 
${}^{\text{e}}$ LAPTH\footnote[6]{Laboratoire d'Annecy-le-Vieux de Physique Th\'{e}orique, UMR 5108},   Universit\'{e} de Savoie, CNRS, 
B.P. 110,  F-74941 Annecy-le-Vieux, France\\[0.5cm]
\end{center}
\begin{abstract}
In arXiv:0905.3629 we described a new class of $\N=2$ topological
amplitudes that depends both on vector and hypermultiplet moduli. Here
we find that this class is actually a 
particular case of much more general topological amplitudes which
appear at higher loops in heterotic string theory compactified on
$K3\times T^2$. 
We analyze their effective field theory interpretation and derive
particular (first order) differential equations as a consequence of supersymmetry
Ward identities and the 1/2-BPS nature of the corresponding effective
action terms. In string theory the latter get 
modified due to anomalous world-sheet boundary contributions,
generalizing in a non-trivial way the familiar holomorphic 
and harmonicity anomalies studied in the past. We prove by direct
computation that the subclass of topological amplitudes studied in
arXiv:0905.3629 forms a closed set under these anomaly equations
and that these equations are integrable. 
\end{abstract}
\thispagestyle{empty}
\end{titlepage}

\tableofcontents

\vspace{.0in}

\section{Introduction}
\setcounter{equation}{0}
Over the last couple of decades, considerable effort has been put into
the study of $1/2$-BPS effective couplings in four-dimensional 
extended supersymmetric theories. Such couplings depend only on half
of the superspace, generalizing the well-known chiral F-terms of 
$\N=1$ supersymmetry. They usually enjoy particular
non-renormalization theorems which is the reason why they are easier
to study 
from a field theoretic point of view. Moreover, these couplings play a
predominant role in many very interesting applications, 
ranging from string phenomenology to entropy corrections of
supersymmetric black holes. It is therefore of considerable 
interest to classify different kinds of 1/2-BPS couplings and study their properties in as much generality as possible.  

In the string theory effective action, $1/2$-BPS couplings are
expected to be captured by topological amplitudes, i.e.~amplitudes
which only depend on the zero-mode structure of the internal
manifold~\cite{Witten:1992fb,Antoniadis:1993ze,Bershadsky:1993cx}. 
The best studied example in this respect is the celebrated series $F_g
\mathcal{W}^{2g}$ where $\mathcal{W}$ is the chiral $\N=2$ 
Weyl superfield of the supergravity multiplet. The coefficients $F_g$
are computed as $g$ loop amplitudes in type~II string theory 
compactified on a Calabi-Yau
threefold~\cite{Antoniadis:1993ze,Bershadsky:1993cx} and are functions
of the vector multiplet moduli 
in the Coulomb phase of the theory. In particular, their independence
of $\N=2$ hypermultiplet moduli 
(and therefore also of the type~II dilaton) implies a
non-renormalization theorem that $F_g$ is $g$-loop exact.
Moreover, classically their BPS nature dictates that the $F_g$ are
holomorphic functions. This property is broken at the quantum level,
however, in a controlled manner 
captured by a first order differential equation, termed the \emph{holomorphic anomaly equation}~\cite{Bershadsky:1993cx}. The latter takes the 
form of a recursion relation in $g$ which in fact allows in some cases to find explicit expressions for $F_g$ up to considerably high genus~\cite{Huang:2006si,Huang:2006hq,Grimm:2007tm}.

The heterotic versions of the $F_g$'s have been considered in
\cite{Antoniadis:1996qg}. They are semi-topological expressions 
(i.e.~topological only in the supersymmetric sector), related to
F-terms of the form $\hat{\mathcal{W}}^{2g}$ 
where $\hat{\mathcal{W}}$ is now the $\N=1$ gauge superfield. Again
the BPS-nature of these couplings 
classically implies a (first order) holomorphicity condition for the $F_g$ which is
broken at the quantum level. 
The difference to the type~II side is, however, that the corresponding
(integrable) holomorphic anomaly 
equation, which captures this breaking, does no longer close on the
class of functions $F_g$, but introduces 
new (semi)-topological objects. The latter have been understood to be
physically related to BPS couplings of 
the form $\Pi^n\hat{\mathcal{W}}^{2g}$ where $\Pi$'s are chiral projections of non-holomorphic functions  of chiral superfields.

More recently, after studying a particular world-sheet involution of a
certain class of $\N=4$ topological amplitudes~
\cite{Antoniadis:2006mr,Antoniadis:2007cw} (see also
\cite{Antoniadis:2009tr} as well as the review
\cite{Antoniadis:2007ta}), 
we have presented a completely new class of $\N=2$ topological
amplitudes~\cite{Antoniadis:2009nv}. The latter compute a 
corresponding class of 1/2-BPS terms in the low energy effective
action of the form $\mathcal{F}_gK_-^{2g}$, where $K_-$ is a 
particular superdescendant of the vector superfield in $\N=2$ harmonic
superspace~\cite{Galperin:1984av}. The novel property 
is that their coupling coefficients depend on both holomorphic vector
multiplet as well as particular (Grassmann) 
analytic hypermultiplet moduli, despite the common wisdom of mutual
decoupling. Classically, this particular analytic 
dependence --- which is again a direct consequence of the BPS nature
of the couplings --- can be captured by 
differential equations, namely a \emph{holomorphicity} relation for
the vector moduli and a \emph{harmonicity relation} 
together with another second order equation for the hypermultiplet
moduli. As before, at the quantum level 
these differential equations get modified by anomalous string
world-sheet boundary contributions. 

In this work, we show that these new topological objects are in fact a
particular subclass of an even further generalized set of $\N=2$
topological amplitudes. We make a systematic analysis of the latter on
the heterotic string side (compactified on $K3\times T^2$) by first
computing them as genus-$g$ amplitudes involving an appropriate number
of fermions (gauginos and hyperinos) and establish explicitly their
connection with the (semi)-topological 
theory obtained by twisting only the supersymmetric (left-moving)
sector. We then show that this generic 
amplitude corresponds to a new series of higher order couplings in the
effective action, involving both 
vector multiplets and neutral hypermultiplets. They depend on the
moduli (both vectors and hypermultiplets) 
in a quite generic manner, which particularly means that there is no
immediate direct generalization of the 
holomorphicity or  harmonicity equation. However, the BPS nature of
these couplings manifests itself in another 
very useful property: The particular (Grassmann)-analytic projection
which is necessary for consistency of these 
couplings leads us to establish relations between different component
couplings. On the string theory side, 
these relations turn out to be highly non-trivial (first order) differential
equations which again get modified by anomalous 
world-sheet boundary contributions. 
We explicitly compute these
anomaly terms for these more general amplitudes (up to possible
curvature dependent contact terms) as well as for the subclass considered
in ~\cite{Antoniadis:2009nv}  and show that the equations close 
in the sense that no new topological objects appear. This indicates
that the amplitude we consider captures the 
most generic expression in this particular class of BPS-couplings.  
We study the
integrability of these equations and show that for the latter, the
equations we have obtained are integrable including also the curvature
terms.

The paper is organized as follows. In Section~\ref{Sect:EffectAction},
we compute a particular multi-fermion 
genus-$g$ physical amplitude and we show that it acquires a
(semi)-topological expression as a correlation 
function in the twisted $K3\times T^2$ heterotic $\sigma$-model. This
amplitude has four types of indices, 
labeling the two different helicities (i.e. Weyl spinor components) of
the antichiral gauginos and hyperinos 
involved. Contracting these indices (which have particular symmetry
properties) with four fermionic variables, 
one can introduce a generating functional, in terms of which the
results take a compact form and are greatly 
simplified. In Section~\ref{Sect:HarmonicDescription}, we describe the
interpretation of the above amplitudes 
as a particular class of 1/2-BPS terms in the string effective action. 
In Sections~\ref{Sect:AnomalyEquation}
and~\ref{Sect:AnomalyEquationM20}, we derive differential equations 
for respectively the generalized amplitudes and the reduced ones that were obtained
in~\cite{Antoniadis:2009nv}. As we show in
Section~\ref{Sect:FieldTheoryEqu}, these relations follow from the particular 
(Grassmann)-analytic projection which we used in formulating the
couplings. These equations are a direct 
consequence of the 1/2-BPS structure of these couplings and relate the
anti-holomorphic dependence on the 
vector moduli with the dependence on the hypermultiplets of `wrong'
harmonicity in different component 
couplings. In string theory we find that (as usual) these equations
get modified by an anomaly due to 
world-sheet boundary contributions. As a result, one obtains recursion
relations for the 
non-holomorphic/harmonic moduli dependence of the above
amplitudes. In Sections \ref{Sect:AnomalyEquation}
and~\ref{Sect:AnomalyEquationM20} we also study the integrability
conditions of the anomaly equations. Our conclusions are presented in 
Section~\ref{conclusions}. Although the notation and conventions of this paper are those of
Ref.~\cite{Antoniadis:2009nv}, to make it self-contained, 
we include several appendices. Appendix~\ref{append:SCA} gives a brief
review of $\N=2$ and $\N=4$ superconformal algebras, 
Appendix~\ref{App:HarmSuperspace} summarizes the main properties of
$\N=2$ harmonic superspace, Appendix~\ref{App:HetK3comp} 
presents some essential features of the heterotic $K3\times T^2$ compactification, Appendix~\ref{App:GaugeFreedomePotential} 
describes the gauge freedom associated to one of the two general series of couplings,
and Appendix~\ref{stringdiffeqs} contains the 
direct string derivation of the differential equations, as well as the
computation of a particular world-sheet boundary contribution which we deemed too lengthy to be presented in the main body of the paper.
\section{Generalized $\cN=2$ Topological Amplitudes}\label{Sect:EffectAction}
\setcounter{equation}{0}
\subsection{A Further Extension of a Class of Topological Amplitudes}
In the paper \cite{Antoniadis:2009nv} by studying higher derivative
couplings of holomorphic $\N=2$ vector multiplets with 
hypermultiplets of a particular analyticity in heterotic string theory
compactified on $K3\times T^2$, a new class of $\N=2$ 
topological amplitudes was discovered. They can be expressed as the following (semi-)topological correlators 
\begin{align}
\mathcal{F}^g=\int_{\mathcal{M}_{g}}\langle\gtmin^g\gkmpin^{2g-4}(\mu\cdot J^{--}_{K3})\psi_3(\alpha)\rangle\cdot (\text{det}Q_1)(\text{det}Q_2)\,.\label{OriginalAmplitude}
\end{align}
Here $\int_{\mathcal{M}_g}$ denotes the integral over the moduli space
of Riemann surfaces of genus $g$. Deformations of the 
latter are parametrized by a total of $3g-3$ Beltrami differentials
$\mu$. The operators sewed with these Beltrami 
differentials in (\ref{OriginalAmplitude}) are part of a twisted
$\N=2$ superconformal algebra, which we have briefly 
reviewed in appendix~\ref{append:SCA}. The free fermion $\psi_3$,
which is inserted at an arbitrary position $\alpha$ on 
the Riemann surface, serves the purpose of soaking up the holomorphic
zero mode of the $T^2$ torus. Finally, $Q_{1,2}$ are the 
zero modes of the corresponding right-moving currents $\bar{J}_{1,2}$ associated with the gauge group in the heterotic theory.

The correlators (\ref{OriginalAmplitude}), however, are not yet the most
generic objects one might envisage. Indeed, 
already in \cite{Antoniadis:2009nv} a generalization was worked out by
allowing also a dependence on non-holomorphic 
vector multiplets. Indeed, the following semi-topological expression
was found in a $g$-loop heterotic string amplitude for 
$n\geq 0$ (for details of the notation see \cite{Antoniadis:2009nv})
\begin{align}
\mathcal{F}_{g,n}=\int_{\mathcal{M}_{(g,n)}}\hspace{-0.25cm}\langle(\mu\cdot
G^-_{T^2})^{g-n}(\mu\cdot G^-_{K3})^{2g+2n-4}(\mu\cdot J^{--}_{K3})
\prod_{j=1}^n\left[\int_{w_j}(\bar{\psi}_3\bar{J})(J^{++}_{K3}\bar{J})(v_j)\right]\psi_3(p)(\text{det}Q)^2\rangle\nonumber
\end{align}
where $\int_{\mathcal{M}_{(g,n)}}$ denotes the integral over the
moduli space of a genus $g$ Riemann surface with $n$ punctures 
$v_j$. Note that deformations of the latter are parametrized by
$3g-3+n$ Beltrami differentials $\mu$. Moreover, it 
was also noted that the hypermultiplets of the 'wrong' analyticity
(\emph{i.e.} different than the one induced 
by the F-term like measure in harmonic superspace) are associated with
operators of the form $(\Xi^{A},\bar{\Xi}^A)$. 
The latter are labeled by an $Sp(2n)$ index $A$ and transform as
doublets under the $SU(2)$ current algebra. 
Besides they are $\N=4$ primary operators in the sense that each of them is annihilated by half of the supercharges
\begin{align}
&\oint G^+_{K3,i}\Xi^A=0\,,&&\text{and} &&\oint G^-_{K3,i}\bar{\Xi}^A=0\,.
\end{align}
These fields have left and right conformal dimension $(1/2,1)$ and are $SU(2)$ highest weight states
\begin{align}
&\oint J^{++}_{K3}\Xi^A=\oint J^{--}_{K3}\bar{\Xi}^A=0\,,&&\text{and}&&\oint J^{++}_{K3}\bar{\Xi}^A=\Xi^A\,,\label{HighestWeightRel}
\end{align}
with $U(1)$ charges $\pm 1$ respectively.

This suggests that the most generic amplitude, with a dependence on
holomorphic and anti-holomorphic vector multiplets 
and hypermultiplets of all analyticities, should structurally be of
the form\footnote{Notice that we have turned the 
arbitrary position $p$ here into an additional puncture
$v_{n+1}$. Moreover, for reasons which will become clearer 
later on, we have denoted this expression with a tilde.}
\begin{align}
&\tilde{\mathcal{H}}^{g,A_1\ldots A_{m_1},B_1\ldots B_{m_2+1}}_{\bar{I}_1\ldots\bar{I}_{n-m_1},\bar{J}_1\ldots\bar{J}_{n-m_2}}:=\frac{1}{(3g+n-3)!}\cdot
\int_{\mathcal{M}_{(g,n+1)}}\langle(\mu\cdot G^-)^{3g+n-3}(\mu\cdot J^{--}_{K3})\prod_{a=1}^{m_1}\int \bar{\Xi}^{A_a}\cdot\nonumber\\
&\hspace{1cm}\cdot\prod_{b=1}^{m_2+1}(\psi_3\Xi^{B_b}) \prod_{c=1}^{n-m_1}\int\bar{\psi}_3\bar{J}_{\bar{I}_c}\prod_{d=1}^{n-m_2}J^{++}_{K3}\bar{J}_{\bar{J}_d}\rangle\,.\label{GenTopObject}
\end{align}
with $g\geq n\geq m_1$ and $n>m_2$ arbitrary integers. As we will see
in the remainder of this work, this is essentially correct and we will 
prove it by showing that there is indeed a string amplitude, which
(after some manipulations) gives rise to this expression. More 
specifically, we will calculate in the next subsection a heterotic
$g$-loop amplitude with insertions of $2g$ chiral gauginos, 
$2n-m_1-m_2$ anti-chiral gauginos, two chiral and $m_1+m_2+2$ anti-chiral hyperfermions and further discuss its relation to (\ref{GenTopObject}).
\subsection{Heterotic BPS-Saturated String Amplitudes}
\subsubsection{BPS-Saturated Amplitudes}\label{Sect:GenericAmplitudeString}
Following the discussion in \cite{Antoniadis:2009nv}, we will compute
the heterotic amplitude at a generic point in the moduli space of the 
$K3\times T^2$ compactification. For a brief review of the notation as
well as for explicit expressions of the vertex operators see 
appendix~\ref{App:HetK3comp}. Here, for convenience, we just give
table~\ref{Tab:Vertex} containing the fermion charges with 
respect to space-time and torus fermions (bosonized in terms of scalars $\phi_1$, $\phi_2$ and $\phi_3$ respectively) and the
$H$-charge of the vertex operators for the physical fields as well as
the picture changing operators (PCO). The last column \textbf{RM} 
denotes the right moving (bosonic) piece of the vertex operator. We
have not displayed the superghost part in the table, however, all 
matter fields are understood to be inserted in the $(-1/2)$ ghost
picture, i.e. their vertices come with $e^{-\phi/2}$, while the PCO 
come with a factor of $e^{\phi}$. 
\begin{table}[h]
\begin{center}
\begin{tabular}{|c|c|c||c|c||c||c||c|}\hline
\textbf{field} & \textbf{pos.} & \textbf{number} &
\parbox{0.5cm}{\vspace{0.2cm}$\phi_1$\vspace{0.2cm}}&
\parbox{0.5cm}{\vspace{0.2cm}$\phi_2$\vspace{0.2cm}} &
\parbox{0.5cm}{\vspace{0.2cm}$\phi_3$\vspace{0.2cm}} &
\parbox{0.5cm}{\vspace{0.2cm}$H$\vspace{0.2cm}} &
\parbox{0.8cm}{\vspace{0.2cm}\textbf{RM}\vspace{0.2cm}} \\\hline\hline
gaugino ${\lambda}_-$& \parbox{0.35cm}{\vspace{0.2cm}$x_i$\vspace{0.2cm}}
& $g$ & \parbox{0.7cm}
{\vspace{0.2cm}$+\frac{1}{2}$\vspace{0.2cm}} & \parbox{0.7cm}
{\vspace{0.2cm}$+\frac{1}{2}$\vspace{0.2cm}} & \parbox{0.7cm}
{\vspace{0.2cm}$+\frac{1}{2}$\vspace{0.2cm}} & \parbox{0.8cm}
{\vspace{0.2cm}$+\frac{1}{\sqrt{2}}$\vspace{0.2cm}} & \parbox{0.5cm}
{\vspace{0.2cm}$\bar{J}_{I_{i}}$\vspace{0.2cm}} \\\hline
 & \parbox{0.35cm}{\vspace{0.2cm}$y_i$\vspace{0.2cm}} & $g$ & \parbox{0.7cm}
{\vspace{0.2cm}$-\frac{1}{2}$\vspace{0.2cm}} & \parbox{0.7cm}
{\vspace{0.2cm}$-\frac{1}{2}$\vspace{0.2cm}} & \parbox{0.7cm}
{\vspace{0.2cm}$+\frac{1}{2}$\vspace{0.2cm}} & \parbox{0.8cm}
{\vspace{0.2cm}$+\frac{1}{\sqrt{2}}$\vspace{0.2cm}} & \parbox{0.55cm}
{\vspace{0.2cm}$\bar{J}_{J_{i}}$\vspace{0.2cm}} \\\hline\hline
gaugino $\bar{\lambda}^+$&
\parbox{0.45cm}{\vspace{0.2cm}$w_j$\vspace{0.2cm}} & $n-m_1$ & \parbox{0.7cm}
{\vspace{0.2cm}$+\frac{1}{2}$\vspace{0.2cm}} & \parbox{0.7cm}
{\vspace{0.2cm}$-\frac{1}{2}$\vspace{0.2cm}} & \parbox{0.7cm}
{\vspace{0.2cm}$-\frac{1}{2}$\vspace{0.2cm}} & \parbox{0.8cm}
{\vspace{0.2cm}$+\frac{1}{\sqrt{2}}$\vspace{0.2cm}} & \parbox{0.5cm}
{\vspace{0.2cm}$\bar{J}_{\bar{I}_{j}}$\vspace{0.2cm}} \\\hline
 & \parbox{0.35cm}{\vspace{0.2cm}$v_j$\vspace{0.2cm}} & $n-m_2$ & \parbox{0.7cm}
{\vspace{0.2cm}$-\frac{1}{2}$\vspace{0.2cm}} & \parbox{0.7cm}
{\vspace{0.2cm}$+\frac{1}{2}$\vspace{0.2cm}} & \parbox{0.7cm}
{\vspace{0.2cm}$-\frac{1}{2}$\vspace{0.2cm}} & \parbox{0.8cm}
{\vspace{0.2cm}$+\frac{1}{\sqrt{2}}$\vspace{0.2cm}} & \parbox{0.75cm}
{\vspace{0.2cm}$c \bar{J}_{\bar{J}_j}$\vspace{0.2cm}} \\\hline\hline
hyperino $\chi^{C_1}$ & \parbox{0.35cm}{\vspace{0.2cm}$z_1$\vspace{0.2cm}}
& 1 & $+\frac{1}{2}$ & $+\frac{1}{2}$ & $-\frac{1}{2}$ & $0$ &
\parbox{0.65cm}
{\vspace{0.2cm}$\bar{\Xi}^{C_1}$\vspace{0.2cm}} \\\hline
hyperino $\chi^{C_2}$& \parbox{0.35cm}{\vspace{0.2cm}$z_2$\vspace{0.2cm}} & 1 &
$-\frac{1}{2}$ & $-\frac{1}{2}$ & $-\frac{1}{2}$ & $0$  & \parbox{0.65cm}
{\vspace{0.2cm}$\bar{\Xi}^{C_2}$\vspace{0.2cm}}\\\hline
hyperino $\bar{\psi}^{A_k}$& \parbox{0.4cm}{\vspace{0.2cm}$u_k$\vspace{0.2cm}} & $m_1+1$
& $+\frac{1}{2}$ & $-\frac{1}{2}$ & $+\frac{1}{2}$ & $0$  & \parbox{0.7cm}
{\vspace{0.2cm}$\bar{\Xi}^{A_k}$\vspace{0.2cm}}\\\hline
hyperino $\bar{\psi}^{B_k}$ & \parbox{0.35cm}{\vspace{0.2cm}$t_k$\vspace{0.2cm}} &
$m_2+1$ & $-\frac{1}{2}$ & $+\frac{1}{2}$ & $+\frac{1}{2}$ & $0$  &
\parbox{0.85cm}
{\vspace{0.2cm}$c\Xi^{B_k}$\vspace{0.2cm}}\\\hline\hline
PCO & \parbox{0.8cm}{\vspace{0.2cm}$\{s_3\}$\vspace{0.2cm}} & $g+m_1+m_2-n$ & $0$ &
$0$ & $-1$ & $0$ & \parbox{0.75cm}
{\vspace{0.2cm}$\partial X_3$\vspace{0.2cm}} \\\hline
& \parbox{0.9cm}{\vspace{0.2cm}$\{s_H\}$\vspace{0.2cm}} & $2g+2n-m_1-m_2$ & $0$ &
$0$ & $0$ & $-\frac{1}{\sqrt{2}}$  & \parbox{0.75cm}
{\vspace{0.2cm}$G^-_{K3}$\vspace{0.2cm}}\\\hline
\end{tabular}
\end{center}
\caption{Fermion and $H$-charges for all vertex operators and picture 
changing operators (PCO). All physical vertex operators are inserted in the $-\tfrac{1}{2}$ picture, while all PCO contribute a factor of $e^{\phi}$.}
\label{Tab:Vertex}
\end{table}
The PCO's are arranged in such a way that $g+m_1+m_2-n$ of them at
$r_a\in\{s_3\}$ contribute the torus part 
while $2g+2n-m_1-m_2$ of them at $r_a\in\{s_H\}$ the $K3$ part. In
fact, by charge conservation, this is the only 
possible grouping of all PCOs --- of course we still have to take into
account all possible distributions of the 
total number $3g-2+n$ of the picture changing operators into these two
classes. Notice moreover that we have put 
$c$-ghosts together with the vertex operators at the points $v_j$ and
$t_{k}$ which means that the latter will not 
be integrated over the world-sheet. Instead, however, the genus $g$ world-sheet will have $n+1$ punctures. 

With this table we can write down the amplitude in a straight-forward
manner (for our notation concerning 
the spin structure dependence see appendix~\ref{App:HetK3compSpin}). In order to save writing we define the following world-sheet positions
{\allowdisplaybreaks
\begin{align}
&a_1=\frac{1}{2}\sum_{i=1}^g(x_i-y_i)-\frac{1}{2}\sum_{j=1}^{n-m_1}w_j+\frac{1}{2}\sum_{j=1}^{n-m_2}v_j+\frac{1}{2}(z_1-z_2)+\frac{1}{2}\sum_{k=1}^{m_2+1}t_k-\frac{1}{2}\sum_{k=1}^{m_1+1}u_k\,,
\nonumber\\
&a_2=\frac{1}{2}\sum_{i=1}^g(x_i+y_i)-\frac{1}{2}\sum_{j=1}^{n-m_1}w_j-\frac{1}{2}\sum_{j=1}^{n-m_2}v_j-\frac{1}{2}(z_1+z_2)+\frac{1}{2}\sum_{k=1}^{m_2+1}t_k+\frac{1}{2}
\sum_{k=1}^{m_1+1}u_k-\sum_a^{\{s_3\}}r_a\,,\nonumber\\
&a_3=\frac{1}{\sqrt{2}}\sum_{i=1}^g(x_i+y_i)+\frac{1}{\sqrt{2}}\sum_{j=1}^{n-m_1}w_j+\frac{1}{\sqrt{2}}\sum_{j=1}^{n-m_2}v_j-\frac{1}{\sqrt{2}}\sum_{a}^{\{s_H\}}r_a\,,\nonumber
\end{align}}
with which the amplitude can be written in the following form
{\allowdisplaybreaks
\begin{align}
&\mathcal{A}^{g,C_1C_2,A_1\ldots A_{m_1+1},B_1\ldots
  B_{m_2+1}}_{I_1\ldots I_g,J_1\ldots
  J_g,\bar{I}_1\ldots\bar{I}_{n-m_1},\bar{J}_1\ldots\bar{J}_{n-m_2}}=F_{\{\Lambda\},s}(a_1,a_2,a_3)
\,G_{\{\Lambda\}}(x_i,y_i,w_j,v_j,z_1,z_2,t_k,u_k,\{s_3\},\{s_H\})\cdot\nonumber\\
&\cdot\frac{\vartheta_s\left(\frac{1}{2}\left(\sum_{i}^g(x_i-y_i)+\sum_{j}^{n-m_1}w_j-\sum_{j}^{n-m_2}v_j+z_1-z_2-\sum_{k}^{m_2+1}t_k+
\sum_k^{m_1+1}u_k\right)\right)}{\parbox{15.92cm}{\footnotesize$\vartheta_s\left(\frac{1}{2}\left(\sum_{i}^g(x_i+y_i)+\sum_{j}^{n-m_1}w_j+
\sum_{j}^{n-m_2}v_j+z_1+z_2+\sum_{k}^{m_2+1}t_k+\sum_k^{m_1+1}u_k\right)-\sum_a^{\{s_3,s_{H}\}}r_a-2\Delta\right)$}}\nonumber\\
&\cdot\frac{\prod_{i<j}^gE(x_i,x_j)E(y_i,y_j)\prod_{i<j}^{n-m_1}E(w_i,w_j)\prod_{i<j}^{n-m_2}(v_i,v_j)\prod_{i<j}^{m_2+1}\!E(t_i,t_j)\prod_{i<j}^{m_1+1}
\!E(u_i,u_j)}{\prod_a^{\{s_3,s_H\}}E(r_a,r_b)\prod_i^gE(x_i,z_2)E(y_i,z_1)\prod_j^{n-m_1}\prod_{k}^{m_2+1}E(w_j,t_k)\prod_{j}^{n-m_2}\prod_k^{m_1+1}E(v_j,u_k)}\nonumber\\
&\cdot\frac{\prod_a^{\{s_3\}}E(r_a,z_1)E(r_a,z_2)\prod_j^{n-m_1}E(r_a,w_j)\prod_j^{n-m_2}E(r_a,v_j)\prod_a^{\{s_H\}}E^{\frac{1}{2}}(r_a,z_1)E^{\frac{1}{2}}(r_a,z_2)}
{E^{\frac{1}{2}}(z_1,z_2)\prod_{k}^{m_2+1}E^{\frac{1}{2}}(z_1,t_k)E^{\frac{1}{2}}(z_2,t_k)\prod_k^{m_1+1}E^{\frac{1}{2}}(z_1,u_k)E^{\frac{1}{2}}(z_2,u_k)\prod_{j}^{m_2+1}E^{\frac{1}{2}}(t_j,u_k)}\nonumber\\
&\cdot\frac{\prod_a^{\{s_H\}}\prod_k^{m_2+1}E^{\frac{1}{2}}(r_a,t_k)\prod_k^{m_1+1}E^{\frac{1}{2}}(r_a,u_k)}{\prod_{a<b}^{\{s_3\}}E(r_a,r_b)\prod_{a<b}^{\{s_H\}}
E(r_a,r_b)}\cdot\langle\prod_i^g\bar{J}_{I_i}\bar{J}_{J_i}\prod_j^{n-m_1}\bar{J}_{\bar{I}_j}\prod_{j}^{n-m_2}\bar{J}_{\bar{J}_j}\rangle\nonumber\\
&\cdot\langle\prod_a^{\{s_H\}}G^-_{K3,+}(r_a)\,\Xi^{C_1}(z_1)\,\Xi^{C_2}(z_2)\,\prod_k^{m_2+1}\Xi^{B_k}(t_k)\,\prod_{k}^{m_1+1}\bar{\Xi}^{A_k}(u_k)
\prod_{k=1}^{m_2+1}c(t_k)\prod_{j}^{n-m_2}c(v_j)\rangle\,.\label{AmplitudeFullExpression}
\end{align}}
For simplicity we will consider in the following differences of vector
multiplet gauge groups in which case all of the operators $\bar{J}_I$
and $\bar{J}_J$ 
with holomorphic indices can only contribute zero modes. Since
therefore their contribution is trivial, we will mostly 
drop them in the following. Furthermore, we are still free to choose a gauge and by picking the condition 
\begin{align}
\sum_a^{\{s_3\}}r_a+\sum_a^{\{s_H\}}r_a=\sum_{i=1}^gy_i+\sum_{j=1}^{n-m_2}v_j+z_2+\sum_{k=1}^{m_2+1}t_k-2\Delta\,,\label{gaugechoice}
\end{align}
essentially both the $\vartheta$-functions cancel in
(\ref{AmplitudeFullExpression}). For the remaining expression we can
directly perform the 
spin-structure sum using (\ref{sssum}) with the result that
$F_{\{\Lambda\},s}(a_1,a_2,a_3)$ is replaced by
$F_{\{\Lambda\}}(a'_1,a'_2,a'_3)$ 
where, after using the gauge condition (\ref{gaugechoice}), we obtain 
\begin{align}
&a'_1=\sum_{i=1}^gx_i-z_2-\Delta\,,\hspace{3cm}a'_2=\sum_a^{\{s_3\}}r_a+\sum_{j=1}^{n-m_1}w_j-\sum_{k=1}^{m_2+1}t_k-\Delta\,,\\
&a'_3=\frac{1}{\sqrt{2}}\left(\sum_a^{\{s_4\}}r_a-2\sum_{j=1}^{n-m_2}v_j-z_1-z_2-\sum_{k=1}^{m_2+1}t_k+\sum_{k=1}^{m_1+1}u_k-2\Delta\right)\,.
\end{align}
Using the same manipulations as in \cite{Antoniadis:2009nv} we can bring the amplitude into the following form
\begin{align}
&\mathcal{A}^{g,C_1C_2,A_1\ldots A_{m_1+1},B_1\ldots
  B_{m_2+1}}_{\bar{I}_1\ldots\bar{I}_{n-m_1},\bar{J}_1\ldots\bar{J}_{n-m_2}}=\int_{\mathcal{M}_{(g,n)}}
\hspace{-0.7cm}(\mu\cdot b)^{3g-2+n}\,\frac{\text{det}\omega_i(x_j)\text{det}\omega_i(y_j)}{(\text{det}(\text{Im}\tau))^2}\cdot\nonumber\\
&\hspace{0.3cm}\cdot\frac{\langle\prod_a^{\{s_H\}}G^-_{K3,+}(r_a)\,\Xi^{C_1}(z_1)\,\Xi^{C_2}(z_2)\,\prod_k^{m_2+1}c\Xi^{B_k}(t_k)\,
\prod_{k}^{m_1+1}\bar{\Xi}^{A_k}(u_k)\prod_{j}^{n-m_2}ce^{i\sqrt{2}H}(v_j)\rangle_{K3}}{\langle\prod_a^{3g+n}b(r_a)\prod_j^{n-m_2}c(v_j)c(z_1)c(z_2)\prod_k^{m_2+1}c(t_k)\rangle_{bc}}\cdot\nonumber\\
&\hspace{0.3cm}\cdot\langle\prod_a^{\{s_3\}}G^-_{T^2}(r_a)\prod_j^{n-m_1}\bar{\psi}_3(w_j)\prod_{k}^{m_2+1}\psi_3(t_k)\rangle_{T^2}\cdot\langle
\prod_j^{n-m_1}\bar{J}_{\bar{I}_j}\prod_{j}^{n-m_2}\bar{J}_{\bar{J}_j}\rangle\,.\nonumber
\end{align}
We can further simplify this expression by collapsing two of the
PCO-insertion points with $z_1$ and $z_2$ respectively. Just as in
\cite{Antoniadis:2009nv}, 
this converts these insertions to $(0)$-picture hyperscalar vertex operators which we can write as covariant derivatives in the following manner
\begin{align}
\mathcal{A}^{g,C_1C_2,A_1\ldots A_{m_1+1},B_1\ldots
  B_{m_2+1}}_{\bar{I}_1\ldots\bar{I}_{n-m_1},\bar{J}_1\ldots\bar{J}_{n-m_2}}=\mathcal{D}_+^{C_1}\mathcal{D}_+^{C_2}
\mathcal{H}^{g,A_1\ldots A_{m_1+1},B_1\ldots B_{m_2+1}}_{\bar{I}_1\ldots\bar{I}_{n-m_1},\bar{J}_1\ldots\bar{J}_{n-m_2}}\,.
\end{align}
Performing now also the integration over $x_i$ and $y_i$ gives a
factor of $(\text{det}(\text{Im}\tau))^2$ which cancels the one
already present. 
The left-over expression therefore becomes
\begin{align}
&\mathcal{H}^{g,A_1\ldots A_{m_1+1},B_1\ldots B_{m_2+1}}_{\bar{I}_1\ldots\bar{I}_{n-m_1},\bar{J}_1\ldots\bar{J}_{n-m_2}}=\nonumber\\
&=\int_{\mathcal{M}_{(g,n+1)}}\hspace{-0.7cm}(\mu\cdot
b)^{3g-2+n}\,\langle\prod_a^{\{s_3\}}G^-_{T^2}(r_a)\prod_j^{n-m_1}\bar{\psi}_3(w_j)
\prod_{k}^{m_2+1}\psi_3(t_k)\rangle_{T^2}\cdot\langle\prod_i^g\bar{J}_{I_i}\bar{J}_{J_i}\prod_j^{n-m_1}\bar{J}_{\bar{I}_j}\prod_{j}^{n-m_2}\bar{J}_{\bar{J}_j}\rangle\nonumber\\
&\hspace{0.3cm}\cdot\frac{\langle\prod_a^{2g+2n-m_1-m_2-2}G^-_{K3,+}(r_a)\,\prod_k^{m_2+1}c\Xi^{B_k}(t_k)\,\prod_{k}^{m_1+1}\bar{\Xi}^{A_k}(u_k)
\prod_{j}^{n-m_2}ce^{i\sqrt{2}H}(v_j)\rangle_{K3}}{\langle\prod_a^{3g+n}b(r_a)\prod_j^{n-m_2}c(v_j)\prod_k^{m_2+1}c(t_k)\rangle_{bc}}\,.\nonumber
\end{align}
In a final step we can again follow \cite{Antoniadis:2009nv} and commute the positions of the PCO to the Beltrami differentials yielding 
\begin{align}
&\mathcal{H}^{g,A_1\ldots A_{m_1+1},B_1\ldots B_{m_2+1}}_{\bar{I}_1\ldots\bar{I}_{n-m_1},\bar{J}_1\ldots\bar{J}_{n-m_2}}=\frac{1}{(g+m_1+m_2-n)!(2g+2n-m_1-m_2-2)!}\cdot\nonumber\\
&\hspace{1cm}\cdot\int_{\mathcal{M}_{(g,n+1)}}\hspace{-0.7cm}\langle\gtmin^{g+m_1+m_2-n}\gkmpin^{2g+2n-m_1-m_2-2}\prod_{a=1}^{m_1+1}\int \bar{\Xi}^{A_a}\prod_{b=1}^{m_2+1}(\psi_3\Xi^{B_b})\cdot\nonumber\\
&\hspace{1cm}\cdot\prod_{c=1}^{n-m_1}\int\bar{\psi}_3\bar{J}_{\bar{I}_c}\prod_{d=1}^{n-m_2}(J^{++}_{K3}\bar{J}_{\bar{J}_d})\rangle\,.\label{GenericLoopHet}
\end{align}
Here we have introduced numerical normalization factors which will
turn out helpful in reducing unnecessary writing in some of the later
computations. 
Since this topological expression has four different types of
insertions we have compiled the dimensions and $U(1)$ charges for all
of them in 
table~\ref{Tab:ChargeDim} for the reader's convenience.
\begin{table}
\begin{center}
\begin{tabular}{|c|c|c|c|}\hline
\textbf{operator} & \textbf{dimension} & \textbf{charge} & \textbf{statistics} \\\hline
&&&\\[-10pt]
$\bar{\Xi}^A$ & $1$ & $-1$ & fermionic\\
&&&\\[-10pt]\hline 
&&&\\[-10pt]
$\psi_3\Xi^B$ & $0$ & $+2$ & bosonic \\
&&&\\[-10pt]\hline
&&&\\[-10pt]
$\bar{\psi}_3\bar{J}_{\bar{I}}$ & $1$ & $-1$ & fermionic \\
&&&\\[-10pt]\hline
&&&\\[-10pt]
$J^{++}_{K3}\bar{J}_{\bar{J}}$ & $0$ & $+2$ & bosonic \\[5pt]\hline
\end{tabular}
\end{center}
\caption{Dimensions and $U(1)$ charges of the operator insertions in (\ref{GenericLoopHet}). The last column denotes whether the operator is of fermionic or bosonic nature.}
\label{Tab:ChargeDim}
\end{table}
Notice that all insertions with dimension $1$ are integrated over the
world-sheet, while those with dimension $0$ are inserted at fixed
points 
associated to the $n+1$ punctures. We also note that this is not quite
the object anticipated in (\ref{GenTopObject}), however, 
we shall see in section~\ref{Sect:ChiExact} that indeed it is very closely related.
\subsubsection{Notation and Simplifications}
Due to the great number of insertions, (\ref{GenericLoopHet}) is
unfortunately a rather complicated expression. We therefore 
would like to introduce certain simplifications (and shorthand
notations) in order to save writing and moreover also make 
the computations more transparent. Let us first introduce the operator
\begin{align}
G^-=G^-_{T^2}+G^-_{K3,+}\,.\label{shorthand}
\end{align}
Although this operator does not have a well defined harmonic
$U(1)$-charge, there is no actual inconsistency because of the 
following reason. Upon writing (\ref{GenericLoopHet}) with the help of (\ref{shorthand})
\begin{align}
&\mathcal{H}^{g,A_1\ldots A_{m_1+1},B_1\ldots B_{m_2+1}}_{\bar{I}_1\ldots\bar{I}_{n-m_1},\bar{J}_1\ldots\bar{J}_{n-m_2}}=\nonumber\\
&=\frac{1}{(3g+n-2)!}\int_{\mathcal{M}_{(g,n+1)}}\hspace{-0.7cm}\langle
(\mu\cdot G^-)^{3g+n-2}\prod_{a=1}^{m_1+1}\int \bar{\Xi}^{A_a}
\prod_{b=1}^{m_2+1}(\psi_3\Xi^{B_b})\prod_{c=1}^{n-m_1}\int\bar{\psi}_3\bar{J}_{\bar{I}_c}\prod_{d=1}^{n-m_2}(J^{++}_{K3}\bar{J}_{\bar{J}_d})\rangle\,,\label{TopAmpSimple}
\end{align}
it is clear that in order to be able to provide the correct number of
contractions with $\bar{\psi}_3$ (and to soak up the torus zero mode) 
from the $3g+n-2$ operators $G^-$ precisely $(g+m_1+m_2-n)$ have to
contribute the $G^-_{T^2}$ piece. All the remaining ones 
have to contribute the $G^-_{K3,+}$ part which makes sure that 
$\mathcal{H}^{g,A_1\ldots A_{m_1+1},B_1\ldots
  B_{m_2+1}}_{\bar{I}_1\ldots\bar{I}_{n-m_1},\bar{J}_1\ldots\bar{J}_{n-m_2}}$ 
has exactly the correct harmonic charge $2g+2n-m_1-m_2-2$. Therefore,
(\ref{shorthand}) provides us with a 
helpful shorthand notation without loosing any information about the amplitude (\ref{GenericLoopHet}).

Eq.(\ref{TopAmpSimple}) is in fact identical to the one appearing in
\cite{Antoniadis:1996qg} in the 
context of $\cN=1$ heterotic topological amplitudes where there are
$n+1$ charge $(-1)$ operators and $n+1$ 
charge $(+2)$ operators together with the required $(3g+n-2)$ $G^-$
insertions folded with the Beltrami differentials. 
These charge $(-1)$ and $(+2)$ operators correspond to $\cN=1$
anti-chiral matter multiplets (with two different helicities). 
When the $\cN=1$ theory is embedded inside an $\cN=2$ theory, the $\cN=1$
anti chiral multiplets can come from $\cN=2$ vector 
multiplets or the hypermultiplets. In (\ref{TopAmpSimple}), the
$(n+1)$ charge $(-1)$ (or charge $(+2)$) 
operators are correspondingly split into $m_1+1$ (or $m_2+1$) that
come from hypermultiplets and $n-m_1$ 
(or $n-m_2$) that come from vector multiplets. In other words, in
(\ref{TopAmpSimple}), the specific 
features of the $\cN=2$ theory have not been used apart from splitting
the operators in terms of their 
vector or hyper origins. In the next subsection we will define certain
reduced topological amplitudes 
that are obtained from (\ref{TopAmpSimple}) by using the special
properties of the $\cN=2$ theory namely 
the $\cN=4$ world-sheet superconformal structure and as a result they will not have an $\cN=1$ counterpart. 

In \cite{Antoniadis:1996qg}, it was also also pointed out that this
class of amplitudes extends to genus zero surfaces, 
where they correspond to $\Pi^{n+1}$ terms. On genus zero, there must
be a net charge $+3$ and three of the dimension 
zero operators are unintegrated which implies that the number of $G^-$
is $n-2$. It is clear that also in the $\cN=2$ case 
under consideration we will have the genus zero counterpart of (\ref{TopAmpSimple}) with the number of $G^-$ being $n-2$.

To proceed further, it will turn out to be very useful to write
(\ref{TopAmpSimple}) in a totally integrated form. 
To this end, for $g>1$, we will localize the $n+1$ operators $G^-$,
which are sewed with the Beltrami differentials 
corresponding to the punctures of the Riemann surface, with the unintegrated insertions.\footnote{As mentioned before,
for the case of $g=1$ and $g=0$ there must be respectively one and
three unintegrated punctures. This means that in these 
cases we can convert only $n$ and $n-2$ charge $+2$ vertices into
integrated form by localizing the $G^-$ sewed with the 
corresponding Beltrami differentials.} Using the OPE relations
\begin{align}
&\oint G^-J^{++}_{K3}\bar{J}_{\bar{J}}=G^+_{K3,+}\bar{J}_{\bar{J}}\,,\\
&\oint G^-(\psi_3\Xi^{B})=:\Phi^B=\partial X_3\Xi^B-\psi_3\oint G^-_{K3,+}\Xi^B\,,
\end{align}
we can rewrite the topological amplitude (\ref{TopAmpSimple}) as
\begin{align}
&\mathcal{H}^{g,A_1\ldots A_{m_1+1},B_1\ldots B_{m_2+1}}_{\bar{I}_1\ldots\bar{I}_{n-m_1},\bar{J}_1\ldots\bar{J}_{n-m_2}}=\nonumber\\
&=\frac{1}{(3g-3)!}\int_{\mathcal{M}_{g}}\langle (\mu\cdot
G^-)^{3g-3}\prod_{a=1}^{m_1+1}\int \bar{\Xi}^{A_a}\prod_{b=1}^{m_2+1}
\int\Phi^{B_b}\prod_{c=1}^{n-m_1}\int\bar{\psi}_3\bar{J}_{\bar{I}_c}\prod_{d=1}^{n-m_2}\int G^+_{K3,+}\bar{J}_{\bar{J}_d}\rangle\,.\label{TopAmpSimpleInteg}
\end{align}
For convenience we have again listed dimensions and charges of all insertions of this expression in table~\ref{Tab:ChargeDimInteg}.
\begin{table}
\begin{center}
\begin{tabular}{|c|c|c|c|}\hline
\textbf{operator} & \textbf{dimension} & \textbf{charge} & \textbf{statistics} \\\hline
&&&\\[-10pt]
$\bar{\Xi}^A$ & $1$ & $-1$ & fermionic\\
&&&\\[-10pt]\hline 
&&&\\[-10pt]
$\Phi^B$ & $1$ & $+1$ & fermionic \\
&&&\\[-10pt]\hline
&&&\\[-10pt]
$\bar{\psi}_3\bar{J}_{\bar{I}}$ & $1$ & $-1$ & fermionic \\
&&&\\[-10pt]\hline
&&&\\[-10pt]
$G^+_{K3,+}\bar{J}_{\bar{J}}$ & $1$ & $+1$ & fermionic \\[5pt]\hline
\end{tabular}
\end{center}
\caption{Dimensions and $U(1)$ charges of the operator insertions in (\ref{TopAmpSimpleInteg}). The last column denotes whether the operator is of fermionic or bosonic nature.}
\label{Tab:ChargeDimInteg}
\end{table}
Notice that all operators have now dimension $1$ and are integrated
over the world-sheet as it is appropriate. 
Let us also remind again that the operator $\Phi^B$ does not have well
defined harmonic $U(1)$ charge. However, 
just as before, after performing all contractions of the $\psi_3$ free
fermions, only those contributions will 
survive in (\ref{TopAmpSimpleInteg}) which have a well defined charge.
\subsubsection{Symmetry Properties of the Amplitude}\label{Sect:PropertyAmplitude}
For later use we will consider now the symmetry properties of the
topological expression (\ref{GenericLoopHet}). 
First of all, we notice that the latter is anti-symmetric in each set
of indices $A$, $B$, $\bar{I}$ and $\bar{J}$ 
separately. This is manifest in the expression (\ref{GenericLoopHet})
for the indices $A$ and $\bar{I}$ and follows 
from the fermionic properties of the operators $\bar{\Xi}^A$ and
$\bar{\psi}_3$ respectively. As we can see from the 
fully integrated expression (\ref{TopAmpSimpleInteg}), in fact the
same is also true for the indices $B$ and $\bar{J}$ 
as it is expected from the physical component amplitude which we
started from in section~\ref{Sect:GenericAmplitudeString}. 

Although not manifestly visible from the expression
(\ref{TopAmpSimple}), the amplitudes $\mathcal{H}^{g}$ also have 
symmetries upon exchanging, say, $\bar{I}$ with $\bar{J}$ indices. To
see this, we consider the expression (\ref{TopAmpSimpleInteg}) 
with integrated insertions and rewrite for example
\begin{align}
&G^+_{K3,+}\bar{J}_{\bar{J}_{1}}=\oint \left(\partial X_3J^{++}_{K3}-\psi_3G^+_{K3,+}\right)(\bar{\psi}_3\bar{J}_{\bar{J}_{1}})\,.
\end{align}
Deforming the contour integral the only non-vanishing residues are
\begin{align}
&\oint \left(\partial X_3J^{++}_{K3}-\psi_3G^+_{K3,+}\right)\bar{\Xi}^A=\partial X_3\Xi^A-\psi_3\oint G^+_{K3,+}\bar{\Xi}^A=\Phi^A\,,\\
&\oint \left(\partial X_3J^{++}_{K3}-\psi_3G^+_{K3,+}\right)(\bar{\psi}_3\bar{J}_{\bar{I}})=G^+_{K3,+}\bar{J}_{\bar{I}}\,,
\end{align}
such that we immediately get
\begin{align}
&\mathcal{H}^{g,A_1\ldots A_{m_1+1},B_1\ldots B_{m_2+1}}_{\bar{I}_1\ldots\bar{I}_{n-m_1},\bar{J}_1\ldots\bar{J}_{n-m_2}}=\frac{1}{(3g-3)!}\int_{\mathcal{M}_{g}}
\bigg\langle (\mu\cdot G^-)^{3g-3}\prod_{a=1}^{m_1}\int \bar{\Xi}^{[A_a}\int \Phi^{A_{m_1+1}]}\prod_{b=1}^{m_2+1}\int\Phi^{B_b}\cdot\nonumber\\
&\hspace{0.5cm}\cdot
\prod_{c=1}^{n-m_1}\int\bar{\psi}_3\bar{J}_{\bar{I}_c} \int
\bar{\psi}_3\bar{J}_{\bar{J}_1}\prod_{d=2}^{n-m_2} 
\int G^+_{K3,+}\bar{J}_{\bar{J}_d}\bigg\rangle+\frac{1}{(3g-3)!}\int_{\mathcal{M}_{g}}\bigg\langle (\mu\cdot G^-)^{3g-3}\prod_{a=1}^{m_1+1}\int \bar{\Xi}^{A_a}\cdot\nonumber\\
&\hspace{0.5cm}\cdot\prod_{b=1}^{m_2+1}\int\Phi^{B_b}\prod_{c=1}^{n-m_1-1}\int\bar{\psi}_3\bar{J}_{\bar{I}_c}\int
\bar{\psi}_3\bar{J}_{\bar{J}_1}\int
G^+_{K3,+}\bar{J}_{\bar{I}_{n-m_1}}
\prod_{d=2}^{n-m_2}\int G^+_{K3,+}\bar{J}_{\bar{J}_d}\bigg\rangle\,.
\end{align}
We can rewrite the right hand side of this equation to obtain
\begin{align}
\mathcal{H}^{g,A_1\ldots A_{m_1+1},B_1\ldots B_{m_2+1}}_{\bar{I}_1\ldots\bar{I}_{n-m_1},\bar{J}_1\ldots\bar{J}_{n-m_2}}=
\mathcal{H}^{g,A_1\ldots A_{m_1},A_{m_1+1}B_1\ldots B_{m_2+1}}_{\bar{I}_1\ldots\bar{I}_{n-m_1}\bar{J}_1,\bar{J}_2\ldots\bar{J}_{n-m_2}}+
\mathcal{H}^{g,A_1\ldots A_{m_1+1},B_1\ldots B_{m_2+1}}_{\bar{I}_1\ldots\bar{I}_{n-m_1-1}\bar{J}_1,\bar{I}_{n-m_1}\bar{J}_2\ldots\bar{J}_{n-m_2}}\,.\label{SymmRel}
\end{align}
Notice that if we formally combine the indices $(\bar{I},A)$ into a new (multi) index $M$ and similarly $(\bar{J},B)$ into $N$, the relation is simply the statement
\begin{align}\label{SymmRel2}
\mathcal{H}^g_{[M_1\ldots M_n,N_1]\ldots N_n}=0\,,
\end{align}
which is the direct generalization of identity (3.7) in \cite{Antoniadis:1996qg}.
\subsubsection{Generating Functional}\label{Sect:GeneratingFunctional}
In view of the symmetry properties derived in the previous section, we can
introduce a shorthand notation which will allow us to 
greatly simplify many of the following computations. Indeed, observing
the particular fermionic nature of all insertions in 
table~\ref{Tab:ChargeDimInteg} it seems appropriate to introduce four
types of Grassmann valued quantities 
$(\grass ^{\bar{I}},\eta^{\bar{J}},\chi_A,\xi_B)$ and to define the
following generating functional --- just like in 
\cite{Antoniadis:1996qg}
\begin{align}
\mathbb{H}^g:=\sum_{n=0}^\infty\sum_{m_1=0}^n\sum_{m_2=0}^n&\frac{\grass ^{\bar{I}_1}\ldots\grass ^{\bar{I}_{n-m_1}}\chi_{A_1}\ldots 
\chi_{A_{m_1+1}}\eta^{\bar{J}_1}\ldots\eta^{\bar{J}_{n-m_2}}\xi_{B_1}\ldots \xi_{B_{m_2+1}}}{(n-m_1)!(n-m_2)!(m_1+1)!(m_2+1)!}\cdot\nonumber\\
&\cdot\mathcal{H}^{g,A_1\ldots A_{m_1+1},B_1\ldots B_{m_2+1}}_{\bar{I}_1\ldots\bar{I}_{n-m_1},\bar{J}_1\ldots\bar{J}_{n-m_2}}\,.\label{GeneratingFunct}
\end{align}
In this form, every summand in $\mathbb{H}^g$ can be thought of as a
differential form with respect to 
$(\grass ^{\bar{I}},\eta^{\bar{J}},\chi_A,\xi_B)$ of degree
$(n-m_1,n-m_2,m_1+1,m_2+1)$. In order to also capture the particular
case $m_2=0$, 
which will be discussed in more detail in
section~\ref{Sect:PartCasem20}, let us also introduce a reduced
version of $\mathbb{H}^g$. 
By restricting to the term $m_2=0$ in the summation of (\ref{GeneratingFunct}) we can define
\begin{align}
\mathbb{H}^g_\text{red}:=\sum_{n=0}^\infty\sum_{m_1=0}^n&\frac{\grass ^{\bar{I}_1}\ldots\grass ^{\bar{I}_{n-m_1}}\chi_{A_1}
\ldots \chi_{A_{m_1+1}}\eta^{\bar{J}_1}\ldots\eta^{\bar{J}_{n}}\xi_{B}}{(n-m_1)!\,n!\,(m_1+1)!}\,\mathcal{H}^{g,A_1\ldots A_{m_1+1},B}_{\bar{I}_1\ldots\bar{I}_{n-m_1},\bar{J}_1\ldots\bar{J}_{n}}\,.\label{GeneratingFunctRed}
\end{align}
Notice that every summand of this expression can be interpreted as a differential form with respect to $(\grass ^{\bar{I}},\eta^{\bar{J}},\chi_A,\xi_B)$ of degree $(n-m_1,n,m_1+1,1)$.

In order to demonstrate the usefulness of this notation let us just note that the symmetry property (\ref{SymmRel}) can be written more compactly as 
\begin{align}
\left[\grass ^{\bar{I}}\frac{\partial}{\partial\eta^{\bar{I}}}+\chi_A\frac{\partial}{\partial\xi_A}\right]\mathbb{H}^g=0\,,&&\text{and} 
&&\left[\grass ^{\bar{I}}\frac{\partial}{\partial\eta^{\bar{I}}}+\chi_A\frac{\partial}{\partial\xi_A}\right]\mathbb{H}^g_{\text{red}}=0\,.
\end{align}
Here we have used the fact that (\ref{SymmRel}) is valid for every
single summand in the sum over $m_2$ in (\ref{GeneratingFunct}). 
Therefore it follows that the equation holds for $\mathbb{H}^g$ as well as for the reduced generating functional (\ref{GeneratingFunctRed}). 
\subsection{$\chi$-Exactness and Potential $\tilde{\mathcal{H}}$}\label{Sect:ChiExact}
An observation which will turn out important for the remainder of this
work is the fact that $\mathbb{H}^g$ is in a certain sense 
an exact form with respect to $\chi_A$. A first hint towards this is to realize that $\mathbb{H}^g$ is closed, i.e.
\begin{align}
\chi_A\mathcal{D}^A_+\,\mathbb{H}^g=0\,.\label{chiExactH}
\end{align}
To see this, we consider the anti-symmetrized derivative
\begin{align}
&\mathcal{D}_+^{[A_{1}}\mathcal{H}^{g,A_2\ldots A_{m_1+2}],B_1\ldots B_{m_2+1}}_{\bar{I}_1\ldots\bar{I}_{n-m_1},\bar{J}_1\ldots\bar{J}_{n-m_2}}=\frac{1}{(g+m_1+m_2-n)!(2g+2n-m_1-m_2-2)!}\cdot\nonumber\\
&\hspace{0.5cm}\cdot\,\int_{\mathcal{M}_{(g,n+1)}}\hspace{-0.7cm}\langle\gtmin^{g+m_1+m_2-n}\gkmpin^{2g+2n-m_1-m_2-2}\int \oint G^+_{K3,+}\bar{\Xi}^{[A_1}\prod_{a=2}^{m_1+2}\int \bar{\Xi}^{A_a]}\cdot\nonumber\\
&\hspace{0.5cm}\cdot\prod_{b=1}^{m_2+1}(\psi_3\Xi^{B_b})\prod_{c=1}^{n-m_1}\int\bar{\psi}_3\bar{J}_{\bar{I}_c}\prod_{d=1}^{n-m_2}(J^{++}_{K3}\bar{J}_{\bar{J}_d})\rangle\,.
\end{align}
Deforming the contour integral $\oint G^+_{K3,+}$, however, it
produces no pole with any of the other operators and therefore
vanishes. 
This proves relation (\ref{chiExactH}), however, we can derive an even
stronger statement by writing $G^-_{K3,+}=\oint G^+_{K3,+}J^{--}_{K3}$
in 
(\ref{GenericLoopHet}) and deforming the contour integration
\begin{align}
&\mathcal{H}^{g,A_1\ldots A_{m_1+1},B_1\ldots B_{m_2+1}}_{\bar{I}_1\ldots\bar{I}_{n-m_1},\bar{J}_1\ldots\bar{J}_{n-m_2}}=-\frac{1}{(g+m_1+m_2-n)!(2g+2n-m_1-m_2-2)!}\cdot\nonumber\\
&\cdot \int_{\mathcal{M}_{(g,n+1)}}\hspace{-0.7cm}\langle\gtmin^{g+m_1+m_2-n}\gkmpin^{2g+2n-m_1-m_2-3}(\mu\cdot J^{--}_{K3})\int\oint  G^+_{K3,+}\bar{\Xi}^{[A_1}\cdot\nonumber\\
&\hspace{1cm}\cdot\prod_{a=2}^{m_1+1}\int \bar{\Xi}^{A_a]}\prod_{b=1}^{m_2+1}(\psi_3\Xi^{B_b})\prod_{c=1}^{n-m_1}\int\bar{\psi}_3\bar{J}_{\bar{I}_c}\prod_{d=1}^{n-m_2}(J^{++}_{K3}\bar{J}_{\bar{J}_d})\rangle\,.
\end{align}
The insertion of $\int\oint G^+_{K3,+}\bar{\Xi}^{A_{1}}$ corresponds
to a derivative with respect to $\mathcal{D}_+^{A_{1}}$ which we can
pull out. Writing the 
remaining expression in terms of the $G^-$ we find 
\begin{align}
&\mathcal{H}^{g,A_1\ldots A_{m_1+1},B_1\ldots B_{m_2+1}}_{\bar{I}_1\ldots\bar{I}_{n-m_1},\bar{J}_1\ldots\bar{J}_{n-m_2}}=-\frac{1}{(2g+2n-m_1-m_2-2)(3g+n-3)!}\,\mathcal{D}^{[A_{1}}_+
\int_{\mathcal{M}_{(g,n+1)}}\hspace{-0.7cm}\langle(\mu\cdot G^-)^{3g+n-3}\cdot\nonumber\\
&\hspace{1cm}\cdot(\mu\cdot J^{--}_{K3})\prod_{a=2}^{m_1+1}\int
\bar{\Xi}^{A_a]}\prod_{b=1}^{m_2+1}(\psi_3\Xi^{B_b})
\prod_{c=1}^{n-m_1}
\int\bar{\psi}_3\bar{J}_{\bar{I}_c}\prod_{d=1}^{n-m_2}(J^{++}_{K3}\bar{J}_{\bar{J}_d})\rangle\,.
\end{align}
Upon introducing the quantity (which is indeed exactly the object we had anticipated in (\ref{GenTopObject}))
\begin{align}
&\tilde{\mathcal{H}}^{g,A_1\ldots A_{m_1},B_1\ldots B_{m_2+1}}_{\bar{I}_1\ldots\bar{I}_{n-m_1},\bar{J}_1\ldots\bar{J}_{n-m_2}}:=\frac{1}{(3g+n-3)!}
\cdot\int_{\mathcal{M}_{(g,n+1)}}\langle(\mu\cdot G^-)^{3g+n-3}(\mu\cdot J^{--}_{K3})\prod_{a=1}^{m_1}\int \bar{\Xi}^{A_a}\cdot\nonumber\\
&\hspace{1cm}\cdot\prod_{b=1}^{m_2+1}(\psi_3\Xi^{B_b}) \prod_{c=1}^{n-m_1}\int\bar{\psi}_3\bar{J}_{\bar{I}_c}\prod_{d=1}^{n-m_2}J^{++}_{K3}\bar{J}_{\bar{J}_d}\rangle\,.\label{DefTildeH}
\end{align}
we obtain the relation 
\begin{align}
-(2g+2n-m_1-m_2-2)\,\mathcal{H}^{g,A_1\ldots A_{m_1+1},B_1\ldots B_{m_2+1}}_{\bar{I}_1\ldots\bar{I}_{n-m_1},\bar{J}_1\ldots\bar{J}_{n-m_2}}=\mathcal{D}^{[A_{1}}_+
\tilde{\mathcal{H}}^{g,A_2\ldots A_{m_1+1}],B_1\ldots B_{m_2+1}}_{\bar{I}_1\ldots\bar{I}_{n-m_1},\bar{J}_1\ldots\bar{J}_{n-m_2}}\,.
\end{align}
Comparing, however, to (\ref{GenericLoopHet}) we notice that the
prefactor on the left hand side is identical to the harmonic charge of 
$\mathcal{H}$ (i.e. it counts the total number of $G^+_{K3,+}$ operators). We can thus replace it by 
\begin{align}
D_0\,\mathcal{H}^{g,A_1\ldots A_{m_1+1},B_1\ldots B_{m_2+1}}_{\bar{I}_1\ldots\bar{I}_{n-m_1},\bar{J}_1\ldots\bar{J}_{n-m_2}}=\mathcal{D}^{[A_{1}}_+
\tilde{\mathcal{H}}^{g,A_2\ldots A_{m_1+1}],B_1\ldots B_{m_2+1}}_{\bar{I}_1\ldots\bar{I}_{n-m_1},\bar{J}_1\ldots\bar{J}_{n-m_2}}\,,\label{IntroPotential}
\end{align}
where $D_0$ is given in (\ref{subhd}). Introducing the generating functional
\begin{align}
&\tilde{\mathbb{H}}^g:=\sum_{n=0}^\infty\sum_{{m_1=0}\atop{m_2=0}}^n\frac{\grass ^{\bar{I}_1}\ldots\grass ^{\bar{I}_{n-m_1}}\chi_{A_1}\ldots 
\chi_{A_{m_1}}\eta^{\bar{J}_1}\ldots\eta^{\bar{J}_{n-m_2}}\xi_{B_1}\ldots
\xi_{B_{m_2+1}}}{(n-m_1)!(n-m_2)!(m_1+1)!(m_2+1)!}
\tilde{\mathcal{H}}^{g,A_1\ldots A_{m_1},B_1\ldots B_{m_2+1}}_{\bar{I}_1\ldots\bar{I}_{n-m_1},\bar{J}_1\ldots\bar{J}_{n-m_2}}\,,\nonumber
\end{align}
equation (\ref{IntroPotential}) can be more compactly written in the following form
\begin{align}
D_0\,\mathbb{H}^g=\chi_A\mathcal{D}^A_+\,\tilde{\mathbb{H}}^g\,.\label{RelHtildeH}
\end{align}
Notice that this relation is in a certain sense equivalent to stating that $\mathbb{H}^g$ is an exact form as concerning the $\chi_A$ variables.
One should realize that similar to $\mathbb{H}^g$ in
(\ref{GeneratingFunct}), also every summand of $\tilde{\mathbb{H}}^g$
can be 
interpreted as a differential form in
$(\grass ^{\bar{I}},\eta^{\bar{J}},\chi_A,\xi_B)$. The important
difference, however, is that 
the corresponding degrees are $(n-m_1,n-m_2,m_1,m_2+1)$, i.e. the degree in $\chi_A$ is shifted by one relative to (\ref{GeneratingFunct}).

\subsection{$\xi$-Exactness and the Particular Case $m_2=0$}\label{Sect:GenAmplitudem20}
The discussion of the previous sub-section focused entirely on the
properties of $\mathbb{H}^g$ as a function of the $\chi_A$
variables. Since in the amplitude computation of
section~\ref{Sect:GenericAmplitudeString} the indices $A$ and $B$ (an
therefore also the variables $\chi_A$ and $\xi_B$) enter on the same
footing, one would expect similar properties of $\mathbb{H}^g$ also
with respect to the $\xi_B$ variables. This is essentially the case,
although, as we will discover now, there is a slight subtlety. Let us
again begin by proving that $\mathbb{H}^g$ is closed with respect to $\xi_B$, i.e.
\begin{align}
\xi_B\mathcal{D}^B_+\,\mathbb{H}^g=0\,.\label{xiExactH}
\end{align}
To this end, we consider the anti-symmetrized derivative
\begin{align}
&\mathcal{D}_+^{[B_{1}}\mathcal{H}^{g,A_1\ldots A_{m_1+1},B_2\ldots B_{m_2+2}]}_{\bar{I}_1\ldots\bar{I}_{n-m_1},\bar{J}_1\ldots\bar{J}_{n-m_2}}=
\frac{1}{(3g+n-1)!}\cdot\,\int_{\mathcal{M}_{(g,n+2)}}\hspace{-0.7cm}\langle(\mu\cdot G^-)^{3g+n-1}\prod_{a=1}^{m_1+1}\int \bar{\Xi}^{A_a}\cdot\nonumber\\
&\hspace{1cm}\cdot\Xi^{[B_1}\prod_{b=2}^{m_2+2}(\psi_3\Xi^{B_b]})\prod_{c=1}^{n-m_1}\int\bar{\psi}_3\bar{J}_{\bar{I}_c}\prod_{d=1}^{n-m_2}(J^{++}_{K3}\bar{J}_{\bar{J}_d})\rangle\,.
\end{align}
Here we have inserted the vertex operator in its un-integrated form at
an additional puncture of the genus $g$ Riemann surface and added a
further 
Beltrami differential. Writing now $\Xi^{B_1}=\oint
\bar{\psi}_3(\psi_3\Xi^{B_1})$ and deforming the $\bar{\psi}_3$
contour integral it produces no 
pole with any of the other insertions. Therefore this expression vanishes, which proves equation (\ref{xiExactH}).
\subsubsection{The Case $m_2=0$}\label{Sect:PartCasem20}
In order to understand in how far $\mathcal{H}^g$ is also an exact
form with respect to $\xi_A$ (in the sense of the previous section), 
we will first discuss the particular case $m_2=0$, which has some interesting properties on its own. The corresponding $\mathcal{H}$ takes the form of
\begin{align}
&\mathcal{H}^{g,A_1\ldots  A_{m_1+1},B}_{\bar{I}_1\ldots\bar{I}_{n-m_1},\bar{J}_1\ldots\bar{J}_{n}}=\frac{1}{(g+m_1-n)!(2g+2n-m_1-2)!}\int_{\mathcal{M}_{(g,n+1)}}
\hspace{-0.7cm}\langle(\mu\cdot G^-_{T^2})^{g+m_1-n}\cdot\nonumber\\
&\hspace{1cm}\cdot (\mu\cdot
G^-_{K3,+})^{2g+2n-m_1-2}\psi_3(\alpha)\prod_{a=1}^{m_1+1}\int
\bar{\Xi}^{A_a}\prod_{c=1}^{n-m_1}
\int\bar{\psi}_3\bar{J}_{\bar{I}_c}\prod_{d=1}^{n}(J^{++}_{K3}\bar{J}_{\bar{J}_d})\,\Xi^{B}\rangle\,.\label{TopAmpM20rev}
\end{align}
In writing this expression we have used the fact that the only
left-over $\psi_3$ (originally situated at $t_1$) is not able to
contract with 
any other operator but is forced to soak the torus zero mode. It can
therefore be inserted at an arbitrary position on the world-sheet 
which we call $\alpha$. We can now proceed to transform the $\Xi^B$ insertion into its integrated form by localizing one of the Beltrami differentials 
\begin{align}
&\mathcal{H}^{g,A_1\ldots  A_{m_1+1},B}_{\bar{I}_1\ldots\bar{I}_{n-m_1},\bar{J}_1\ldots\bar{J}_{n}}=\frac{1}{(g+m_1-n)!(2g+2n-m_1-3)!}\int_{\mathcal{M}_{(g,n)}}
\hspace{-0.7cm}\langle(\mu\cdot G^-_{T^2})^{g+m_1-n}\cdot\nonumber\\
&\hspace{1cm}\cdot (\mu\cdot
G^-_{K3,+})^{2g+2n-m_1-3}\psi_3(\alpha)\prod_{a=1}^{m_1+1}\int
\bar{\Xi}^{A_a}\prod_{c=1}^{n-m_1}
\int\bar{\psi}_3\bar{J}_{\bar{I}_c}\prod_{d=1}^{n}(J^{++}_{K3}\bar{J}_{\bar{J}_d})\int\oint G^-_{K3,+}\Xi^{B}\rangle\,.\nonumber
\end{align}
The insertion of $\oint G^-_{K3,+}\Xi^B$ is, however, just a zero picture vertex operator, which we can pull out in the form of a moduli derivative
\begin{align}
\mathcal{H}^{g,A_1\ldots  A_{m_1+1},B}_{\bar{I}_1\ldots\bar{I}_{n-m_1},\bar{J}_1\ldots\bar{J}_{n}}=\mathcal{D}^{B}_+
\hat{\mathcal{H}}^{g,A_1\ldots A_{m_1+1}}_{\bar{I}_1\ldots\bar{I}_{n-m_1},\bar{J}_1\ldots\bar{J}_{n}}\,,\label{HyperDer2}
\end{align}
with the new topological expression 
\begin{align}
&\hat{\mathcal{H}}^{g,A_1\ldots A_{m_1+1}}_{\bar{I}_1\ldots\bar{I}_{n-m_1},\bar{J}_1\ldots\bar{J}_{n}}=\nonumber\\
&=\frac{1}{(3g+n-3)!}\int_{\mathcal{M}_{(g,n)}}\hspace{-0.7cm}\langle(\mu\cdot
G^-)^{3g+n-3}\psi_3(\alpha)
\prod_{a=1}^{m_1+1}\int \bar{\Xi}^{A_a}\prod_{c=1}^{n-m_1}\int\bar{\psi}_3\bar{J}_{\bar{I}_c}\prod_{d=1}^{n}(J^{++}_{K3}\bar{J}_{\bar{J}_d})\rangle=\nonumber\\
&=\frac{1}{(3g-3)!}\int_{\mathcal{M}_{g}}\langle(\mu\cdot
G^-)^{3g-3}\psi_3(\alpha)\prod_{a=1}^{m_1+1}\int
\bar{\Xi}^{A_a}\prod_{c=1}^{n-m_1}\int
\bar{\psi}_3\bar{J}_{\bar{I}_c}\prod_{d=1}^{n}\int (G^+_{K3,+}\bar{J}_{\bar{J}_d})\rangle\,.
\end{align}
We can also find a similar interpretation for equations
(\ref{HyperDer2}) and (\ref{HyperDer3}) involving the reduced
generating functional. 
Indeed, equation (\ref{HyperDer2}) can be rewritten in the following form
\begin{align}
\mathbb{H}^g_{\text{red}}=\xi_A\mathcal{D}^A_+\hat{\mathbb{H}}^g_{\text{red}}
\end{align}
with the newly introduced potential
\begin{align}
&\hat{\mathbb{H}}^g_\text{red}:=\sum_{n=0}^\infty\sum_{m_1=0}^n\frac{\grass ^{\bar{I}_1}\ldots\grass ^{\bar{I}_{n-m_1}}\chi_{A_1}\ldots 
\chi_{A_{m_1+1}}\eta^{\bar{J}_1}\ldots\eta^{\bar{J}_{n}}}{(n-m_1)!\,n!\,(m_1+1)!}\,\mathcal{H}^{g,A_1\ldots A_{m_1+1}}_{\bar{I}_1\ldots\bar{I}_{n-m_1},\bar{J}_1\ldots\bar{J}_{n}}\,.
\end{align}
For completeness, let us also mention that $\mathbb{H}^g_{\text{red}}$
and $\hat{\mathbb{H}}^g_{\text{red}}$ are of course still exact with 
respect to $\chi_A$ in the following sense 
\begin{align}
&D_0\mathbb{H}^g_{\text{red}}=\chi_A\mathcal{D}_+^A\tilde{\mathbb{H}}^g_{\text{red}}=\chi_A\xi_B\mathcal{D}_+^A\mathcal{D}_+^B\,\hat{\tilde{\mathbb{H}}}^g_{\text{red}}\,,\label{Exactnessm20}\\
&D_0\hat{\mathbb{H}}^g_{\text{red}}=\chi_A\mathcal{D}_+^A\hat{\tilde{\mathbb{H}}}^g_{\text{red}}\,,\label{Exactnessm202}
\end{align}
with the corresponding potentials
\begin{align}
&\tilde{\mathbb{H}}^g_{\text{red}}:=\sum_{n=0}^\infty\sum_{m_1=0}^n\frac{\grass^{\bar{I}_1}\ldots\grass^{\bar{I}_{n-m_1}}\chi_{A_1}\ldots 
\chi_{A_{m_1}}\eta^{\bar{J}_1}\ldots\eta^{\bar{J}_{n}}\xi_{B}}{(n-m_1)!\,n!\,(m_1+1)!}\,\tilde{\mathcal{H}}^{g,A_1\ldots A_{m_1},B}_{\bar{I}_1\ldots\bar{I}_{n-m_1},\bar{J}_1\ldots\bar{J}_{n}}\,,\label{GeneratingFunctTildeRed}\\
&\hat{\tilde{\mathbb{H}}}^g_{\text{red}}:=\sum_{n=0}^\infty\sum_{m_1=0}^n\frac{\grass ^{\bar{I}_1}\ldots\grass ^{\bar{I}_{n-m_1}}\chi_{A_1}\ldots
  \chi_{A_{m_1}}\eta^{\bar{J}_1}\ldots\eta^{\bar{J}_{n}}}{(n-m_1)!
\,n!\,(m_1+1)!}\,\hat{\tilde{\mathcal{H}}}^{g,A_1\ldots A_{m_1}}_{\bar{I}_1\ldots\bar{I}_{n-m_1},\bar{J}_1\ldots\bar{J}_{n}}\,,
\end{align}
where we have introduced the topological object
\begin{align}
&\hat{\tilde{\mathcal{H}}}^{g,A_1\ldots  A_{m_1}}_{\bar{I}_1\ldots\bar{I}_{n-m_1},\bar{J}_1\ldots\bar{J}_{n}}=\int_{\mathcal{M}_{(g,n)}}\hspace{-0.5cm}
\langle(\mu\cdot G^-)^{3g+n-4}(\mu\cdot
J^{--}_{K3})\psi_3(\alpha)\prod_{a=1}^{m_1}\int
\bar{\Xi}^{A_a}\prod_{b=1}^{n-m_1}\int\bar{\psi}_3\bar{J}_{\bar{I}_b}
\prod_{c=1}^{n}(J^{++}_{K3}\bar{J}_{\bar{J}_c})\rangle=\nonumber\\
&=\int_{\mathcal{M}_{g}}\langle(\mu\cdot G^-)^{3g-4}(\mu\cdot
J^{--}_{K3})\psi_3(\alpha)\prod_{a=1}^{m_1}\int \bar{\Xi}^{A_a}
\prod_{b=1}^{n-m_1}\int\bar{\psi}_3\bar{J}_{\bar{I}_b}\prod_{c=1}^{n}\int(G^+_{K3,+}\bar{J}_{\bar{J}_c})\rangle\,,\label{M20RedCase}
\end{align}
which satisfies the following identity
\begin{align}
&-(2g+2n-m_1-m_2-3)\,\hat{\mathcal{H}}^{g,A_1\ldots  A_{m_1+1}}_{\bar{I}_1\ldots\bar{I}_{n-m_1},\bar{J}_1\ldots\bar{J}_{n}}=
\mathcal{D}^{[A_1}\hat{\tilde{\mathcal{H}}}^{g,A_2\ldots A_{m_1+1}]}_{\bar{I}_1\ldots\bar{I}_{n-m_1},\bar{J}_1\ldots\bar{J}_{n}}\,.\label{HyperDer3}
\end{align}
The degrees of every single summand in (\ref{Exactnessm20}) are respectively
\begin{align}
&\text{deg}(\tilde{\mathbb{H}}^g_{\text{red}})=(n-m_1,n-m_2,m_1,1)\,,\\
&\text{deg}(\hat{\mathbb{H}}^g_{\text{red}})=(n-m_1,n-m_2,m_1+1,0)\,,\\
&\text{deg}(\hat{\tilde{\mathbb{H}}}^g_{\text{red}})=(n-m_1,n-m_2,m_1,0)\,.
\end{align}
We remark that (\ref{M20RedCase}) is precisely the topological
expression conjectured in (\ref{GenTopObject}). 
Hence we have shown that there is also an actual string amplitude
related to (\ref{M20RedCase}). This raises 
the question what the corresponding term in the string effective
action looks like and whether it is possible 
to extract from it information about the $1/2$-BPS nature of
$\mathcal{H}^{g,A_1\ldots A_{m_1},B_1\ldots  B_{m_2+1}}_{\bar{I}_1\ldots\bar{I}_{n-m_1},\bar{J}_1\ldots\bar{J}_{n-m_2}}$ 
along the lines of \cite{Antoniadis:2009nv,Antoniadis:2006mr,Antoniadis:2007ta}. We will return to this question in section~\ref{Sect:HarmonicDescription}.
\subsubsection{The Case $m_2>0$ and Gauge Freedom}
It is important to point out that all manipulations of the previous
section were only possible for $m_2=0$ since only then the string
theory 
amplitude allows to put the $\psi_3$ insertion at some arbitrary
position. In the case of $m_2>0$, there would be additional
contractions which invalidate these steps. However, independent of its origin in string theory, even for $m_2>0$ we can consider the following object
\begin{align}
&\hat{\mathcal{H}}^{g,A_1\ldots A_{m_1+1},B_1\ldots B_{m_2}}_{\bar{I}_1\ldots\bar{I}_{n-m_1},\bar{J}_1\ldots\bar{J}_{n-m_2}}(\alpha)=\nonumber\\
&\frac{1}{(3g+n-3)!}\int_{\mathcal{M}_{(g,n)}}\hspace{-0.7cm}\langle(\mu\cdot
G^-)^{3g+n-3}\psi_3(\alpha)\prod_{a=1}^{m_1+1}\int
\bar{\Xi}^{A_a}\prod_{b=1}^{n-m_1}
\int\bar{\psi}_3\bar{J}_{\bar{I}_b}\prod_{c=1}^{m_2}(\psi_3\Xi^{B_c})\prod_{d=1}^{n-m_2}(J^{++}_{K3}\bar{J}_{\bar{J}_d})\rangle\,.
\end{align}
Notice in particular that this quantity depends on the position
$\alpha$ (which is a clear indication that it cannot come from a
string theory computation). 
However, as we show in appendix~\ref{App:GaugeFreedomePotential},
changing the position $\alpha$ just amounts to adding an expression
which is a 
derivative with respect to $\xi_B\mathcal{D}_+^B$
\begin{align}
\hat{\mathbb{H}}^g(\alpha)-\hat{\mathbb{H}}^g(\beta)=\xi_B\mathcal{D}_+^B\mathbb{G}^g(\alpha,\beta)\,,\label{GaugeFreedome}
\end{align}
where $\mathbb{G}^g(\alpha,\beta)$ is written in (\ref{GeneratingFunctG}). Therefore, taking an anti-symmetrized derivative of $\hat{\mathcal{H}}$ we in fact obtain
\begin{align}
&\mathcal{D}_+^{[B_1}\hat{\mathcal{H}}^{g,A_1\ldots  A_{m_1+1},B_2\ldots  B_{m_2+1}]}_{\bar{I}_1\ldots\bar{I}_{n-m_1},\bar{J}_1\ldots\bar{J}_{n}}(\alpha)=\frac{1}{(3g+n-2)!}\int_{\mathcal{M}_{(g,n+1)}}
\hspace{-0.7cm}\langle(\mu\cdot G^-)^{3g+n-2}\psi_3(\alpha)\prod_{a=1}^{m_1+1}\int \bar{\Xi}^{A_a}\cdot\nonumber\\
&\hspace{1cm}\cdot\prod_{b=1}^{n-m_1}\int(\bar{\psi}_3\bar{J}_{\bar{I}_b})\,\Xi^{[B_1}\prod_{c=2}^{m_2+1}(\psi_3\Xi^{B_c]})\prod_{d=1}^{n-m_2}(J^{++}_{K3}\bar{J}_{\bar{J}_d})\rangle\,.
\end{align}
Notice that this expression is independent of $\alpha$: According to
(\ref{GaugeFreedome}) changing the position of $\alpha$ results in an
extra contribution which lies in the 
kernel of $\xi_B\mathcal{D}_+^B$. Thus we can put $\alpha$ at the position of $\Xi^{B_1}$, such that 
\begin{align}
&\mathcal{H}^{g,A_1\ldots A_{m_1+1},B_1\ldots  B_{m_2+1}}_{\bar{I}_1\ldots\bar{I}_{n-m_1},\bar{J}_1\ldots\bar{J}_{n}}=\mathcal{D}_+^{[B_1}
\hat{\mathcal{H}}^{g,A_1\ldots A_{m_1+1},B_2\ldots B_{m_2+1}]}_{\bar{I}_1\ldots\bar{I}_{n-m_1},\bar{J}_1\ldots\bar{J}_{n}}(\alpha)\,,
\end{align}
which in the language of the generating functionals is written as
\begin{align}
\mathbb{H}^g=\xi_B\mathcal{D}_+^B\,\hat{\mathbb{H}}^g(\alpha)\,.
\end{align}
Therefore, $\mathbb{H}^g$ is also an exact form with respect to $\xi$, however, the corresponding potential has no nice interpretation within the framework of BPS-saturated string amplitudes. We note in passing that the subtle difference between $\chi_A$ and $\xi_A$ is due to the particular choice of setup (and gauge fixing condition) in section~\ref{Sect:GenericAmplitudeString}, where we have computed the topological amplitude.
\section{Harmonic description and Effective Action Coupling}\label{Sect:HarmonicDescription}
\setcounter{equation}{0}
In the previous Section we have considered particular amplitudes in
heterotic string theory which are captured by correlation functions in
a twisted two-dimensional theory. 
In this Section we would like to understand which terms in the
heterotic effective action these amplitudes correspond to and whether
they have any interesting properties 
with respect to their moduli dependence. It turns out that the
effective action is best formulated in $\cN=2$ harmonic superspace
\cite{Galperin:1984av}, for which we 
have reviewed our conventions and discussed the relevant Grassmann analytic on-shell superfields in appendix~\ref{App:HarmSuperspace}.

Using the G-analytic superfields (\ref{01}) and (\ref{17}), we now
want to construct effective action couplings which correspond to the
superstring amplitudes 
computed in section~\ref{Sect:GenericAmplitudeString}. As it turns
out, just as the amplitudes themselves were generalizations of the
expressions in \cite{Antoniadis:2009nv}, 
also the corresponding effective action couplings will be a
generalization of those discussed in \cite{Antoniadis:2009nv}. The
coupling discussed in the 
latter work was of the form
\begin{align}
S = \gameas\ (K_{-} \cdot K_{-})^{g}\, F^{-2(g-2)}(q^{+}_{{\hat A}a},u)\ , \label{19'}
\end{align}
where we have introduced the G-analytic superspace measure
\begin{align}
\gameas:=\int \ d^4x \ du\ (\bar{D}_+\cdot \bar{D}_+) (D^-\cdot D^-)=\int \ d^4x \ du\ d^2\q^+ d^2\bq_{-}\,,\label{IntegMeasure}
\end{align}
$K_{-} \cdot K_{-} \equiv K_{-}^\a\ep_{\a\b} K_{-}^\b$, and  ${\hat A}=1,\ldots ,n$ is an $SO(n)$ vector index
labeling the coordinates of the coset of physical scalars (for a
discussion see \cite{Antoniadis:2009nv}), while $a={1,2}$ is an
$SU(2)$ index. 
For a closer description of these superfields see
appendix~\ref{App:Superfields}. Notice finally, the superfield
$K_{-}^\a$ being a fermion, 
one needs a gauge group of sufficiently high rank, so that $(K_{-}^\a\ep_{\alpha\beta} K_{-}^\b)^{g} \neq 0$. We will assume that this requirement is met.

We can generalize (\ref{19'}) in various ways. The most general is to
allow the coupling function to depend on holomorphic $W_I$ and
anti-holomorphic 
vector multiplets $\bar{W}_{\bar{I}}$ as well as hypermultiplets of
both analyticities i.e. $(q^+,\tilde q_-) \leftrightarrow q^+_a$
and $(q^-,\tilde q_+) \leftrightarrow q^-_a$, $a=1,2$. 
To make this coupling manifestly G-analytic, however,  we need to project it with four spinor derivatives
\begin{align}
(D_- \cdot D_-) (\bar D^+\cdot \bar D^+)\ F^{-2(g+1)}(q^+,q^-,W, \bar W;u)\,, \label{gen4}
\end{align}
where we have a great number of possibilities to distribute the
derivatives, as we shall encounter in
section~\ref{Sect:FieldTheoryEqu}. Note also 
that the $U(1)$ charge of the function $F$ has changed, to compensate
for the extra charges brought along by the derivatives. In addition to
(\ref{gen4}) 
we can also put a number of fermionic factors such that the most generic coupling takes the form
\begin{align}
S = \gameas& \ (K_{-} \cdot K_{-})^{g-1}\, (D_- \cdot D_-) (\bar D^+\cdot \bar D^+)\bigg[(\bar\Psi_{A(+)} \bar K^{+\bar J}_{(-)})^{m_1} (\bar\Psi_{B(-)} \bar K^{+\bar I}_{(+)})^{m_2}\cdot\nonumber\\
& \cdot(\bar K^{+\bar I}_{(+)} \bar K^{+\bar J}_{(-)})^{n-m_1-m_2-1} H^{(2-d)A_1 \cdots A_{m_1},B_1 \cdots B_{m_2}}_{\bar I_1 \cdots \bar I_{n-m_1-1}, \bar J_1 \cdots \bar J_{n-m_2-1}}(q^+,q^-,W, \bar W; u)\bigg]\,,\label{gen44}
\end{align}
where $d=2g+2n-m_1-m_2-2$ and $2-d$ is the total harmonic charge of
the function $F$. As usual, the operator $(D_- \cdot D_-) (\bar
D^+\cdot \bar D^+)$ is the 
projector on the G-analytic superspace, in order to match the
properties of the measure. Moreover, the newly introduced $\bar\Psi$
denotes the antichiral hyperino. 
Since we now have both types of hypermultiplets, G-analytic and anti-analytic, we can obtain the same hyperino in two ways,
\begin{align}
    \bar\Psi_\da = \bar D^+_\da\ q^- \quad {\rm or} \quad \bar\Psi_\da = \bar D^-_\da\ q^+\,,&&\da=(\pm)\label{hyp}
\end{align}
(both superfields start with the same spinor component). The labels
$(\pm)$ in \p{gen44} indicate the helicities of the corresponding
spinors. In addition, in 
order to match the string theory computation of the previous section,
we make a distinction between the labels $A,B$ of $\bar\Psi$ and $\bar
I, \bar J$ of $\bar K$, 
in association with the helicity, to indicate their symmetry
properties. The origin of this helicity splitting of the coupling
\p{gen44} can be understood as follows. 
Let us start by denoting the two antichiral fermions collectively as
\begin{equation}\label{coll}
    \Phi^M_\da =(\bar\Psi_{A\da}, \bar K^{+ \bar I}_\da)\,.
\end{equation}
Then we can write the expression in the square brackets in \p{gen44} in the form
\begin{equation}\label{symb}
    (\Phi^{M_1}\cdot \Phi^{M_2})\ldots (\Phi^{M_{n-2}}\cdot \Phi^{M_{n-1}})\ F_{M_1 M_2 \ldots M_{n-2} M_{n-1}}\,,
\end{equation}
where 
\begin{equation}\label{not}
    (\Phi^{M}\cdot \Phi^{N})= (\Phi^{N}\cdot \Phi^{M})= \ep^{\da\db}\Phi^{M}_\da  \Phi^{N}_\db = \Phi^{M}_{(+)} \Phi^{N}_{(-)} - \Phi^{M}_{(-)} \Phi^{N}_{(+)}
\end{equation}
is the Lorentz invariant contraction of two spinors. Here $(\pm)$
denotes the two projections of the spinor index $\da =((+),(-))$, or
two helicities. 
Note that the expression in \p{symb} must have a total harmonic charge $-2g+2$ to match the charge of the first line in \p{gen44}.

The coefficient function $H$ in \p{symb} has two obvious
symmetries. One of them reflects the $M \leftrightarrow N$ symmetry of
the contraction in \p{not} 
(recall that $\Phi$ are fermions), the other corresponds to swapping
two such contractions $(\Phi^{M_{k-1}}\cdot \Phi^{M_k})$ and
$(\Phi^{M_{l-1}}\cdot \Phi^{M_l})$. 
A third, less obvious symmetry is due to the two-component spinor cyclic identity
\begin{equation}\label{cyc}
    (\Phi^{M}\cdot \Phi^{N}) \Phi^K_\da + (\Phi^{N}\cdot \Phi^{K}) \Phi^M_\da + (\Phi^{K}\cdot \Phi^{M}) \Phi^N_\da = 0\,.
\end{equation}
These symmetries do not make the rank $(n-1)$ tensor $H_{M_1 M_2
  \ldots M_{n-2} M_{n-1}}$ fully irreducible. The reduction is
achieved by breaking up the 
contractions \p{not} in \p{symb} into pieces, first according to the
two ingredients $\bar\Psi, \bar K$ in \p{coll}, and then according to
helicity.  
At the first step we get three types of contractions,
\begin{equation}\label{3types}
    (\bar\Psi\cdot\bar\Psi)^a\ (\bar\Psi\cdot\bar K^+)^b\ (\bar K^+ \cdot \bar K^+)^c\ F^{(d)}\,, \qquad  a+b+c=n-1\,,
\end{equation}
where $2n-2$ is the total number of fermions and $d=-2g+2-b-2c$ is the harmonic charge of the function $H$.
Assuming that $a \leq c$, i.e. $d \leq 3-2g-n$, and using the identity
\p{cyc}, we can rewrite all contractions of the first type as
contractions 
of the second type. So, from now on we set $a=0$, $b+c=n-1$. Further,
we split each contraction into helicity states according to \p{not},
and 
expand the resulting binomials. This gives rise to a number of terms
of the type in the square brackets in \p{gen44}, with $b=m_1+m_2$ and
$c=n-m_1-m_2-1$. 
In the process, each term of the new type picks a particular
irreducible projection of the tensor $H_{M_1 M_2 \ldots M_{n-2}
  M_{n-1}}$. As an example, consider 
the following simple case:
\begin{equation}\label{exa}
    (\bar\Psi_A\cdot\bar K^{+\bar I})(\bar\Psi_B\cdot\bar K^{+\bar J})\ H^{AB}_{\bar I \bar J}\,.
\end{equation}
The function $H$ is symmetric under the exchange of pairs of indices
${}^{\bar I}_A$ and ${}^{\bar J}_B$. This symmetry can be realized in
two ways, 
$H_{(\bar I \bar J)}^{(AB)}$ and $H_{[\bar I \bar J]}^{[AB]}$, each of which is associated with a particular helicity distribution:
\begin{eqnarray}
&&\left[ (\bar\Psi_{A(+)} \bar K^{+\bar I}_{(-)})(\bar\Psi_{B(-)} \bar K^{+\bar J}_{(+)}) + (\bar\Psi_{B(+)} \bar K^{+\bar J}_{(-)})(\bar\Psi_{A(-)} 
\bar K^{+\bar I}_{(+)}) \right] H_{(\bar I \bar J)}^{(AB)} \nn \\
&&  \left[ (\bar\Psi_{A(+)} \bar K^{+\bar I}_{(-)})(\bar\Psi_{B(+)} \bar K^{+\bar J}_{(-)}) + (\bar\Psi_{A(-)} \bar K^{+\bar I}_{(+)})(\bar\Psi_{B(-)} \bar K^{+\bar J}_{(+)}) \right] H_{[\bar I \bar J]}^{[AB]}\,.
\end{eqnarray}
\section{Differential Equations from String Theory for $m_2>0$}\label{Sect:AnomalyEquation}
\setcounter{equation}{0}
In this section we will derive differential equations for the
amplitudes $\mathcal{H}^{g,A_1\ldots A_{m_1+1},B_1\ldots  B_{m_2+1}}_{\bar{I}_1\ldots\bar{I}_{n-m_1},\bar{J}_1\ldots\bar{J}_{n-m_2}}$
in 
(\ref{GenericLoopHet}) and its 'potential'
$\tilde{\mathcal{H}}^{g,A_1\ldots A_{m_1},B_1\ldots  B_{m_2+1}}_{\bar{I}_1\ldots\bar{I}_{n-m_1},\bar{J}_1\ldots\bar{J}_{n-m_2}}$ 
with respect to the vector multiplet and hypermultiplet moduli. All
the equations in this section will be true only modulo possible
contact terms that could give rise to moduli space curvature terms. In
fact we have not been able to find the structure of the contact terms
in these equations 
that would satisfy the integrability conditions in the most generic setting. Violation of the
integrability conditions therefore turns out to be proportional to curvature
terms. Even so the fact that the equations are integrable in the flat
limit is highly non-trivial and this will be proven in Section
(4.3). However in Section 5 we will obtain differential equations for
the reduced topological amplitudes $\hat{\tilde{\mathcal{H}}}$ and
$\hat{\mathcal{H}}$ where we have been able to get the correct contact
terms so that the resulting equations are exactly integrable even in
the the case of general curved moduli spaces. 

Based
on the relations found in \cite{Antoniadis:2009nv}, 
we start out be applying the 'harmonicity operator' to
$\tilde{\mathbb{H}}^g$. As we will see, this equation will lead us 
directly to further interesting equations. In particular, since
$\tilde{\mathbb{H}}^g$ is related to $\mathbb{H}^g$, application 
of a projected derivative $\chi_A\mathcal{D}_+^A$ will yield a
(weaker) condition for $\mathbb{H}^g$.\footnote{Notice, since the kernel of $\chi_A\mathcal{D}_+^A$ is non-trivial, the equation for $\mathbb{H}^g$ follows from the equation for $\tilde{\mathbb{H}}^g$ but not the other way round.}
\subsection{Differential Equation for $\tilde{\mathbb{H}}^g$}\label{Sect:DiffEqTildeH}
We will start by applying the harmonicity operator to the
'$\chi$-potential' $\tilde{\mathcal{H}}^{g,A_1\ldots A_{m_1},B_1\ldots  B_{m_2+1}}_{\bar{I}_1\ldots\bar{I}_{n-m_1},\bar{J}_1\ldots\bar{J}_{n-m_2}}$. 
The covariant form of the harmonicity operator was given in \cite{Antoniadis:2009nv} to be of the form
\begin{align}
\nabla^A_-=-\mathcal{D}_-^AD_0-\mathcal{D}_+^A{D_-}^+\,.
\end{align}
However, since $\tilde{\mathcal{H}}^{g,A_1\ldots A_{m_1},B_1\ldots  B_{m_2+1}}_{\bar{I}_1\ldots\bar{I}_{n-m_1},\bar{J}_1\ldots\bar{J}_{n-m_2}}$
only depends on 
$\bar{u}^i_+$, we can also use the more practical expression involving --- strictly speaking illegal --- partial harmonic derivatives
\begin{align}
\nabla^A_-=\epsilon^{ij}\frac{\partial}{\partial\bar{u}^i_+}\frac{\partial}{\partial f^A_j}\,.\label{harop}
\end{align}
Applying this operator to $\tilde{\mathbb{H}}^g$, we obtain in the
language of the generating functionals the following equation 
(the explicit calculation is rather lengthy and technical and we have therefore relegated it to appendix~\ref{DiffTildeH})
\begin{align}
\left[(-D_0+2)\grass ^{\bar{I}}\mathcal{D}_{\bar{I}}-\chi_A\mathcal{D}_-^AD_0-\chi_A\mathcal{D}_+^A{D_-}^+\right]\tilde{\mathbb{H}}^g=(-D_0+2)\tilde{\mathbb{C}}^{\text{bdy}}\,,\label{DiffEqTildeHD0}
\end{align}
where (as already mentioned in appendix~\ref{DiffTildeH}) we have
converted the prefactor $(2g+2n-m_1-m_2-2)$, which accompanies the
boundary terms, 
into the action of the operator $(-D_0+2)$. We should mention at this
point, that this relation is correct up to possible contact 
terms which stem from two of the punctures coming close
together. Since these terms are notoriously difficult to handle, they
have not 
been considered. We note, however, for completeness, that these terms
will correspond to curvature contributions and are a consequence 
of the fact that the moduli space of the heterotic compactification is not flat.

In fact, it will be much more convenient to slightly rewrite the harmonicity operator by moving the (projected) hyper-derivatives to directly hit $\tilde{\mathbb{H}}^g$
\begin{align}
\left[(-D_0+2)\grass ^{\bar{I}}\mathcal{D}_{\bar{I}}+(-D_0+2)\chi_A\mathcal{D}_-^A-{D_-}^+\chi_A\mathcal{D}_+^A\right]\tilde{\mathbb{H}}^g=(-D_0+2)\tilde{\mathbb{C}}^{\text{bdy}}\,.
\end{align}
Using relation (\ref{RelHtildeH}), we may further write
\begin{align}
(-D_0+2)\left[\grass ^{\bar{I}}\mathcal{D}_{\bar{I}}+\chi_A\mathcal{D}_-^A\right]\tilde{\mathbb{H}}^g+(-D_0+2){D_-}^+\mathbb{H}^g=(-D_0+2)\tilde{\mathbb{C}}^{\text{bdy}}\,.
\end{align}
Notice that here $(-D_0+2)$ acts on the whole equation. Since none of
the topological amplitudes lies in the kernel of $(-D_0+2)$ we can
drop it and formulate the 
\emph{first order} differential equation
\begin{align}
&\mathfrak{D}\tilde{\mathbb{H}}^g+{D_-}^+\mathbb{H}^g=\tilde{\mathbb{C}}^{\text{bdy}}\,,\label{FirstOrderTildeH}
\end{align}
where for later convenience we have introduced the first order differential operator
\begin{align}
\mathfrak{D}:=\grass ^{\bar{I}}\mathcal{D}_{\bar{I}}+\chi_A\mathcal{D}_-^A\,,\label{DefTildeDop}
\end{align}
and $\tilde{\mathbb{C}}^{\text{bdy}}$ as given in Appendix E is
\begin{align}
&\tilde{\mathbb{C}}^{\text{(bdy)}}=\sum_{g_s=0}^{g}\left[\left(\mathcal{D}_+^A\tilde{\mathbb{H}}^{g_s}\right)\Omega_{AB}\left(\frac{\partial\mathbb{H}^{g-g_s}}{\partial\xi_B}\right)-\left(\mathcal{D}_+^A\mathbb{H}^{g_s}\right)\Omega_{AB}\left(\frac{\partial\tilde{\mathbb{H}}^{g-g_s}}{\partial\xi_B}\right)+\right.\nonumber\\
&+\left(\frac{\partial\mathbb{H}^{g_s}}{\partial\chi_A}\right)\Omega_{AB}\left(\frac{\partial\mathbb{H}^{g-g_s}}{\partial\xi_B}\right)+\left.\left(\mathcal{D}_I\tilde{\mathbb{H}}^{g_s}\right)G^{I\bar{J}}\left(\frac{\partial \mathbb{H}^{g-g_s}}{\partial\eta^{\bar{J}}}\right)-\left(\mathcal{D}_I\mathbb{H}^{g_s}\right)G^{I\bar{J}}\left(\frac{\partial \tilde{\mathbb{H}}^{g-g_s}}{\partial\eta^{\bar{J}}}\right)\right]\,.
\end{align}
\subsection{Differential Equation for $\mathbb{H}^g$}\label{Sect:DiffEqH}
After the discussion of $\tilde{\mathbb{H}}^g$ we will now also derive
a differential equation for $\mathbb{H}^g$. We will show that 
combining the equations we have already obtained so far results in a
new non-trivial differential equation for $\mathbb{H}^g$. 
In appendix~\ref{Sect:DiffEqHstring}, we have shown that the same
equation can also be obtained from a direct computation within 
string theory.

We will start from equation (\ref{FirstOrderTildeH}) and act with the
differential $\chi_A\mathcal{D}^A_+$ on it. Inserting the explicit 
form of the boundary contribution (\ref{CbdyGenFunc}) we get
\begin{align}
&\chi_C\mathcal{D}^C_+\left[\mathfrak{D}\tilde{\mathbb{H}}^g+{D_-}^+\mathbb{H}^g\right]=\nonumber\\
&=\chi_C\mathcal{D}^C_+\sum_{g_s=0}^{g}\left[\left(\mathcal{D}_+^A\tilde{\mathbb{H}}^{g_s}\right)\Omega_{AB}
\left(\frac{\partial\mathbb{H}^{g-g_s}}{\partial\xi_B}\right)-\left(\mathcal{D}_+^A\mathbb{H}^{g_s}\right)
\Omega_{AB}\left(\frac{\partial\tilde{\mathbb{H}}^{g-g_s}}{\partial\xi_B}\right)+\right.\nonumber\\
&+\left(\frac{\partial\mathbb{H}^{g_s}}{\partial\chi_A}\right)\Omega_{AB}
\left(\frac{\partial\mathbb{H}^{g-g_s}}{\partial\xi_B}\right)+\left.\left(\mathcal{D}_I
\tilde{\mathbb{H}}^{g_s}\right)G^{I\bar{J}}\left(\frac{\partial \mathbb{H}^{g-g_s}}{\partial\eta^{\bar{J}}}\right)-
\left(\mathcal{D}_I\mathbb{H}^{g_s}\right)G^{I\bar{J}}\left(\frac{\partial \tilde{\mathbb{H}}^{g-g_s}}{\partial\eta^{\bar{J}}}\right)\right]\,.
\end{align}
Our strategy is now to commute the $\chi_C\mathcal{D}^C_+$ through
until it hits either $\mathbb{H}^g$ or 
$\tilde{\mathbb{H}}^g$ and then use either relation
(\ref{chiExactH}) or (\ref{RelHtildeH}). 
Using the definition (\ref{DefTildeDop}) of the operator $\mathfrak{D}$, we find
\begin{align}
&-D_0\grass ^{\bar{I}}\mathcal{D}_{\bar{I}}\mathbb{H}^g-\chi_A\mathcal{D}_-^A(D_0-1)
\mathbb{H}^g+\chi_C\chi_A[\mathcal{D}_+^C,\mathcal{D}_-^A]\tilde{\mathbb{H}}^g-\chi_A
\mathcal{D}_-^A\mathbb{H}^g=\nonumber\\
&\hspace{0.3cm}=\sum_{g_s=0}^{g}D_0\left[\left(\mathcal{D}_+^A\mathbb{H}^{g_s}\right)
\Omega_{AB}\left(\frac{\partial\mathbb{H}^{g-g_s}}{\partial\xi_B}\right)+
\left(\mathcal{D}_I\mathbb{H}^{g_s}\right)G^{I\bar{J}}\left(\frac{\partial \mathbb{H}^{g-g_s}}{\partial\eta^{\bar{J}}}\right)\right]+\nonumber\\
&\hspace{0.8cm}+\sum_{g_s=0}^{g}\chi_C\left[\left([\mathcal{D}_+^C,\mathcal{D}_+^A]\,
\tilde{\mathbb{H}}^{g_s}\right)\Omega_{AB}\left(\frac{\partial\mathbb{H}^{g-g_s}}{\partial\xi_B}\right)-
\left([\mathcal{D}_+^C,\mathcal{D}_+^A]\,\mathbb{H}^{g_s}\right)\Omega_{AB}\left(\frac{\partial\tilde{\mathbb{H}}^{g-g_s}}{\partial\xi_B}\right)\right]\,.
\end{align}
The commutator terms can be computed using (\ref{firstt}) following the outline of quaternionic geometry in appendix~\ref{App:QuatGeom}
\begin{align}
&\chi_C\chi_A[\mathcal{D}_+^C,\mathcal{D}_-^A]=-2\chi^2\, R\, \tz_0\,,&&\text{and} &&\chi_C[\mathcal{D}_+^C,\mathcal{D}_+^A]=2\chi_C\Omega^{CA}\,\tz_{++}\,,
\end{align}
where $\chi^2:=\chi_A\chi_B\Omega^{AB}$ and we have used that the
contribution of the $Sp(n)$ curvature is symmetric in $(AC)$. 
Using moreover the identification $\tz_0=D_0$ and $\tz_{++}={D_+}^-$,
we obtain the simpler relation\footnote{Notice that this identification is not in contradiction to our previous paper \cite{Antoniadis:2009nv}: 
There the charge $\tz_0$ had been constrained by the presence of a
G-analytic integral measure of the harmonic 
superspace to be slightly shifted with respect to $D_0$. In the case
at hand, due to the 
presence of the hypermultiplets with 'wrong' analyticity, the integral
measure is no longer 
analytic and there is no constraint fixing $\tz_0$. Thus we are free to identify the latter with $D_0$.}
\begin{align}
&-D_0\left(\grass ^{\bar{I}}\mathcal{D}_{\bar{I}}+\chi_A\mathcal{D}_-^A\right)\mathbb{H}^g-2\chi^2\,R\,D_0\tilde{\mathbb{H}}^g=\nonumber\\
&\hspace{2cm}=\sum_{g_s=0}^{g}D_0\left[\left(\mathcal{D}_+^A\mathbb{H}^{g_s}\right)\Omega_{AB}
\left(\frac{\partial\mathbb{H}^{g-g_s}}{\partial\xi_B}\right)+\left(\mathcal{D}_I\mathbb{H}^{g_s}\right)G^{I\bar{J}}\left(\frac{\partial \mathbb{H}^{g-g_s}}{\partial\eta^{\bar{J}}}\right)\right]\,.
\end{align}
Here we have also used that
${D_+}^-\,\mathbb{H}^g={D_+}^-\,\tilde{\mathbb{H}}^g=0$, since
$\mathbb{H}^g$ and 
$\tilde{\mathbb{H}}^g$ only depend on $\bar{u}^i_+$.  As we can see, $D_0$ becomes an 
overall operator which can be dropped, thus implying the relation
\begin{align}
&\mathfrak{D}\mathbb{H}^g+2\chi^2\,R\,\tilde{\mathbb{H}}^g=-\sum_{g_s=0}^{g}\left[\left(\mathcal{D}_+^A
\mathbb{H}^{g_s}\right)\Omega_{AB}\left(\frac{\partial\mathbb{H}^{g-g_s}}{\partial\xi_B}\right)+
\left(\mathcal{D}_I\mathbb{H}^{g_s}\right)G^{I\bar{J}}\left(\frac{\partial \mathbb{H}^{g-g_s}}{\partial\eta^{\bar{J}}}\right)\right]\,.\label{DiffEqH}
\end{align}
We would like to point out that this is also a first order
differential equation, just as (\ref{FirstOrderTildeH}). Even though
we have neglected possible contact terms in (\ref{FirstOrderTildeH}),
we have shown the additional curvature dependent term $R$ in (\ref
{DiffEqH}).  This is because the derivation of the equation for $\hat{\mathbb{H}}$
starting from that for $\hat{\tilde{\mathbb{H}}}$ in Section 5, will 
follow the same steps as above and there we will indeed keep track of
all the contact terms correctly.


\subsection{Integrability}\label{Sect:IntegrabilityH}
An important consistency check for our differential equations (in
particular the boundary terms) is integrability. 
For simplicity we will consider here only equation (\ref{DiffEqH}) and
check its self-consistency by computing 
$\mathfrak{D}^2\mathbb{H}^g$. Since the square of the operator $\mathfrak{D}$ is given by
\begin{align}
\mathfrak{D}^2=\grass ^{\bar{I}}\grass ^{\bar{J}}[\mathcal{D}_{\bar{I}},\mathcal{D}_{\bar{J}}]+\chi_A\chi_B[\mathcal{D}_-^A,\mathcal{D}_-^B]=2\chi^2R{D_-}^+\,,
\end{align}
and since in this Section we are ignoring curvature terms, the relation we have to prove is
\begin{align}
0{\begin{array}{c}  \vspace{-0.3cm}\text{\tiny ?} \\ = \\ \vspace{-0.3cm}
\phantom{\text{\tiny  ?}}\end{array}}\mathfrak{D}\left[\,\sum_{g_s=0}^{g}\left(\left(\mathcal{D}_+^A\mathbb{H}^{g_s}\right)\Omega_{AB}\left(\frac{\partial\mathbb{H}^{g-g_s}}
{\partial\xi_B}\right)+\left(\mathcal{D}_I\mathbb{H}^{g_s}\right)G^{I\bar{J}}\left(\frac{\partial \mathbb{H}^{g-g_s}}{\partial\eta^{\bar{J}}}\right)\right)\right]
\end{align}
Ignoring the curvature terms we can further move $\mathfrak{D}$
across various covariant derivatives which results in the following equation.
\begin{align}
0{\begin{array}{c} \vspace{-0.3cm}\text{\tiny ?} \\ = \\ \vspace{-0.3cm}\phantom{\text{\tiny ?}}\end{array}}&\sum_{g_s=0}^{g}\left[\left(\mathcal{D}_+^A\,\mathfrak{D}\,\mathbb{H}^{g_s}\right)\Omega_{AB}\left(\frac{\partial\mathbb{H}^{g-g_s}}{\partial\xi_B}\right)-\left(\mathcal{D}_+^A\mathbb{H}^{g_s}\right)\Omega_{AB}\left(\frac{\partial}{\partial\xi_B}\,\mathfrak{D}\mathbb{H}^{g-g_s}\right)\right]+\nonumber\\
+&\sum_{g_s=0}^g\left[\left(\mathcal{D}_I\,\mathfrak{D}\mathbb{H}^{g_s}\right)G^{I\bar{J}}\left(\frac{\partial \mathbb{H}^{g-g_s}}{\partial\eta^{\bar{J}}}\right)-\left(\mathcal{D}_I\mathbb{H}^{g_s}\right)G^{I\bar{J}}\left(\frac{\partial }{\partial\eta^{\bar{J}}}\,\mathfrak{D}\mathbb{H}^{g-g_s}\right)\right]\,.
\end{align}
In order to further streamline this expression, we will use equation (\ref{DiffEqH})
\begin{align}
0{\begin{array}{c} \vspace{-0.3cm}\text{\tiny ?} \\ = \\ \vspace{-0.3cm}\phantom{\text{\tiny ?}}\end{array}}&\sum_{g_s=0}^{g}\sum_{h_s=0}^{g_s}\mathcal{D}_+^A\left(\mathcal{D}_+^C\mathbb{H}^{h_s}\Omega_{CD}\frac{\partial\mathbb{H}^{g_s-h_s}}{\partial\xi_D}+\left(\mathcal{D}_I\mathbb{H}^{h_s}\right)G^{I\bar{J}}\frac{\partial\mathbb{H}^{g_s-h_s}}{\partial\eta^{\bar{J}}}\right)\Omega_{AB}\left(\frac{\partial\mathbb{H}^{g-g_s}}{\partial\xi_B}\right)\nonumber\\
-&\sum_{g_s=0}^{g}\sum_{h_s=0}^{g-g_s}\left(\mathcal{D}_+^A\mathbb{H}^{g_s}\right)\Omega_{AB}\frac{\partial}{\partial\xi_B}\,\left[\left(\mathcal{D}_+^C\mathbb{H}^{h_s}\Omega_{CD}\frac{\partial\mathbb{H}^{g-g_s-h_s}}{\partial\xi_D}+\left(\mathcal{D}_I\mathbb{H}^{h_s}\right)G^{I\bar{J}}\frac{\partial\mathbb{H}^{g-g_s-h_s}}{\partial\eta^{\bar{J}}}\right)\right]\nonumber\\
+&\sum_{g_s=0}^g\sum_{h_s=0}^{g_s}\left[\mathcal{D}_I\left(\mathcal{D}_+^A\mathbb{H}^{h_s}\Omega_{AB}\frac{\partial\mathbb{H}^{g_s-h_s}}{\partial\xi_B}+\left(\mathcal{D}_J\mathbb{H}^{h_s}\right)G^{J\bar{K}}\frac{\partial\mathbb{H}^{g_s-h_s}}{\partial\eta^{\bar{K}}}\right)G^{I\bar{J}}\left(\frac{\partial \mathbb{H}^{g-g_s}}{\partial\eta^{\bar{J}}}\right)\right]\nonumber\\
-&\sum_{g_s=0}^g\sum_{h_s=0}^{g-g_s}\left[\left(\mathcal{D}_I\mathbb{H}^{g_s}\right)G^{I\bar{J}}\frac{\partial }{\partial\eta^{\bar{J}}}\left(\mathcal{D}_+^A\mathbb{H}^{h_s}\Omega_{AB}\frac{\partial\mathbb{H}^{g-g_s-h_s}}{\partial\xi_B}+\left(\mathcal{D}_J\mathbb{H}^{h_s}\right)G^{J\bar{K}}\frac{\partial\mathbb{H}^{g-g_s-h_s}}{\partial\eta^{\bar{K}}}\right)\right]\,.\nonumber
\end{align}
Explicitly expanding this expression (and using anti-symmetry of $\Omega_{AB}$) we find the following terms
{\allowdisplaybreaks
\begin{align}
0{\begin{array}{c} \vspace{-0.3cm}\text{\tiny ?} \\ = \\ \vspace{-0.3cm}\phantom{\text{\tiny ?}}\end{array}}\sum_{g_s=0}^{g}\sum_{h_s=0}^{g_s}\bigg[&\mathcal{D}_+^C\mathbb{H}^{h_s}\Omega_{CD}\mathcal{D}_+^A\frac{\partial\mathbb{H}^{g_s-h_s}}{\partial\xi_D}\Omega_{AB}\frac{\partial\mathbb{H}^{g-g_s}}{\partial\xi_B}+\mathcal{D}_+^A\mathcal{D}_I\mathbb{H}^{h_s}G^{I\bar{J}}\frac{\partial\mathbb{H}^{g_s-h_s}}{\partial\eta^{\bar{J}}}\Omega_{AB}\frac{\partial\mathbb{H}^{g-g_s}}{\partial\xi_B}+\nonumber\\
+&\mathcal{D}_I\mathbb{H}^{h_s}G^{I\bar{J}}\mathcal{D}_+^A\frac{\partial\mathbb{H}^{g_s-h_s}}{\partial\eta^{\bar{J}}}\Omega_{AB}\frac{\partial\mathbb{H}^{g-g_s}}{\partial\xi_B}+\mathcal{D}_I\mathcal{D}_+^A\mathbb{H}^{h_s}\Omega_{AB}\frac{\partial\mathbb{H}^{g_s-h_s}}{\partial\xi_B}G^{I\bar{J}}\frac{\partial\mathbb{H}^{g-g_s}}{\partial\eta^{\bar{J}}}+\nonumber\\
+&\mathcal{D}_+^A\mathbb{H}^{h_s}\Omega_{AB}\mathcal{D}_I\frac{\partial\mathbb{H}^{g_s-h_s}}{\partial\xi_B}G^{I\bar{J}}\frac{\partial\mathbb{H}^{g-g_s}}{\partial\eta^{\bar{J}}}+\mathcal{D}_J\mathbb{H}^{h_s}G^{J\bar{K}}\mathcal{D}_I\frac{\partial\mathbb{H}^{g_s-h_s}}{\partial\eta^{\bar{K}}}G^{I\bar{J}}\frac{\partial\mathbb{H}^{g-g_s}}{\partial\xi_B}\bigg]-\nonumber\\
-\sum_{g_s=0}^{g}\sum_{h_s=0}^{g-g_s}\bigg[&\mathcal{D}_+^A\mathbb{H}^{g_s}\Omega_{AB}\mathcal{D}_+^C\frac{\partial\mathbb{H}^{h_s}}{\partial\xi_B}\Omega_{CD}\frac{\partial\mathbb{H}^{g-g_s-h_s}}{\partial\xi_D}+\mathcal{D}_+^A\mathbb{H}^{g_s}\Omega_{AB}\mathcal{D}_I\frac{\partial\mathbb{H}^{h_s}}{\partial\xi_B}G^{I\bar{J}}\frac{\partial\mathbb{H}^{g-g_s-h_s}}{\partial\eta^{\bar{J}}}+\nonumber\\
+&\mathcal{D}_+^A\mathbb{H}^{g_s}\Omega_{AB}\mathcal{D}_I\mathbb{H}^{h_s}G^{I\bar{J}}\frac{\partial^2\mathbb{H}^{g-g_s-h_s}}{\partial\xi_B\partial\eta^{\bar{J}}}+\mathcal{D}_I\mathbb{H}^{g_s}G^{I\bar{J}}\mathcal{D}_+^A\frac{\partial\mathbb{H}^{h_s}}{\partial\eta^{\bar{J}}}\Omega_{AB}\frac{\partial\mathbb{H}^{g-g_s-h_s}}{\partial\xi_B}+\nonumber\\
+&\mathcal{D}_I\mathbb{H}^{g_s}G^{I\bar{J}}\mathcal{D}_+^A\mathbb{H}^{h_s}\Omega_{AB}\frac{\partial^2\mathbb{H}^{g-g_s-h_s}}{\partial\eta^{\bar{J}}\partial\xi_B}+\mathcal{D}_I\mathbb{H}^{g_s}G^{I\bar{J}}\mathcal{D}_J\frac{\partial\mathbb{H}^{h_s}}{\partial\eta^{\bar{J}}}G^{J\bar{K}}\frac{\partial\mathbb{H}^{g-g_s-h_s}}{\partial\eta^{\bar{K}}}\bigg]\,.\nonumber
\end{align}}
Indeed, upon reshuffling the sums over the genera, all the terms on the right hand side of this expression mutually cancel, which proves integrability of (\ref{DiffEqH}) (up to possible curvature terms). 
\section{Differential Equations from String Theory for $m_2=0$}\label{Sect:AnomalyEquationM20}
\setcounter{equation}{0}
Since the case $m_2=0$ is of particular interest, we will
discuss the differential equations for
$\hat{\mathbb{H}}^g_{\text{red}}$ and
$\hat{\tilde{\mathbb{H}}}^g_{\text{red}}$ separately 
and show some interesting relations to the earlier work performed in
\cite{Antoniadis:2009nv}. Contrary to the previous Section, here we will keep all the
curvature terms and show that the resulting equations are integrable.
\subsection{Differential Equation for $\hat{\tilde{\mathbb{H}}}^g_{\text{red}}$}\label{Sect:DiffEqHm20}
Generically, the case $m_2=0$ can be directly read off from the differential equations (\ref{DiffEqFin}) and (\ref{GenHarmonicityComponent}).\footnote{Care has to be taken, however, that certain contributions are only non-zero for the underlying surface being a torus or sphere. In this case, some insertions should be understood in the non-integrated form. For a discussion of this subtlety, see e.g.~\cite{Antoniadis:2009nv}.} For completeness, however, we will also work out the equations explicitly for the particular object $\hat{\tilde{\mathcal{H}}}^{g,A_1\ldots A_{m}}_{\bar{I}_1\ldots\bar{I}_{n-m_1},\bar{J}_1\ldots\bar{J}_{n}}$ defined in (\ref{M20RedCase}). Here, in order to save writing, we simply relabel $m_1$ by $m$. The explicit computation is very similar to the case $m_2>0$ and only some salient features are outlined in appendix~\ref{Sect:StringDerhattilde}. The final answer can again be written in the language of the generating functionals
\begin{align}
\left[(-D_0+2)\grass ^{\bar{I}}\mathcal{D}_{\bar{I}}-\chi_A\mathcal{D}_-^AD_0-\chi_A\mathcal{D}_+^A{D_+}^-\right]\hat{\tilde{\mathbb{H}}}^g_{\text{red}}&=(-D_0+2)\hat{\tilde{\mathbb{C}}}^g_{\text{bdy}}\,,
\end{align}
with the boundary contribution as given in (\ref{Bdytildehat}). Following the same steps as in section~\ref{Sect:DiffEqTildeH}, this equation can in fact be rewritten as
\begin{align}
(-D_0+2)\left[\grass ^{\bar{I}}\mathcal{D}_{\bar{I}}+\chi_A\mathcal{D}_-^A\right]\hat{\tilde{\mathbb{H}}}^g_{\text{red}}+(-D_0+2){D_-}^+\hat{\mathbb{H}}^g_{\text{red}}&=(-D_0+2)\hat{\tilde{\mathbb{C}}}^g_{\text{bdy}}\,,
\end{align}
which entails the first order differential equation
\begin{align}
\mathfrak{D}\hat{\tilde{\mathbb{H}}}^g_{\text{red}}+{D_-}^+\hat{\mathbb{H}}^g_{\text{red}}&=\sum_{g_s=1}^{g-1}\left[\mathcal{D}_+^A\hat{\mathbb{H}}^g_{\text{red}}\,\Omega_{AB}\,\mathcal{D}^B_+\hat{\tilde{\mathbb{H}}}^g_{\text{red}}+\left(\frac{\partial\hat{\mathbb{H}}^{g_s}_{\text{red}}}{\partial\chi_A}\right)\Omega_{AB}\mathcal{D}_+^B\hat{\mathbb{H}}^{g-g_s}_{\text{red}}\right]\,.\label{DiffEquhattildeH}
\end{align}
\subsection{Differential Equation for  $\hat{\mathbb{H}}^g_{\text{red}}$}
As in the case $m_2\neq 0$ the equations (\ref{Exactnessm202}) and (\ref{DiffEquhattildeH}) put together, allow us to derive a further equation for $\hat{\mathbb{H}}^g_{\text{red}}$. To this end we act with $\chi_C\mathcal{D}_+^C$ on (\ref{DiffEquhattildeH}). Following the same procedures as in section~\ref{Sect:AnomalyEquation} we obtain
\begin{align}
D_0(\mathfrak{D}\hat{\mathbb{H}}^g_{\text{red}})+2\chi^2 RD_0\hat{\tilde{\mathbb{H}}}^g_{\text{red}}&=-\chi_C\mathcal{D}^C_+\sum_{g_s=1}^{g-1}\left[\mathcal{D}_+^A\hat{\mathbb{H}}^g_{\text{red}}\,\Omega_{AB}\,\mathcal{D}^B_+\hat{\tilde{\mathbb{H}}}^g_{\text{red}}+\left(\frac{\partial\hat{\mathbb{H}}^{g_s}_{\text{red}}}{\partial\chi_A}\right)\Omega_{AB}\mathcal{D}_+^B\hat{\mathbb{H}}^{g-g_s}_{\text{red}}\right]\nonumber\\
&=\sum_{g_s=1}^{g-1}\Bigg[\mathcal{D}_+^A\hat{\mathbb{H}}^g_{\text{red}}\,\Omega_{AB}\,\mathcal{D}^B_+D_0\hat{\mathbb{H}}^g_{\text{red}}-2R\chi_B({D_+}^-\hat{\mathbb{H}}^{g_s}_{\text{red}})\mathcal{D}_+^B\hat{\tilde{\mathbb{H}}}^{g-g_s}_{\text{red}}-\nonumber\\
&\hspace{1cm}-\mathcal{D}_+^A\hat{\mathbb{H}}^{g_s}_{\text{red}}\Omega_{AB}\mathcal{D}_+^B\hat{\mathbb{H}}^{g-g_s}_{\text{red}}-2R\chi_A\left(\frac{\partial\hat{\mathbb{H}}^{g_s}_{\text{red}}}{\partial\chi_A}\right)({D_+}^-\hat{\mathbb{H}}^{g-g_s}_{\text{red}})\Bigg]\,.\nonumber
\end{align}
Using the fact that $\hat{\mathbb{H}}^{g-g_s}_{\text{red}}$ is only a function of $\bar{u}^i_+$, we can simplify this expression
\begin{align}
D_0(\mathfrak{D}\hat{\mathbb{H}}^g_{\text{red}})+2\chi^2 R\,D_0\,\hat{\tilde{\mathbb{H}}}^g_{\text{red}}&=\sum_{g_s=1}^{g-1}\Bigg[\mathcal{D}_+^A\hat{\mathbb{H}}^g_{\text{red}}\,\Omega_{AB}\,(D_0+1)\mathcal{D}^B_+\hat{\mathbb{H}}^g_{\text{red}}-\mathcal{D}_+^A\hat{\mathbb{H}}^{g_s}_{\text{red}}\Omega_{AB}\mathcal{D}_+^B\hat{\mathbb{H}}^{g-g_s}_{\text{red}}\Bigg]\nonumber\\
&=\frac{1}{2}D_0\sum_{g_s=1}^{g-1}\mathcal{D}_+^A\hat{\mathbb{H}}^g_{\text{red}}\,\Omega_{AB}\,\mathcal{D}^B_+\hat{\mathbb{H}}^g_{\text{red}}\,,\nonumber
\end{align}
which implies the differential equation
\begin{align}
\mathfrak{D}\hat{\mathbb{H}}^g_{\text{red}}+2\chi^2 R\,\hat{\tilde{\mathbb{H}}}^g_{\text{red}}=\frac{1}{2}\sum_{g_s=1}^{g-1}\mathcal{D}_+^A\hat{\mathbb{H}}^g_{\text{red}}\,\Omega_{AB}\,\mathcal{D}^B_+\hat{\mathbb{H}}^g_{\text{red}}\,.\label{DiffEqhatH}
\end{align}

\subsection{Comparison to Earlier Results}
From the differential equations derived in the previous subsections we
can make direct contact to similar equations worked out in
\cite{Antoniadis:2009nv}. The first equation derived in the latter
work was a holomorphicity relation with respect to the
vector-multiplet moduli. Translated into our language, this is the
relation\footnote{In fact in \cite{Antoniadis:2009nv} it was 
erroneously stated that there would be additional boundary
contributions on the right hand side of this equation, which would
come from the exchange of 
vector multiplets in the limit when the genus $g$ Riemann surface
factorizes into a genus $g-1$ surface and a torus. However,  this 
contribution should be cancelled by a further exchange of a state of charge
$3$, which gives the same contribution but must appear with opposite sign. 
In fact we have checked that this (highly non-trivial) cancellation is
indeed necessary to guarantee integrability (and thus consistency) of
the differential equations derived in the previous section, as we shall see in the next subsection.}
\begin{align}
\grass ^{\bar{I}}\partial_{\bar{I}}\hat{\tilde{\mathbb{H}}}^g_{\text{red}}=0\,.\label{DiffComp1}
\end{align}
Indeed, this equation is precisely the term in (\ref{DiffEquhattildeH}) which is independent of $\chi_A$
\begin{align}
\mathfrak{D}\hat{\tilde{\mathbb{H}}}^g_{\text{red}}+{D_-}^+\hat{\mathbb{H}}^g_{\text{red}}\bigg|_{\chi_A=0}&=\sum_{g_s=1}^{g-1}\left[\mathcal{D}_+^A\hat{\mathbb{H}}^g_{\text{red}}\,\Omega_{AB}\,\mathcal{D}^B_+\hat{\tilde{\mathbb{H}}}^g_{\text{red}}+\left(\frac{\partial\hat{\mathbb{H}}^{g_s}_{\text{red}}}{\partial\chi_A}\right)\Omega_{AB}\mathcal{D}_+^B\hat{\mathbb{H}}^{g-g_s}_{\text{red}}\right]\bigg|_{\chi_A=0}\,.
\end{align}
Here we have used that $\hat{\mathbb{H}}^g_{\text{red}}$ is at least linear in $\chi_A$ and the only part of $\mathfrak{D}$ which is independent of $\chi_A$ is exactly the vector derivative, as in (\ref{DiffComp1}). 

Moreover, a second order differential equation with respect to the hypermultiplet scalars has been derived in \cite{Antoniadis:2009nv}. In preparation of reproducing this equation, we consider the commutator
{\allowdisplaybreaks
\begin{align}
&\left[\mathfrak{D}\chi_A\mathcal{D}_+^A-\chi_A\mathcal{D}_+^A\mathfrak{D}\right]\hat{\tilde{\mathbb{H}}}^g_{\text{red}}=\mathfrak{D}D_0\hat{\mathbb{H}}^g_{\text{red}}+\chi_A\mathcal{D}_+^A({D_-}^+\hat{\mathbb{H}}^g_{\text{red}})-\nonumber\\*
&\hspace{5cm}-\chi_C\mathcal{D}_+^C\sum_{g_s=1}^{g-1}\left[\mathcal{D}_+^A\hat{\mathbb{H}}^g_{\text{red}}\,\Omega_{AB}\,\mathcal{D}^B_+\hat{\tilde{\mathbb{H}}}^g_{\text{red}}+\left(\frac{\partial\hat{\mathbb{H}}^{g_s}_{\text{red}}}{\partial\chi_A}\right)\Omega_{AB}\mathcal{D}_+^B\hat{\mathbb{H}}^{g-g_s}_{\text{red}}\right]\nonumber\\
&\hspace{1cm}=(D_0-2)\mathfrak{D}\hat{\mathbb{H}}^g_{\text{red}}+2\grass ^{\bar{I}}\mathcal{D}_{\bar{I}}\hat{\mathbb{H}}^g_{\text{red}}+\frac{1}{2}D_0\sum_{g_s=1}^{g-1}\mathcal{D}_+^A\hat{\mathbb{H}}^g_{\text{red}}\,\Omega_{AB}\,\mathcal{D}^B_+\hat{\mathbb{H}}^g_{\text{red}}=\nonumber\\
&\hspace{1cm}=2\grass ^{\bar{I}}\mathcal{D}_{\bar{I}}\hat{\mathbb{H}}^g_{\text{red}}-2\chi^2\,R\,(D_0-2)\hat{\tilde{\mathbb{H}}}^g_{\text{red}}+\sum_{g_s=1}^{g-1}\mathcal{D}_+^A\hat{\mathbb{H}}^g_{\text{red}}\,\Omega_{AB}\,\mathcal{D}^B_+\hat{\mathbb{H}}^g_{\text{red}}\,.
\end{align}}
Singling out the term which is independent of $\grass ^{\bar{I}}$ we obtain
\begin{align}
\left[\mathfrak{D}\chi_A\mathcal{D}_+^A-\chi_A\mathcal{D}_+^A\mathfrak{D}\right]\hat{\tilde{\mathbb{H}}}^g_{\text{red}}\bigg|_{\grass ^{\bar{I}}=0}=\left[-2\chi^2\,R\,(D_0-2)\hat{\tilde{\mathbb{H}}}^g_{\text{red}}+\sum_{g_s=1}^{g-1}\mathcal{D}_+^A\hat{\mathbb{H}}^g_{\text{red}}\,\Omega_{AB}\,\mathcal{D}^B_+\hat{\mathbb{H}}^g_{\text{red}}\right]\bigg|_{\grass ^{\bar{I}}=0}\nonumber
\end{align}
which can equivalently be rewritten in the following form
\begin{align}
\chi_A\chi_B\left[\mathcal{D}_+^A\mathcal{D}_-^B-\mathcal{D}_-^A\mathcal{D}_+^B\right]\hat{\tilde{\mathbb{H}}}^g_{\text{red}}\bigg|_{\grass ^{\bar{I}}=0}&=\left[2\chi^2\,R\,(D_0-2)\hat{\tilde{\mathbb{H}}}^g_{\text{red}}-\sum_{g_s=1}^{g-1}\mathcal{D}_+^A\hat{\mathbb{H}}^g_{\text{red}}\,\Omega_{AB}\,\mathcal{D}^B_+\hat{\mathbb{H}}^g_{\text{red}}\right]\bigg|_{\grass ^{\bar{I}}=0}\nonumber\\
\epsilon^{ij}\chi_A\chi_B\mathcal{D}_i^A\mathcal{D}_j^B\hat{\tilde{\mathbb{H}}}^g_{\text{red}}\bigg|_{\grass ^{\bar{I}}=0}&=\left[2\chi^2\,R\,(D_0-2)\hat{\tilde{\mathbb{H}}}^g_{\text{red}}-\sum_{g_s=1}^{g-1}\mathcal{D}_+^A\hat{\mathbb{H}}^g_{\text{red}}\,\Omega_{AB}\,\mathcal{D}^B_+\hat{\mathbb{H}}^g_{\text{red}}\right]\bigg|_{\grass ^{\bar{I}}=0}
\end{align}
This is precisely the equation derived in \cite{Antoniadis:2009nv}, including also the boundary terms. Notice particularly the operator $(D_0-2)$ in the first term on the right hand side. Translated to the setup in \cite{Antoniadis:2009nv}, this becomes $2(g-1)$, which is exactly the prefactor found there.
\subsection{Integrability}
We would also like to briefly check integrability of the differential equations derived in the previous section. For simplicity we will focus on equation (\ref{DiffEqhatH}) for which we have to prove
\begin{align}
\mathfrak{D}^2\hat{\mathbb{H}}^g_{\text{red}}=2\chi^2\,R\,{D_-}^+\hat{\mathbb{H}}^g_{\text{red}}{\begin{array}{c} \vspace{-0.3cm}\text{\tiny ?} \\ = \\ \vspace{-0.3cm}\phantom{\text{\tiny ?}}\end{array}} \mathfrak{D}\left[-2\chi^2\,R\,\hat{\tilde{\mathbb{H}}}^g_{\text{red}}+\frac{1}{2}\sum_{g_s=0}^g\mathcal{D}_+^A\hat{\mathbb{H}}^{g_s}_{\text{red}}\Omega_{AB}\mathcal{D}_+^B\hat{\mathbb{H}}^{g-g_s}_{\text{red}}\right]\,.
\end{align}
For the first term on the right hand side we can use (\ref{DiffEquhattildeH}), while the 
remaining terms will also contain commutator terms of covariant derivatives. The latter will result in curvature contributions, in particular
\begin{align}
{[\mathfrak{D},\mathcal{D}_+^A]^C}_D=\!2\chi_B\left[\Omega^{BA}\delta^C_D R D_0-{R^{(BAC}}_{D)}+R\left(\Omega^{BC}\delta^A_D+\Omega^{AC}\delta^B_D\right)\right]\!\!\left(\chi_C\frac{\partial}{\partial\chi_D}+\xi_C\frac{\partial}{\partial\xi_D}\right)\,.
\end{align}
After some algebra we get
\begin{align}
0{\begin{array}{c} \vspace{-0.3cm}\text{\tiny ?} \\ = \\ \vspace{-0.3cm}\phantom{\text{\tiny ?}}\end{array}}& -2\chi^2\,R\,\sum_{g_s=0}^g\left[\mathcal{D}_+^A\hat{\mathbb{H}}^{g_s}_{\text{red}}\Omega_{AB}\mathcal{D}_+^B\hat{\tilde{\mathbb{H}}}^{g-g_s}_{\text{red}}+\left(\frac{\partial\hat{\mathbb{H}}^{g_s}_{\text{red}}}{\partial\chi_A}\right)\Omega_{AB}\mathcal{D}_+^B\hat{\mathbb{H}}^{g-g_s}_{\text{red}}\right]+\nonumber\\
&+\sum_{g_s=0}^g\mathcal{D}_+^A\left[-2\chi^2\,R\,\hat{\tilde{\mathbb{H}}}^{g_s}_{\text{red}}+\frac{1}{2}\sum_{g'_s=0}^{g_s}\mathcal{D}_+^C\hat{\mathbb{H}}^{g'_s}_{\text{red}}\,\Omega_{CD}\,\mathcal{D}_+^D\hat{\mathbb{H}}^{g_s-g'_s}_{\text{red}}\right]\Omega_{AB}\mathcal{D}_+^B\hat{\mathbb{H}}^{g-g_s}_{\text{red}}+\nonumber\\
&+2\sum_{g_s=0}^g\Bigg[\chi_E\left[\Omega^{EA}\,R\,D_0\,\delta^{C}_D-\,{R^{(EAC}}_{D)}+\Omega^{DA}\,R\,\left(\Omega^{EC}\delta^A_D+\Omega^{AC}\delta^B_D\right)\right]\cdot\nonumber\\
&\hspace{2cm}\cdot\left(\chi_C\frac{\partial}{\partial\chi_C}+\xi_C\frac{\partial}{\partial\xi_C}\right)\hat{\mathbb{H}}^{g_s}_{\text{red}}\,\Omega_{AB}\,\mathcal{D}_+^B\hat{\mathbb{H}}^{g-g_s}_{\text{red}}\Bigg]\,.
\end{align}
Using at this stage that $\hat{H}^g_{\text{red}}$ is independent of $\xi_A$, we can streamline this condition in the following manner
\begin{align}
0{\begin{array}{c} \vspace{-0.3cm}\text{\tiny ?} \\ = \\ \vspace{-0.3cm}\phantom{\text{\tiny ?}}\end{array}}-2\chi^2\,R\,\sum_{g_s=0}^g\Bigg[&\mathcal{D}_+^A\hat{\mathbb{H}}^{g_s}_{\text{red}}\Omega_{AB}\mathcal{D}_+^B\hat{\tilde{\mathbb{H}}}^{g-g_s}_{\text{red}}+\left(\frac{\hat{\mathbb{H}}^{g_s}_{\text{red}}}{\partial\chi_A}\right)\Omega_{AB}\mathcal{D}^B_+\hat{\mathbb{H}}^{g-g_s}_{\text{red}}+\mathcal{D}_+^A\hat{\tilde{\mathbb{H}}}^{g_s}_{\text{red}}\Omega_{AB}\mathcal{D}_+^B\hat{\mathbb{H}}^{g-g_s}_{\text{red}}-\nonumber\\
-&\left(\frac{\hat{\mathbb{H}}^{g_s}_{\text{red}}}{\partial\chi_A}\right)\Omega_{AB}\mathcal{D}^B_+\hat{\mathbb{H}}^{g-g_s}_{\text{red}}\Bigg]-\sum_{g_s=0}^g{D_+}^-\hat{\mathbb{H}}^{g_s}_{\text{red}}\sum_{g'_s=0}^{g_s}\mathcal{D}_+^A\hat{\mathbb{H}}^{g-g'_s}_{\text{red}}\Omega_{AB}\mathcal{D}_+^B\hat{\mathbb{H}}^{g-g_s+g'_s}_{\text{red}}\,,\nonumber
\end{align}
which is indeed satisfied, thus entailing that equation
(\ref{DiffEqhatH}) is integrable. Notice that integrability holds also
including the curvature contributions, which lends strong evidence to
the fact that we have indeed obtained the full boundary contribution
including the contact terms. We should point out that for $m_2=0$
using the relation $\frac{\partial}{\partial \xi_A}\mathbb{\tilde{H}}|_{\xi=0}=  \mathcal{D}^A_+
\mathbb{\hat{\tilde{H}}}$ we can obtain the differential equation for
former in terms of the equation for the latter
(\ref{DiffEqhatH}).
The resulting equation has an additional
curvature dependent term as compared to the one obtained from
(\ref{FirstOrderTildeH}). Including this additional piece we can prove
integrability for equations for $\mathbb{H}$ and $\mathbb{\tilde{H}}$
to the first order in $\xi$ (i.e. for $m_2=0$) for a general curved
moduli space. This indeed shows that some contact terms are missing in
(\ref{FirstOrderTildeH}). While for $m_2=0$ this extra contact term
can be deduced from (\ref{DiffEquhattildeH}), we have not been able to
extend this to the general case $m_2 \neq 0$.
\section{Field Theory Explanation of Differential Equations}\label{Sect:FieldTheoryEqu}
\setcounter{equation}{0}
After having derived several differential equations for the string theory amplitudes $\mathbb{H}^g$, $\tilde{\mathbb{H}}^g$, $\hat{\mathbb{H}}_{\text{red}}^g$ and $\hat{\tilde{\mathbb{H}}}_{\text{red}}^g$ we would also like to obtain a better understanding from a field theoretic point of view. In the past \cite{Antoniadis:2006mr,Antoniadis:2007cw,Antoniadis:2007ta,Antoniadis:2009nv} such equations could always be explained to root in particular analyticity properties of the effective action couplings with respect to moduli as well as harmonic coordinates. Put differently, these equations were a consequence of the fact that the couplings did not depend on all multiplets in a random manner. As explained in section~\ref{Sect:HarmonicDescription}, the situation now is slightly different in so far as, say, a generic $\mathcal{H}^{g,A_1\ldots A_{m_1+1},B_1\ldots B_{m_2+1}}_{\bar{I}_1\ldots\bar{I}_{n-m_1},\bar{J}_1\ldots\bar{J}_{n-m_2}}$ depends on all projections of the hypermultiplets and holomorphic and anti-holomorphic vector multiplets alike. Thus, all equations derived in the previous sections indeed will not reflect the analyticity properties of a single coupling, but rather encode relations between two or more couplings which are induced by supersymmetry. We will explain this idea in the following for equation (\ref{FirstOrderTildeH}) while a similar argumentation will also hold for the remaining equations.
\subsection{Extracting the Component Coupling}
As a starting point let us first translate equation (\ref{FirstOrderTildeH}) back from the language of generating functionals into the language of effective couplings. To this end we recall that the coefficient of $(\chi_A)^{m_1+1}(\xi_B)^{m_2+1}(\grass ^{\bar{I}})^{n-m_1}(\eta^{\bar{J}})^{n-m_2}$ of the left hand side of (\ref{FirstOrderTildeH}) reads  
\begin{align}
&\mathcal{D}_{[\bar{I}_1}\tilde{\mathcal{H}}^{g,A_1\ldots A_{m_1+1},B_1\ldots B_{m_2+1}}_{\bar{I}_2\ldots\bar{I}_{n-m_1}],\bar{J}_1\ldots\bar{J}_{n-m_2}}+\mathcal{D}_-^{[A_1}\tilde{\mathcal{H}}^{g,A_2\ldots A_{m_1+1}],B_1\ldots B_{m_2+1}}_{\bar{I}_1\ldots\bar{I}_{n-m_1},\bar{J}_1\ldots\bar{J}_{n-m_2}}+{D_-}^+\mathcal{H}^{g,A_1\ldots A_{m_1+1},B_1\ldots B_{m_2+1}}_{\bar{I}_1\ldots\bar{I}_{n-m_1},\bar{J}_1\ldots\bar{J}_{n-m_2}}\,.\label{FirstOrderComp}
\end{align}
We note in passing that this expression is consistent in so far as each of the terms has the same harmonic charge $-(2g+2n-m_1-m_2-4)$. We would now like to prove that this expression is zero.\footnote{In fact (\ref{FirstOrderTildeH}) predicts that it is equal to some boundary contribution which in field theory arises due to integrating out auxiliary fields as has been discussed in \cite{Antoniadis:2007cw}. We will not attempt to do this explicitly here.} To this end we realize that (\ref{FirstOrderComp}) involves three different component couplings: All three of them stem from (\ref{gen44}), however, differ in the way the $(D_-\cdot D_-)(\bar{D}^+\cdot \bar{D}^+)$ spinor derivatives are distributed and how the Grassmannian integral measure (see equation (\ref{IntegMeasure})) is used on the (master) coupling function $H^{(2-d),A_1\ldots A_{m_1},B_1\ldots m_2}_{\bar{I}_1\ldots \bar{I}_{n-m_1-1},\bar{J}_1\ldots\bar{J}_{n-m_2-1}}$ to single out a particular component term. To make things precise, we consider $\mathcal{H}^{g,A_1\ldots A_{m_1+1},B_1\ldots B_{m_2+1}}_{\bar{I}_1\ldots\bar{I}_{n-m_1},\bar{J}_1\ldots\bar{J}_{n-m_2}}$ to correspond to the component coupling
\begin{align}
\int d^4x\int du\,&(K_-\cdot K_-)^{g}\,(D^-\cdot D^-)\bigg[(\bar\Psi_{A(+)} \bar K^{+\bar J}_{(-)})^{m_1} (\bar\Psi_{B(-)} \bar K^{+\bar I}_{(+)})^{m_2}(\bar K^{+\bar I}_{(+)} \bar K^{+\bar J}_{(-)})^{n-m_1-m_2-1}\cdot\nonumber\\
&\cdot (\bar D^+ \bar W^{\bar I} \cdot \bar D^+ \bar W^{\bar J})  (\bar D_+ q^+_{A_1} \cdot \bar D_+ q^+_{B_1})\bigg]\ \mathcal{H}^{g,A_1\ldots A_{m_1+1},B_1\ldots B_{m_2+1}}_{\bar{I}_1\ldots\bar{I}_{n-m_1},\bar{J}_1\ldots\bar{J}_{n-m_2}}\bigg|_{\q=\bq =0}\,,\label{SuperExprH}
\end{align}
where $\mathcal{H}^{g,A_1\ldots A_{m_1+1},B_1\ldots B_{m_2+1}}_{\bar{I}_1\ldots\bar{I}_{n-m_1},\bar{J}_1\ldots\bar{J}_{n-m_2}}$ is defined in terms of the `master' coupling
\begin{align}
&\mathcal{H}^{g,A_1\ldots A_{m_1+1},B_1\ldots B_{m_2+1}}_{\bar{I}_1\ldots\bar{I}_{n-m_1},\bar{J}_1\ldots\bar{J}_{n-m_2}}:=\frac{\pa^6 H^{(2-d),A_2\ldots A_{m_1+1},B_2\ldots B_{m_2+1}}_{\bar{I}_2\ldots \bar{I}_{n-m_1},\bar{J}_2\ldots\bar{J}_{n-m_2}}}{\pa W^2 \pa \bar W^{\bar{I}_1} \pa \bar W^{\bar{J}_1} \pa q^+_{A_1} \pa q^+_{B_1}}\,,\label{MasterRel1}
\end{align}
where an appropriate antisymmetrization of the indices is understood. Notice that the coupling $\mathcal{H}^{g,A_1\ldots A_{m_1+1},B_1\ldots B_{m_2+1}}_{\bar{I}_1\ldots\bar{I}_{n-m_1},\bar{J}_1\ldots\bar{J}_{n-m_2}}$ indeed has charge $-2g-2n+m_1+m_2+2$ as required. The two $\tilde{\mathcal{H}}$-couplings in (\ref{FirstOrderComp}) now correspond to slightly different couplings with a slightly modified distribution of the spinor derivatives. To be concrete we may for example consider the following distributions
\begin{align}
\int d^4x\int du\,&\bar{D}_+^{\dot{\alpha}}(D^-\cdot D^-)\bigg[(K_-\cdot K_-)^{g} (\bar\Psi_{A(+)} \bar K^{+\bar J}_{(-)})^{m_1} (\bar\Psi_{B(-)} \bar K^{+\bar I}_{(+)})^{m_2}(\bar K^{+\bar I}_{(+)} \bar K^{+\bar J}_{(-)})^{n-m_1-m_2-1}\cdot\nonumber\\
&\cdot (\bar D^+ q_{A_1}^- \cdot \bar D^+ \bar W^{\bar J})  (\bar{D}_+ q^+_{B_1})_{\dot{\alpha}}\bigg]\ \tilde{\mathcal{H}}^{g,A_1\ldots A_{m_1+1},B_1\ldots B_{m_2+1}}_{\bar{I}_2\ldots\bar{I}_{n-m_1},\bar{J}_1\ldots\bar{J}_{n-m_2}}\bigg|_{\q=\bq =0}\,,\label{SuperExprHt1}\\
\int d^4x\int du\,&\bar{D}_+^{\dot{\alpha}}(D^-\cdot D^-)\bigg[(K_-\cdot K_-)^{g}(\bar\Psi_{A(+)} \bar K^{+\bar J}_{(-)})^{m_1} (\bar\Psi_{B(-)} \bar K^{+\bar I}_{(+)})^{m_2}(\bar K^{+\bar I}_{(+)} \bar K^{+\bar J}_{(-)})^{n-m_1-m_2-1}\cdot\nonumber\\
&\cdot (\bar D^+ \bar W^{\bar I} \cdot \bar D^+ \bar W^{\bar J})  (\bar D_+ q^+_{B_1})_{\dot{\alpha}}\bigg]\ \mathcal{H}^{g,A_2\ldots A_{m_1+1},B_1\ldots B_{m_2+1}}_{\bar{I}_1\ldots\bar{I}_{n-m_1},\bar{J}_1\ldots\bar{J}_{n-m_2}}\bigg|_{\q=\bq =0}\,,\label{SuperExprHt2}
\end{align}
with the following relations to the 'master' coupling
\begin{align}
&\tilde{\mathcal{H}}^{g,A_1\ldots A_{m_1+1},B_1\ldots B_{m_2+1}}_{\bar{I}_2\ldots\bar{I}_{n-m_1},\bar{J}_1\ldots\bar{J}_{n-m_2}}:=\frac{\pa^5 H^{(2-d),A_2\ldots A_{m_1+1},B_2\ldots B_{m_2+1}}_{\bar{I}_2\ldots \bar{I}_{n-m_1},\bar{J}_2\ldots\bar{J}_{n-m_2}}}{\pa W^2  \pa \bar W^{\bar{J}_1} \pa q^-_{A_1} \pa q^+_{B_1}}\,,\label{MasterRel3}\\
&\tilde{\mathcal{H}}^{g,A_2\ldots A_{m_1+1},B_1\ldots B_{m_2+1}}_{\bar{I}_1\ldots\bar{I}_{n-m_1},\bar{J}_1\ldots\bar{J}_{n-m_2}}:=\frac{\pa^5 H^{(2-d),A_2\ldots A_{m_1+1},B_2\ldots B_{m_2+1}}_{\bar{I}_2\ldots \bar{I}_{n-m_1},\bar{J}_2\ldots\bar{J}_{n-m_2}}}{\pa W^2 \pa \bar W^{\bar{I}_1} \pa \bar W^{\bar{J}_1} \pa q^+_{B_1}}\,.\label{MasterRel2}
\end{align}
These couplings have charges $-(2g+2n-m_1-m_2-4)$ and $-(2g+2n-m_1-m_2-3)$ respectively, which is indeed as required. We should note that all expressions are still arbitrary functions with respect to the scalar moduli $f_i^A$ and $\varphi$, as well as the harmonic variables $\bar{u}^i_\pm$. Before proceeding, however, we notice an opportunity for simplifying our notation and saving a lot of writing: Since all superfields which appear in (\ref{SuperExprH}), (\ref{SuperExprHt1}) and (\ref{SuperExprHt2}) are fermionic in nature, all four sets of indices ($A,B,\bar{I}$ and $\bar{J}$) will be separately anti-symmetrized. We can thus, just as in section~\ref{Sect:GeneratingFunctional} introduce Grassmann valued expressions $(\chi_A,\xi_B,\grass ^{\bar{I}},\eta^{\bar{J}})$ and formulate everything in terms of generating functionals $\mathbb{H}^g$ and $\tilde{\mathbb{H}}^g$.  
\subsection{Harmonic Gauge Fixing}\label{Sect:DiffConstraintSupersp}
Our next step after preparing the component couplings is to perform the harmonic $u$-integral which is most easily done after fixing the harmonic gauge freedom. In order to do this, we proceed in a similar fashion as in~\cite{Antoniadis:2009nv} and reduce all couplings to $SU(2)$ irreducible representations. The main novelty, however, is that, due to the dependence on the anti-G-analytic hypermultiplets $q^-$, $\tilde q_+$, the couplings can have an arbitrary scalar dependence. Thus the generic gauge fixed form for the three couplings respectively will look like (as announced, we will be using the language of the generating functionals)
\begin{align}
\mathcal{H}^{g,A_1\ldots A_{m_1+1},B_1\ldots B_{m_2+1}}_{\bar{I}_1\ldots\bar{I}_{n-m_1},\bar{J}_1\ldots\bar{J}_{n-m_2}}\longrightarrow&\left[\mathbb{H}^{g}=\mathfrak{H}^{g}_{(i_1\ldots i_{d})}(\varphi,f_i)\,\bar{u}_+^{(i_1}\ldots \bar{u}_+^{i_{d})}\right]\bigg|_{\chi^{m_1+1}\grass ^{n-m_1}\xi^{m_2+1}\eta^{n-m_2}}\,,\\
\tilde{\mathcal{H}}^{g,A_1\ldots A_{m_1+1},B_1\ldots B_{m_2+1}}_{\bar{I}_2\ldots\bar{I}_{n-m_1},\bar{J}_1\ldots\bar{J}_{n-m_2}}\longrightarrow&\left[\tilde{\mathbb{H}}^g=\tilde{\mathfrak{H}}^{g}_{(i_1\ldots i_{d-2})}(\varphi,f_i)\,\bar{u}_+^{(i_1}\ldots \bar{u}_+^{i_{d-2})}\right]\bigg|_{\chi^{m_1+1}\grass ^{n-m_1-1}\xi^{m_2+1}\eta^{n-m_2}}\,,\label{GaugeFix1}\\
\mathcal{H}^{g,A_2\ldots A_{m_1+1},B_1\ldots B_{m_2+1}}_{\bar{I}_1\ldots\bar{I}_{n-m_1},\bar{J}_1\ldots\bar{J}_{n-m_2}}\longrightarrow&\left[\tilde{\mathbb{H}}^g=\tilde{\mathfrak{H}}^{g}_{(i_1\ldots i_{d-1})}(\varphi,f_i)\,\bar{u}_+^{(i_1}\ldots \bar{u}_+^{i_{d-1})}\right]\bigg|_{\chi^{m_1}\grass ^{n-m_1}\xi^{m_2+1}\eta^{n-m_2}}\,,
\end{align}
where $\big|_{\chi^{a_1}\grass ^{a_2}\xi^{a_3}\eta^{a_4}}$ means extraction of the appropriate coefficient in the power series expansion in terms the Grassmann variables. The precise power also determines the harmonic charge, \emph{i.e.} $d=2g+2n-m_1-m_2-2$. Notice that after taking into account the definition in terms of the master couplings (\ref{MasterRel3}) and (\ref{MasterRel2}), we may immediately state the relation
\begin{align}
\left[\mathfrak{H}^{g}_{(i_1\ldots i_{d})}=-\frac{1}{d}\,\chi_A\mathcal{D}_{(i_d}^{A}\,\tilde{\mathfrak{H}}^{g}_{\,i_1\ldots i_{d-1})}\right]\bigg|_{\chi^{m_1+1}\grass ^{n-m_1}\xi^{m_2+1}\eta^{n-m_2}}\,,\label{RelHtildeHfield}
\end{align}
which is exactly equation (\ref{RelHtildeH}) which we find in string theory. After these preparations we are now in a position to explain equation (\ref{FirstOrderTildeH}). To this end, we consider (the equation is understood to be the coefficient of ${\chi^{m_1+1}\grass ^{n-m_1}\xi^{m_2+1}\eta^{n-m_2}}$ in the Grassmann expansion)
\begin{align}
\chi_A\mathcal{D}_-^{A}\tilde{\mathbb{H}}^{g}&=\bar{u}^j_-\bar{u}_+^{(i_1}\ldots \bar{u}_+^{i_{d-1})}\chi_A\mathcal{D}_j^A\tilde{\mathfrak{H}}^{g}_{(i_1\ldots i_{d-1})}=\nonumber\\
&=\chi_A\mathcal{D}_j^A\tilde{\mathfrak{H}}^{g}_{(i_1\ldots i_{d-1})}\left[\bar{u}^{(j}_-\bar{u}_+^{i_1}\ldots \bar{u}_+^{i_{d-1})}+\epsilon^{j(i_1}\bar{u}_+^{i_2}\ldots \bar{u}_+^{i_{d-1})}\right]=\nonumber\\
&=\frac{1}{d}\,\chi_A\mathcal{D}_{(i_1}^A\tilde{\mathfrak{H}}^{g}_{i_2\ldots i_{d})}\,{D_-}^+\left(\bar{u}^{i_1}_+\ldots \bar{u}_+^{i_{d}}\right)+\chi_A\mathcal{D}^{A,j}\tilde{\mathfrak{H}}^{g}_{(ji_1\ldots i_{d-2})}\bar{u}_+^{i_1}\ldots \bar{u}_+^{i_{d-2}}\,.\label{TowSU2struct}
\end{align}
Using at this stage relation (\ref{RelHtildeHfield}) as well as (\ref{MasterRel3}) and (\ref{MasterRel2}) we may rewrite
\begin{align}
\left[\chi_A\mathcal{D}_-^{A}\tilde{\mathbb{H}}^{g}=-{D_-}^+\mathbb{H}^{g}-\grass ^{\bar{I}}\mathcal{D}_{\bar{I}}\tilde{\mathfrak{H}}^{g}_{(i_1\ldots i_{d-2})}\,\bar{u}_+^{(i_1}\ldots \bar{u}_+^{i_{d-2})}\right]\bigg|_{\chi^{m_1+1}\grass ^{n-m_1}\xi^{m_2+1}\eta^{n-m_2}}\,.
\end{align}
In the last term, however, we recognize again (\ref{GaugeFix1}), such that
\begin{align}
\chi_A\mathcal{D}_-^{A}\tilde{\mathbb{H}}^{g}+{D_-}^+\mathbb{H}^{g}+\grass ^{\bar{I}}\mathcal{D}_{\bar{I}}\tilde{\mathbb{H}}^{g}=0\,,
\end{align}
which is exactly relation (\ref{FirstOrderComp}) that we wished to prove. Notice, if we treat the auxiliary fields properly (instead of putting them to zero in the on-shell approach we are using here), the right-hand side will be modified by additional boundary terms, which is indeed what we find from the explicit string computation in equation (\ref{FirstOrderTildeH}).

There is one further comment we would like to make: The crucial step in this derivation was the second line of (\ref{TowSU2struct}). There we have used that a single $\bar{u}^i_-$ combined with set of (totally symmetric) $\bar{u}^i_+$ gives rise to two irreducible $SU(2)$ tensor structures corresponding to the two terms in the square bracket of the second line in (\ref{TowSU2struct}). Since both of them have different isospin, they cannot be related to each other and thus, from a harmonic-$SU(2)$ point of view, equation (\ref{FirstOrderComp}) decomposes into two distinct equations. This in fact can also be directly verified in (\ref{FirstOrderTildeH}) by hitting it with the two projection operators ${D_+}^-$ and $-D_0+2-{D_-}^+{D_+}^-$ respectively. In the first case we obtain
\begin{align}
{D_+}^-\left[\mathfrak{D}\tilde{\mathbb{H}}^g+{D_-}^+\mathbb{H}^g\right]&={D_+}^-\tilde{\mathbb{C}}^{\text{bdy}}\\
\chi_A\mathcal{D}_+^A\tilde{\mathbb{H}}^g-D_0\mathbb{H}^g&=0\,,
\end{align}
which is just equation (\ref{RelHtildeH}). Here we have used that ${D_+}^-$ annihilates $\mathbb{H}^g$, $\tilde{\mathbb{H}}^g$ as well as every term contained in $\tilde{\mathbb{C}}^{\text{bdy}}$. Applying the operator $-D_0+2-{D_-}^+{D_+}^-$, we in turn obtain 
{\allowdisplaybreaks
\begin{align}
(-D_0+2-{D_-}^+{D_+}^-)\left[\mathfrak{D}\tilde{\mathbb{H}}^g+{D_-}^+\mathbb{H}^g\right]&=(-D_0+2-{D_-}^+{D_+}^-)\tilde{\mathbb{C}}^{\text{bdy}}\nonumber\\
(-D_0+2)\grass ^{\bar{I}}\mathcal{D}_{\bar{I}}\tilde{\mathbb{H}}^g+\chi_A(-D_0+2)\mathcal{D}_-^A\tilde{\mathbb{H}}^g-\chi_A{D_-}^+\mathcal{D}_+^A\tilde{\mathbb{H}}^g&=(-D_0+2)\tilde{\mathbb{C}}^{\text{bdy}}\nonumber\\
\left[(-D_0+2)\grass ^{\bar{I}}\mathcal{D}_{\bar{I}}-\chi_A\mathcal{D}_-^AD_0-\chi_A\mathcal{D}_+^A{D_-}^+\right]\tilde{\mathbb{H}}^g&=(-D_0+2)\tilde{\mathbb{C}}^{\text{bdy}}\,,
\end{align}}
which is just the original harmonicity equation (\ref{DiffEqTildeHD0}). Thus the two $SU(2)$ structures of (\ref{TowSU2struct}) indeed encapsulate all the relevant information about the couplings $\tilde{\mathbb{H}}^g$ and $\mathbb{H}^g$.
\section{Conclusions}\label{conclusions}
In this work we have studied a generalization of the new class of
topological amplitudes introduced in our previous
work~\cite{Antoniadis:2009nv}. These amplitudes compute the
coupling-coefficients of a very broad class of $1/2$-BPS terms in the
string effective action, which mix vector multiplets with neutral
hypermultiplets in a non-trivial manner. In fact, we find a pair of
(series of) amplitudes, which we combine into the generating
functionals $(\mathbb{H}^g,\tilde{\mathbb{H}}^g)$. These two objects
are related through a hyper-moduli derivative and
$\tilde{\mathbb{H}}^g$ plays the role of 'gauge-potential' for
$\mathbb{H}^g$.  In particular the moduli dependence of these coupling
functions satisfies a set of differential equations emerging from
their 1/2-BPS structure, generalizing in a sense the holomorphic
anomaly and harmonicity equations studied
in~\cite{Antoniadis:2009nv,Antoniadis:2007cw}. In fact, the analytic
projection which we use to obtain a consistent BPS action coupling
requires non-trivial relations between different terms at the
component level. The latter predict highly non-trivial differential
equations between different topological amplitudes which we have
directly checked in string theory. The main novelty is that they are
first order differential equations (as compared to the second order
relations derived in \cite{Antoniadis:2009nv,Antoniadis:2007cw}),
however, mixing $\mathbb{H}^g$ with $\tilde{\mathbb{H}}^g$. As usual,
we find that they get violated by world-sheet boundary contributions
which we have evaluated explicitly. In addition, however, due to the
fact that the hypermultiplet and vector multiplet moduli space of the
$\cN=2$ string compactification is not flat, we also obtain contact
terms, which turn out to be very difficult to control for the generic
amplitudes. However, we were able to identify a particular subset of
the series of topological amplitudes (denoted
$(\hat{\mathbb{H}}^g_{\text{red}},\hat{\tilde{\mathbb{H}}}^g_{\text{red}})$) 
for which we managed to control the curvature contributions and prove
integrability (and thus consistency)  
of our equations. We should also note that the equations close on the
subset $(\hat{\mathbb{H}}^g_{\text{red}},\hat{\tilde{\mathbb{H}}}^g_{\text{red}})$ 
in the sense that no new classes of topological objects are
introduced. Thus one might hope, 
taking into account the first order nature of these equations as well
as its iterative structure, 
that it is possible to iteratively solve for
$(\hat{\mathbb{H}}^g_{\text{red}},\hat{\tilde{\mathbb{H}}}^g_{\text{red}})$,
 using similar methods as in the case of pure vector multiplet
 dependence 
(see
e.g.~\cite{Bershadsky:1993cx,Huang:2006si,Huang:2006hq,Grimm:2007tm}). In
this way one might be able to obtain further insight into the
structure of the moduli spaces of $\cN=2$ string
compactifications. From a field theoretic point of view, the origin of
the anomalous terms can probably be also understood via the process of 
integrating out auxiliary fields
(for a discussion see~\cite{Antoniadis:2009nv}). It would be very interesting in the future to make this point more precise.
\section*{Acknowledgements}
We would like to thank Sergio Ferrara and Stephan Stieberger for enlightening discussions. This work was supported in part by the European Commission under the ERC Advanced Grant 226371 and the contract PITN-GA-2009-237920, and in part by the French Agence Nationale de la Recherche, contract ANR-06-BLAN-0142. SH would like to thank the ICTP Trieste and CERN for kind hospitality during completion of this work.

\appendix
\renewcommand{\theequation}{\Alph{section}.\arabic{equation}}
\section{Superconformal Algebras}\label{append:SCA}
\setcounter{equation}{0}
For heterotic string theory compactified on $K3\times T^2$ the world-sheet theory is a product of an $\N=2$ and an $\N=4$ superconformal field theory representing $T^2$ and $K3$ respectively. In the following we will briefly review both theories mainly in order to fix our notation.
\subsection{The $\N=2$ Superconformal Algebra}\label{reviewopentopI}
The (untwisted) $\N=2$ superconformal algebra (SCA) contains besides the energy momentum tensor $T_{T^2}$ two  supercurrents $G^\pm_{T^2}$ which are positively and negatively charged with respect to a $U(1)$ Kac-Moody current $J_{T^2}$. Here we have added the subscript $T^2$ to all operators in order to indicate that they represent the torus piece of the heterotic world-sheet theory. The conformal weights of these operators are given by
\begin{align}
&h_{T_{T^2}}=2,&&h_{G^\pm_{T^2}}=\frac{3}{2}, &&h_{J_{T^2}}=1.
\end{align}
The non-trivial operator product expansions (OPE) of these objects are given by
{\allowdisplaybreaks
\begin{align}
&T_{T^2}(z)T_{T^2}(w)=\frac{2T_{T^2}(w)}{(z-w)^2}+\frac{\partial_wT_{T^2}(w)}{z-w}\,,\label{N2SCAa}\\
&T_{T^2}(z)G_{T^2}^{\pm}(w)=\frac{3G_{T^2}^\pm(w)}{2(z-w)^2}+\frac{\partial_wG_{T^2}^\pm(w)}{z-w}\,,\\
&T_{T^2}(z)J_{T^2}(w)=\frac{J_{T^2}(w)}{(z-w)^2}+\frac{\partial_wJ_{T^2}(w)}{z-w}\,,\\
&G_{T^2}^+(z)G_{T^2}^-(w)=\frac{6}{(z-w)^3}+\frac{2J_{T^2}(w)}{(z-w)^2}+\frac{2T_{T^2}(w)+\partial_wJ_{T^2}(w)}{z-w}\,,\\
&J_{T^2}(z)G_{T^2}^\pm(w)=\pm\frac{G_{T^2}^\pm(w)}{z-w}\,,\\
&J_{T^2}(z)J_{T^2}(w)=\frac{2}{(z-w)^2}\,.\label{N2SCAf}
\end{align}}
An explicit representation of the corresponding algebra in terms of a free complex boson $X_3$ and fermion $\psi_3$ living on $T^2$ can be written as
\begin{align}
&T_{T^2}=\frac{1}{2}\psi_3{\leftrightarrow\atop\displaystyle{\partial\atop~}}\bar{\psi}_3+\partial X_3\partial \bar{X}_3,&&G^-_{T^2}=\bar{\psi}_3\partial X_3, &&G^{+}_{T^2}=\psi_3\partial\bar{X}_3, &&J_{T^2}=\psi_3\bar{\psi}_3\ .
\end{align}
A twisted version of the $\N=2$ SCA is given by redefining the energy momentum tensor in the following manner
\begin{align}
T_{T^2}\to T_{T^2}-\frac{1}{2}\partial J_{T^2}\,.
\end{align}
This in particular has the effect of shifting the conformal weight of all operators by half their $U(1)$ charge. In this way, we find
\begin{align}
&h^{\text{twist}}_{T_{T^2}}=2\,,&&h^{\text{twist}}_{G^-_{T^2}}=2\,, &&h^{\text{twist}}_{G^+_{T^2}}=1\,, &&h^{\text{twist}}_{J_{T^2}}=1\,.
\end{align}
The new conformal weights of the supercurrents allow us to identify $G^+_{T^2}$ with the BRST operator of the twisted theory, while $G^-_{T^2}$ becomes the parametrization anti-ghost thereby defining the measure of the topological string.
\subsection{The $\N=4$ Superconformal Algebra}\label{Sect:SCFT}
An $\N=4$ SCFT contains the energy momentum tensor $T_{K3}$ which is accompanied by two doublets of supercurrents $(G^+_{K3},\tilde{G}^-_{K3})$ and $(\tilde{G}^+_{K3},G^-_{K3})$, transforming under an $SU(2)$ Kac-Moody current algebra formed by $(J^{\pm\pm}_{K3},J_{K3})$. Here, we have added the subscript $K3$ in order to indicate that these operators represent the $K3$-piece of the heterotic world-sheet theory. The conformal weights of these operators are
\begin{align}
&h_{T_{K3}}=2\,,&&h_{G^\pm_{K3}}=h_{\tilde{G}^\pm_{K3}}=\frac{3}{2}\,,&&h_{J^{\pm\pm}_{K3}}=h_{J_{K3}}=1\,.
\end{align}
The non-trivial OPEs of these objects are given by
{\allowdisplaybreaks
\begin{align}
&J^{--}_{K3}(z)G^+_{K3}(0)\sim\frac{\tilde{G}^-_{K3}(0)}{z}, &&J_{K3}^{++}(z)\tilde{G}_{K3}^-(0)
\sim-\frac{G^+_{K3}(0)}{z}\,,\nonumber\\
&J_{K3}^{++}(z)G^-_{K3}(0)\sim \frac{\tilde{G}^+_{K3}(0)}{z}, &&J_{K3}^{--}(z)
\tilde{G}^+_{K3}(0)\sim-\frac{G^-_{K3}(0)}{z}\,,\nonumber
\end{align}
\begin{align}
&G^+_{K3}(z)G^-_{K3}(0)\sim\frac{J_{K3}(0)}{z^2}-\frac{T^B_{K3}(0)-\frac{1}{2}\partial J_{K3}(0)}{z}\,,
\nonumber\\
&\tilde{G}^+_{K3}(z)\tilde{G}^-_{K3}(0)\sim\frac{J_{K3}(0)}{z^2}-\frac{T^B_{K3}(0)-\frac{1}{2}
\partial J_{K3}(0)}{z}\,,\nonumber\\
&\tilde{G}^+_{K3}(z)G^+_{K3}(0)\sim\frac{2J^{++}_{K3}(0)}{z^2}+\frac{\partial J^{++}_{K3}(0)}{z}\,,
\nonumber\\
&\tilde{G}^-_{K3}(z)G^-_{K3}(0)\sim\frac{2J^{--}_{K3}(0)}{z^2}+\frac{\partial J^{--}_{K3}(0)}{z}\,,
\nonumber
\end{align}
}
while for any operator $O^q_{K3}$ of $U(1)$ charge $q$, one has:
\begin{align}
J_{K3}(z)O^q_{K3}(0)\sim q\frac{O^q_{K3}(0)}{z}\,.
\nonumber
\end{align}
A representation in terms of free bosons $X_{4,5}$ and fermions $\psi_{4,5}$ living on a torus-orbifold realization of $K3$ is given by
\begin{align}
T_{K3}=\partial X_4\partial{\bar X}_4+\partial X_5\partial{\bar X}_5+
{1\over 2}(\psi_4{\leftrightarrow\atop\displaystyle{\partial\atop~}}{\bar\psi}_4+
\psi_5{\leftrightarrow\atop\displaystyle{\partial\atop~}}{\bar\psi}_5)\ ,
\end{align}
${}$\vspace{-0.8cm}
\begin{align}
&J_{K3}=\psi_4\bar{\psi}_4+\psi_5\bar{\psi}_5, && J^{++}_{K3}=\psi_4\psi_5, &&J^{--}_{K3}
=\bar{\psi}_4\bar{\psi}_5\,,
\label{N4curs}
\end{align}
\begin{align}
&G^+_{K3}=\psi_4\partial \bar{X}_4+\psi_5\partial\bar{X}_5, &&\tilde{G}^+_{K3}=-\psi_5\partial X_4+\psi_4\partial X_5\,,\label{K3scurrent1}\\
&G^{-}_{K3}=\bar{\psi}_4\partial X_4+\bar{\psi}_5\partial X_5, &&\tilde{G}^-_{K3}=-\bar{\psi}_5\partial \bar{X}_4+\bar{\psi}_4\partial \bar{X}_5\,.
\label{K3scurrent2}
\end{align}
The topologically twisted theory can be defined after specifying an $\N=2$ subalgebra inside the $\N=4$. We can then similarly redefine the energy momentum tensor
\begin{align}
T_{K3}\to T_{K3}-\frac{1}{2}\partial J_{K3}\ .
\end{align}
In this way, just as in the $\N=2$ case, the conformal dimensions of all operators are shifted by half their charge with respect to $J_{K3}$
\begin{align}
&h^{\text{twist}}_{T_{K3}}=2\,,&&h^{\text{twist}}_{G^-_{K3}}=h^{\text{twist}}_{\tilde{G}^-_{K3}}=2\,, &&h^{\text{twist}}_{G^+_{K3}}=h^{\text{twist}}_{\tilde{G}^+_{K3}}=1\,,\nonumber\\
&h^{\text{twist}}_{J^{--}_{K3}}=2\,, &&h^{\text{twist}}_{J^{++}_{K3}}=0\,, &&h^{\text{twist}}_{J_{K3}}=1\,.
\end{align}
In view of the harmonic coordinates introduced in appendix~\ref{App:HarmonicNotation} we will rewrite the $\N=4$ superconformal algebra in an $SU(2)$ covariant manner. To this end, we group the supercurrents into $SU(2)$-doublets in the following manner
\begin{align}
&G^+_{K3,i}\equiv\left(\begin{array}{c}\tilde{G}^+_{K3} \\ G^+_{K3}\end{array}\right),&&\text{and} &&G^-_{K3,i}\equiv \left(\begin{array}{c}G^-_{K3} \\ -\tilde{G}^-_{K3}\end{array}\right)\,.\label{CovarSupercurrents}
\end{align}
For these doublets the OPE relations take the following more compact form
\begin{align}
&G^+_{K3,i}(z)G^+_{K3,j}(0)\sim\epsilon_{ij}\left(\frac{2J^{++}_{K3} (0)}{z^2}+\frac{\partial J^{++}_{K3} (0)}{z}\right),\label{algebraCOV1}\\
&G^-_{K3,i}(z)G^-_{K3,j}(0)\sim \epsilon_{ij}\left(\frac{2J^{--}_{K3} (0)}{z^2}+\frac{\partial J^{--}_{K3} (0)}{z}\right),\\
&G^+_{K3,i}(z)G^-_{K3,j}(0)\sim-\epsilon_{ij}\left(\frac{J(0)}{z^2}-\frac{T_{K3}(0)-\frac{1}{2}\partial J(0)}{z}\right),
\end{align}
\begin{align}
&J^{++}_{K3} (z)G^+_{K3,i}(0)\sim0, &&J^{++}_{K3} (z)G^-_{K3,i}\sim\frac{G^+_{K3,i}}{z}\,,\label{algebraCOV4}\\
&J^{--}_{K3} (z)G^+_{K3,i}(0)\sim-\frac{G^-_{i}}{z}, &&J^{--}_{K3} (z)G^-_{K3,i}\sim0.\label{algebraCOV5}
\end{align}
We are now also free to project the $SU(2)$-indices with harmonic variables to define the following operators
\begin{align}
&G^+_{K3,\pm}=G^+_{K3,i}\bar{u}^i_\pm,&&\text{and} &&G^-_{K3,\pm}=G^-_{K3,i}\bar{u}^i_\pm\,.
\end{align}
\section{$\cN=2$ Harmonic Superspace}\label{App:HarmSuperspace}
\setcounter{equation}{0}
In this section of the appendix we discuss our conventions for the $\cN=2$ harmonic superspace and introduce the Grassmann analytic superfields which will be needed throughout this paper. Our conventions are identical to \cite{Antoniadis:2009nv}.
\subsection{$SU(2)$ Harmonic Variables}\label{App:HarmonicNotation}
Throughout this paper we consider $\cN=2$ supersymmetry in four dimensions whose automorphism group is
$SU(2)$. We introduce harmonic variables \cite{Galperin:1984av} on the coset $SU(2)/U(1)$ in the
form of matrices $(u^+_i, \, u^-_i) \in SU(2)$. They have an index
$i=1,2$ transforming under the fundamental representation of $SU(2)$ and $U(1)$ charges $\pm 1$. Together with their
complex conjugates $\bu^i_+ = \overline{(u^+_i)}, \, \bu^i_- =
\overline{(u^-_i)}$  they satisfy the unitarity relations
\begin{align}
&u^+_i\, \bu^i_+ = u^-_i\, \bu^i_- = 1 \,, && u^+_i\, \bu^i_- = u^-_i\, \bu^i_+ = 0\,, &&u^+_i\, \bu^j_+ + u^-_i\, \bu^j_- = \delta^j_i\label{12'}
\end{align}
and the unit determinant condition
\begin{align}
&\ep^{ij} u^+_i u^-_j = 1\ , &&u^+_i u^-_j - u^-_i u^+_j = \ep_{ij}\,,\label{12}
\end{align}
(with $\ep^{12} = -\ep_{12} = 1$).

The harmonic functions have harmonic expansions homogeneous under the action of the
subgroup $U(1)$. The harmonic expansions are organized in irreps of
$SU(2)$, keeping the balance of projected indices so that the overall $U(1)$ charge is always the same.
An example of a harmonic function is \emph{e.g.} $f^+(u) = f^i u^+_i + f^{ijk} u^+_i u^+_j u^-_k + \cdots$. The first component in this expansion is a doublet of $SU(2)$. The higher components give rise to higher-dimensional irreps.

The harmonic derivatives can be viewed as the covariant derivatives on the harmonic coset  $SU(2)/U(1)$, or equivalently, as the generators of the algebra of $SU(2)$ written in a $U(1)$ basis. This means that they are invariant under the left action of the group $SU(2)$, but covariant under the right action of the subgroup $U(1)$. They can be split into generators of the subalgebra $U(1)$:
\begin{equation}\label{subhd}
   D_0 = u^+_i \frac{\pa}{\pa u^+_i } - \bu_+^i \frac{\pa}{\pa \bu_+^i}
\end{equation}
and of the coset:
\begin{align}
&D_+{}^- = u^-_i \frac{\pa}{\pa u^+_i }  = \bu_+^i \frac{\pa}{\pa \bu_-^i}\,, &&\text{and} &&D_-{}^+ = u^+_i \frac{\pa}{\pa u^-_i } = \bu_-^i \frac{\pa}{\pa \bu_+^i} \ . \label{cosethd}
\end{align}
The harmonic derivatives are differential operators preserving the defining algebraic constraints (\ref{12'}) and (\ref{12}).

The derivative (\ref{subhd}) acts homogeneously on the harmonic functions. For instance, the function $f^{+}(u)$ above has $U(1)$ charge $+1$, hence
\begin{equation}\label{exhf}
    D_0  f^{+}(u) = f^{+}(u)\ .
\end{equation}
The harmonic expansion of this function defines an infinitely reducible representation of $SU(2)$. It can be made irreducible by requiring that the raising operator $D_-{}^+$ annihilates the function:
\begin{equation}\label{ropcon}
    D_-{}^+ f^{+}(u) = 0\ \quad \Rightarrow \quad  f^{+}(u) = f^i u^+_i \ .
\end{equation}
In other words, such a function is a highest-weight state of a doublet of $SU(2)$. The irreducibility condition (\ref{ropcon}) is also called a condition for harmonic (H-) analyticity.
\subsection{Grassmann Analytic On-shell Superfields}\label{App:Superfields}
The introduction of harmonic variables allows us to define `1/2-BPS short' or Grassmann
(G-) analytic superfields.\footnote{The notion of Grassmann analyticity (with breaking of the R symmetry) was first proposed in \cite{Galperin:1980fg} in the context of the $\cN=2$ hypermultiplet. Later on it was made R-symmetry covariant in the framework of $\cN=2$ harmonic superspace in \cite{Galperin:1984av}. } They depend only on half of the Grassmann variables which can
be chosen to be $\q^{+}_\a = \q^i_\a\, u_i^{+}$ and $\bq^\da_{-} =
\bu^i_{-}\,\bq^\da_i $ with $\alpha=(\pm)$ ($\da=(\pm)$) an (anti-)chiral spinor index. In this work we will encounter two different examples which we will briefly review now (for more information see e.g. \cite{Antoniadis:2009nv}).
\subsubsection{Linearized On-shell Hypermultiplet}
The first superfield which we will discuss is the {\it linearized  on-shell} hypermultiplet ($\cN=2$ matter multiplet)
\begin{align}
q^{+}(x^\mu,\q^+,\bq_-,u) = f^{i} u^{+}_{i} + \q^{+}_\a\, \chi^{\a} +\bar\psi_{\da}\, \bq^\da_{-} +\text{derivative terms}.\label{01}
\end{align}
Here $f^{i}$ are the two complex scalars,
$\chi^{\a}$ and $\bar\psi_{\da}$ are the two fermions of the on-shell multiplet. To exhibit manifest G-analyticity, one has to choose the appropriate analytic basis in superspace,
\begin{equation}\label{chbashss}
    x^\mu\ \to \ x^\mu + i \q^{+}\sigma^\mu\bar\q_{+} - i \q^{-}\sigma^\mu\bar\q_{-}\ ,
\end{equation}
analogous to the familiar chiral basis. Then $q^{+}$ satisfies the conditions
\begin{align}
D_-^\alpha q^{+}(\q^+,\bq_-,u)=\bar{D}^+_{\dot{\alpha}} q^{+}(\q^+,\bq_-,u)=0\,.\label{GanalyticityHyper}
\end{align}
Note that the harmonic dependence in (\ref{01}) is cut down to linear. This is typical for on-shell
multiplets which, in addition to the G-analyticity condition (\ref{GanalyticityHyper}), also satisfy the H-analyticity condition
\begin{equation}\label{hansf}
    D_{-}{}^{+} q^{+}(\q^+,\bq_-,u) = 0\ .
\end{equation}
Here the harmonic derivative is supersymmetrized by going to the manifestly G-analytic superspace coordinates (\ref{chbashss}). One can show that the `ultrashort' on-shell superfield (\ref{01}) is a solution to the simultaneous conditions for G- and H-analyticity \cite{Galperin:1984av,Hartwell:1994rp,Andrianopoli:1999vr}.

Note that in the $\cN=2$ G-analytic superspace there exists a special conjugation $\
\widetilde{}\ $ combining complex conjugation with a reflection on the harmonic coset,
such that G-analyticity is preserved. In this sense we can define the conjugate hypermultiplet
\begin{align}
&\tilde q_-(x^\mu,\q^+,\bq_-,u) = \bar f_{i} \bu_-^{i} + \bq_-^{\da}\, \bar\chi_{\da} +\psi^{\a}\, \q^+_\a +\text{derivative terms}.\label{01'}
\end{align}
In what follows it will be convenient to combine the two versions of the hypermultiplet into a doublet of an external $SU(2)$ (not the R symmetry one), $(q^+, \tilde q_-) \ \leftrightarrow \ q^+_a$, $a=1,2$.
\subsubsection{Linearized On-shell Vector Multiplet}
Another example of a G-analytic superfield is the {\it linearized}  on-shell vector
multiplet. It is obtained from the off-shell chiral field strength
\begin{equation}\label{14}
    W(\q_\a^i) = \varphi + \q_\a^i  \la^\a_i + \q_\a^i \q_\b^j\, ( \ep_{ij}F^{(\a\b)}_{(+)} + \ep^{\a\b} S_{(ij)}) \ .
\end{equation}
Here $\varphi$ is the complex physical scalar and $F_{(+)}$ is the self-dual part of the gluon field strengths, while $S$ is a triplet of auxiliary fields. On-shell the latter must vanish, hence
the additional constraint
\begin{equation}\label{15}
   \ep_{\a\b} D_i^\a D_j^\b\ W = 0\,.
\end{equation}
Now, define the superfield (a superdescendant of $W$)
\begin{equation}\label{16}
    K_{-}^\a = D_{-}^\a\ W\,,
\end{equation}
where we have projected the $SU(2)$ index of $D^\a_i$ with the harmonic
$\bu_{-}^i$ . This superfield is annihilated by half of the spinor
derivatives and hence is 1/2 BPS short. Indeed, this is true for the projections $\bar
D^{+}_{\dot\b}$ since $\{\bar D^{+}, D_{-}\}=0$ and $\bar D^i_{\dot\b}W=0$ (chirality).
Further, hitting (\ref{16}) with $D_{-}^\b$ we obtain zero as a consequence of
the projection of the on-shell constraint (\ref{15}) with $\bu_{-}^i \bu_{-}^j$.
We conclude that $K_{-}^\a$ satisfies the same G-analyticity constraints as the on-shell hypermultiplet (\ref{GanalyticityHyper})
\begin{equation}\label{16'}
    D_{-}^\b K_{-}^\a = \bar D^{+}_{\dot\b} K_{-}^\a = 0\,,
\end{equation}
which imply that $K_-$ depends only on half of the $\q$'s:
\begin{equation}\label{17}
    K_{-}^\a(\q^+,\bq_-,u) = \la^\a_i  \bu_{-}^i  + (\sigma^{\m})^{\a\da}\bq_{\da\,-}\ i\pa_\m\varphi + \q^+_\b\, F^{\a\b}_{(+)} +  \mbox{derivative terms}.
\end{equation}
In addition, the harmonic dependence of $K_{-}^\a$ is restricted to be linear. As in (\ref{hansf}), this follows from the condition for H-analyticity
\begin{equation}\label{hwconK}
    D_{-}{}^{+} K_{-}^\a = 0 \ ,
\end{equation}
in turn derived from the harmonic independence of $W$ ($D_{-}{}^{+}  W = 0$) and the commutator $[D_{-}{}^{+} , D_{-}^\a]=0$. This is another example of an ultrashort superfield. Note, however, that it is not a primary object but rather a superdescendant of the chiral  on-shell vector
multiplet.
\subsection{Quaternionic Geometry}\label{App:QuatGeom}
For completeness, we will review in this appendix some relevant aspects of quaternionic manifolds. We will closely follow the discussion in \cite{Galperin:1992pj} and use the same notation and conventions as in \cite{Antoniadis:2009nv}. Following \cite{a9,a12,a13}, a quaternionic manifold $\mathfrak{M}$ is defined as a $4n$ dimensional Riemann manifold whose holonomy group is restricted to a subgroup of $Sp(n)\times Sp(1)$. Therefore, we can locally choose a coordinate frame $\{x^{Mk}\}$ where $M=1,\ldots, 2n$ and $k=1,2$ are indices of $Sp(n)$ and $Sp(1)\sim SU(2)$ respectively. In this coordinate frame, the covariant derivatives can be written as
\begin{align}
{\cal D}_{A i} &= e^{M k}_{A i}\partial_{M k}- \omega_{A i(CD)}\Gamma^{(CD)}- \omega_{A i(lk)}\Gamma^{(lk)}\nonumber\\ 
:&=  \nabla_{A i}- \omega_{A i(CD)}\Gamma^{(CD)}- \omega_{A i(lk)}\Gamma^{(lk)}\,, \label{5.3.1}
\end{align}
where we have introduced the $Sp(n)$ and $Sp(1)$ connections $\omega_{A i(CD)}$ and $\omega_{A i(lk)}$ respectively. Our convention for the $Sp(n)$ and $Sp(1)$ algebra of the generators $\Gamma^{(CD)}$ and $\Gamma^{(lk)}$ follows \cite{Galperin:1992pj} and is given by
\begin{align}
[\Gamma^{(CD)},\Gamma^{(EF)}]&=\tfrac{1}{2}\left(\Omega^{CE}\Gamma^{(DF)}+ \Omega^{CF}\Gamma^{(DE)}  + \Omega^{DE} \Gamma^{(CF)}+ \Omega^{DF}\Gamma^{(CE)}\right)\,,\\
[\Gamma^{(lk)},\Gamma^{(mn)}]&=\tfrac{1}{2}\left(\epsilon^{lm}\Gamma^{(kn)}+ \epsilon^{ln}\Gamma^{(km)}  + \epsilon^{km} \Gamma^{(ln)}+ \epsilon^{kn}\Gamma^{(lm)}\right)\,,
\end{align}
where $\Omega^{AB}$ denotes the $Sp(n)$ invariant tensor. In this paper we will only consider the fundamental spinor representations of $Sp(n)$ and $Sp(1)$, in which case
\begin{align}
{({\cal D}_{A i})_{B n}}^{B' n'} = \delta_B^{B'}\delta_n^{n'}\nabla_{A i} + \delta_n^{n'}\omega_{A i\;B}{}^{B'} + \delta_B^{B'}\omega_{A i\;n}{}^{n'}\;. \label{5.3.2}
\end{align}
Particularly, we will be interested in the commutator of two covariant derivatives leading to the $Sp(n)$ and $Sp(1)$ components of the curvature tensor
\begin{align}
{[{\cal D}_{A i},{\cal D}_{B j}]_{C n}}^{C' n'} = \delta_n^{n'} {R_{A i\;B j\; C}}^{C'}+ \delta_C^{C'}{R_{A i\;B j\;n}}^{n'} =: {R_{A i,\;B j\; C n}}^{C' n'}\;. \label{5.4.1}
\end{align}
According to \cite{Galperin:1992pj}, restricting the holonomy group of $\mathfrak{M}$ to (a subgroup of) $Sp(n) \times Sp(1)$ implies the covariant constraints\footnote{The defining constraint of hyper-K\"ahler manifolds \cite{a2} implies that the right-hand side of (\ref{5.4.2b}) vanishes, while in the quaternionic case it becomes identical to the non-vanishing $Sp(1)$ part of the holonomy group.}
\begin{align}
&R_{A i\;B j\; C}{}^{C'} &= \epsilon_{ij} R_{AB ; C}{}^{C'} \,,&&\text{and} &&R_{A i\;B j\;n}{}^{n'} &= \Omega_{AB}R_{i j\;n}{}^{n'}\;. \label{5.4.2b}
\end{align}
Hence we have for the commutator 
\begin{align}
[{\cal D}_{A i},{\cal D}_{B j}]_{C n}{}^{C' n'} = -2\;\delta_C^{C'}\Omega_{AB}\;R\;\tz_{(ij)\;n}{}^{n'} + \delta_n^{n'}\epsilon_{ij}[ R_{(ABC}{}^{C')}- R\;(\Omega_{BC}\delta_A^{C'}+ \Omega_{AC}\delta_B^{C'})]\,, \label{5.5.1}
\end{align}
where we have introduced
\begin{align}
&\tz_{(ij)\;n}{}^{n'}:={1\over 2}(\epsilon_{in}\delta_j^{n'}+
\epsilon_{jn}\delta_i^{n'})\,,&& \text{and} && R:= {1\over 6}R_{(ij)}{}^{(ij)}\,.\nn
\end{align}
as the $Sp(1)$ generators analog to (\ref{subhd}) and (\ref{cosethd}). Projecting with the corresponding harmonic variables, we particularly obtain
\begin{align}
&[{\cal D}_{A \pm},{\cal D}_{B \pm}]_{C n}{}^{C' n'} = \pm2\delta_C^{C'}\Omega_{AB}\;R\; (\tz_{\pm\pm})_n{}^{n'}\, \label{firstt}\\
&{[}{\cal D}_{A +},{\cal D}_{B -}{]}_{C n}{}^{C' n'} = -2\delta_C^{C'}\Omega_{AB}\;R\; (\tz_{0})_n{}^{n'} + \delta_n^{n'}[ R_{(ABC}{}^{C')}- R\;(\Omega_{BC}\delta_A^{C'}+ \Omega_{AC}\delta_B^{C'})]\,, \label{secondt}
\end{align}
where we have introduced the projected quantities
\begin{align}
&\tz_0 := \tz_{+-} = \tz_{-+}=u^i_+ u^j_-\tz_{ij}, &&\text{and} &&  \tz_{\pm\pm}=\mp\frac{1}{2} u^i_\pm u^j_\pm \tz_{ij}.
\end{align}
\section{Heterotic $K3\times T^2$ Compactification}\label{App:HetK3comp}
\setcounter{equation}{0}
In this appendix we recall some important features of heterotic $K3\times T^2$ compactifications, in particular expressions for various vertex operators and picture changing operators as well as our conventions for the spin-structure sum.
\subsection{Vertex Operators}\label{App:HetK3compVertex}
The space-time and $T^2$ torus piece of all vertex operators is fairly standard and will be written using the free bosons $\phi_{1,2}$ and $\phi_3$ respectively (for more information see e.g.~\cite{Antoniadis:2009nv}). For the $K3$ piece we recall that the $SU(2)$ current algebra inside the $\N=4$ superconformal algebra can be bosonized in terms of a free boson which we call $H$:
\begin{align}
&J_{K3}= i\sqrt{2}\partial H\,, && J_{K3}^{\pm \pm} = e^{i\pm \sqrt{2} H}\,, &&G^{\pm}_{K3,i}= e^{\pm \frac{i}{\sqrt{2}} H} \hat{G}_{K3,i} \,.
\end{align}
Here $\hat{G}_{K3,i}$ are dimension $5/4$ operators which have non-singular OPE with $H$ and have no spin structure dependence. The latter enters through the projections and shifts in the $U(1)$ charge lattice of $J_{K3}$ which in turn is given by the momentum lattice of $H$.  Therefore in the $\N=4$ internal theory, only correlation functions and the partition function containing $H$ depend on the spin-structure. The term in the picture changing operator containing a $\N=4$ superconformal generator is
\begin{equation}
P = e^{\phi} ( G^+_{K3,+} + G^-_{K3,-} )+\ldots\,,
\end{equation}
where dots indicate the remaining terms and $\phi$ bosonizes the superghost.

Concerning the vertex operators of physical fields, the chiral vector multiplet scalar in the $(-1)$-ghost picture is simply $\psi_3$ and in particular does not depend on the $K3$ fields. The gaugino vertex however involves, besides the space-time and torus spin-field, also $e^{\pm \frac{i}{\sqrt{2}} H}$. Thus the vertex operator for the gauginos  $\lambda_{\mp}$ and  $\bar{\lambda}^{\pm}$ in the $(-1/2)$-picture also carry $e^{\pm \frac{i}{\sqrt{2}} H}$. 

Finally the vertex operators (at zero momentum) for the hyperscalar $f_A^{\pm}$ in the $(-1)$ and $(0)$ ghost pictures are given by
\begin{align}
&V^{(-1)}_{f_A^{\pm}}= e^{-\phi}e^{\pm \frac{i}{\sqrt{2}} H} \hat{V}_{A}\,,&&\text{and}&& V^{(0)}_{f_A^{\pm}}=P V^{(-1)}_{f_A^{\pm}}=:\hat{G}_{K3,\pm} \hat{V}_{A}:\,,
\label{hypervertex}
\end{align}
where $\hat{V}_{A}$ have dimension $1/4$ and have non-singular OPE with $H$. The normal ordered expression $:\hat{G}_{K3,\pm} \hat{V}_{A}:$ is defined as the coefficient of the $1/\sqrt{z}$ singularity in the corresponding OPE.

The vertex operators in the $(-1/2)$ ghost picture for the hyperfermions $\chi_A$ and $\bar{\psi}_A$ are
\begin{align}
&V^{(-1/2)}_{\chi^{\alpha}_A} = e^{-\frac{\phi}{2}} S^{\alpha} e^{-i\frac{\phi_3}{2}} \hat{V}_{A}\,,&&\text{and} &&V^{(-1/2)}_{\bar{\psi}^{\dot{\alpha}}_A} = e^{-\frac{\phi}{2}} S^{\dot{\alpha}} e^{i\frac{\phi_3}{2}} \hat{V}_{A}\,,
\end{align}
where $S^{\alpha}$ are the space-time spin fields.

As mentioned above the spin structure dependence enters only through the superghost, spacetime and 
torus fermions and the charge lattice of $H$. It does not depend in particular on the rest of the details of the $\N=4$ superconformal theory. On the other hand, the topological theory (besides shifting the dimensions of the torus fermion) involves precisely twisting  by adding an appropriate background charge for the field $H$ and the rest of the internal $\N=4$ theory is insensitive to this twisting. 

\subsection{Spin Structure}\label{App:HetK3compSpin}
Let $\Gamma$ be the $U(1)$ lattice of $H$ charges. The space-time and torus fermions define an $SO(2)\times SO(2) \times SO(2)$ lattice. If one takes an $SO(2)\times SO(2)$ sublattice thereof and combines it with $\Gamma$, then (see \cite{Lerche:1988np,Lechtenfeld:1989be,Lust:1988yf}), the resulting 3-dimensional lattice is given by the coset $E_7/SO(8)$. The characters are given by the branching functions $F_{\Lambda,s}(\tau)$ and satisfy:
\begin{align}
\chi_{\Lambda}(\tau)= \sum_{s} F_{\Lambda,s}(\tau) \chi_s(\tau)\,,
\end{align}
where $\chi_{\Lambda}$ and $\chi_s$ are the $E_7$ and $SO(8)$ level one characters respectively, $\Lambda$ denotes the two conjugacy classes of $E_7$ and $s$ represent the four conjugacy classes of $SO(8)$ in the spin structure basis. The characters of the internal $\N=4$ superconformal field theory times two free complex fermions can therefore be expressed as $\sum_{\Lambda} F_{\Lambda,s}(\tau) Ch_{\Lambda}(\tau)$ where $Ch_{\Lambda}(\tau)$ is the contribution of the rest of the internal theory and most importantly does not depend on the spin-structure. The generalization to higher genus is obtained by assigning an $E_7$ conjugacy class $\Lambda$ for each loop and we will denote this collection by $\{\Lambda\}$. We can define a more general character $F_{\{\Lambda\},s}(a_1,a_2,a_3)$ by introducing chemical potentials for the three charges; $a_1$ and $a_2$ coupling to the two $SO(2)$ charges and $a_3$ to $H$-charge. For a genus $g$ surface, the couplings $a_1$, $a_2$ and $a_3$ each are $g$-dimensional vectors and represent the coupling to charges going through each loop. In the calculation of the amplitudes $(a_1,a_2,a_3)$ are related to the positions of various vertex operators weighted by the corresponding charges via Abel map. The spin structure sum is given by the formula:
\begin{align}
\sum_{s} F_{\{\Lambda\},s}(a_1,a_2,a_3)=
F_{\{\Lambda\}}(\frac{1}{2}(a_1+a_2+\sqrt{2}a_3),\frac{1}{2}(a_1+a_2-\sqrt{2}a_3),
\frac{1}{\sqrt{2}}(a_1-a_2))\,,\label{sssum}
\end{align}
where we have introduced
\begin{align}
&F_{\{\Lambda\}}(a_1,a_2,a_3)=\vartheta(\tau,a_1)\vartheta(\tau,a_2)\Theta(\tau,a_3)\,,\\
&\Theta(\tau,a_3)=\sum_{n_i \in Z} e^{2\pi i(n_i+\frac{\lambda_i}{2})\tau_{ij}(n_j+\frac{\lambda_j}{2})+2\pi i
\sqrt{2}(n_i+\frac{\lambda_i}{2})(a_3)_i}\,,
\end{align}
where $\lambda_i$ (with $i=1,...,g$) are $0$ and $1$ for the $E_7$ conjugacy classes $\mathbf{1}$ and $\mathbf{56}$ respectively. In fact, apart from the non-zero mode determinant of a scalar,  $\Theta$ is just the character valued genus $g$ partition function of level one $SU(2)$, with the two classes above corresponding to the two representations of level one $SU(2)$ Kac-Moody algebra based on $SU(2)$ representations $\mathbf{1}$ and $\mathbf{2}$. 
\section{Gauge Freedom}\label{App:GaugeFreedomePotential}
To see the gauge freedom of the potential for $\hat{\mathcal{H}}$ we consider the ($\alpha$-dependent) potential function $\hat{\mathcal{H}}^{g,A_1\ldots A_{m_1+1},B_1,\ldots,B_{m_2}}_{\bar{I}_1\ldots\bar{I}_{n-m_1},\bar{J}_1\ldots\bar{J}_{n}}(\alpha)$ but insert the unit operator in the form of $\oint \bar{\psi}_3\psi_3$ at a different position $\beta$. Then we deform the $\oint\bar{\psi}_3$ contour integral to obtain
{\allowdisplaybreaks
\begin{align}
&(3g+n-3)!\hat{\mathcal{H}}^{g,A_1\ldots A_{m_1+1},B_1\ldots B_{m_2}}_{\bar{I}_1\ldots\bar{I}_{n-m_1},\bar{J}_1\ldots\bar{J}_{n-m_2}}(\alpha)=\nonumber\\
&=\int_{\mathcal{M}_{(g,n)}}\hspace{-0.7cm}\langle(\mu\cdot G^-)^{3g+n-3}\psi_3(\alpha)\oint \bar{\psi}_3\psi_3(\beta)\prod_{a=1}^{m_1+1}\int \bar{\Xi}^{A_a}\prod_{b=1}^{n-m_1}\int\bar{\psi}_3\bar{J}_{\bar{I}_b}\prod_{c=1}^{m_2}(\psi_3\Xi^{B_c})\prod_{d=1}^{n-m_2}(J^{++}_{K3}\bar{J}_{\bar{J}_d})\rangle\nonumber\\
&=\int_{\mathcal{M}_{(g,n)}}\hspace{-0.7cm}\langle(\mu\cdot G^-)^{3g+n-3}\oint \bar{\psi}_3\psi_3(\alpha)\psi_3(\beta)\prod_{a=1}^{m_1+1}\int \bar{\Xi}^{A_a}\prod_{b=1}^{n-m_1}\int\bar{\psi}_3\bar{J}_{\bar{I}_b}\prod_{c=1}^{m_2}(\psi_3\Xi^{B_c})\prod_{d=1}^{n-m_2}(J^{++}_{K3}\bar{J}_{\bar{J}_d})\rangle\nonumber\\*
&\hspace{0.5cm}+\int_{\mathcal{M}_{(g,n)}}\hspace{-0.7cm}\langle(\mu\cdot G^-)^{3g+n-3}\psi_3(\alpha)\psi_3(\beta)\prod_{a=1}^{m_1+1}\int \bar{\Xi}^{A_a}\Xi^{[B_1}\prod_{b=1}^{n-m_1}\int\bar{\psi}_3\bar{J}_{\bar{I}_b}\prod_{c=2}^{m_2}(\psi_3\Xi^{B_c]})\prod_{d=1}^{n-m_2}(J^{++}_{K3}\bar{J}_{\bar{J}_d})\rangle\nonumber\\
&=(3g+n-3)!\hat{\mathcal{H}}^{g,A_1\ldots A_{m_1+1},B_1\ldots B_{m_2}}_{\bar{I}_1\ldots\bar{I}_{n-m_1},\bar{J}_1\ldots\bar{J}_{n-m_2}}(\beta)+(3g+n-3)\mathcal{D}_+^{[B_1}\int_{\mathcal{M}_{(g,n)}}\hspace{-0.7cm}\langle(\mu\cdot G^-)^{3g+n-4}\psi_3(\alpha)\psi_3(\beta)\cdot\nonumber\\
&\hspace{0.5cm}\cdot\prod_{a=1}^{m_1+1}\int \bar{\Xi}^{A_a}\prod_{b=1}^{n-m_1}\int\bar{\psi}_3\bar{J}_{\bar{I}_b}\prod_{c=1}^{m_2}(\psi_3\Xi^{B_c]})\prod_{d=1}^{n-m_2}(J^{++}_{K3}\bar{J}_{\bar{J}_d})\rangle
\end{align}}
This indeed shows that
\begin{align}
&\hat{\mathcal{H}}^{g,A_1\ldots A_{m_1+1},B_1\ldots B_{m_2}}_{\bar{I}_1\ldots\bar{I}_{n-m_1},\bar{J}_1\ldots\bar{J}_{n-m_2}}(\alpha)-\hat{\mathcal{H}}^{g,A_1\ldots A_{m_1+1},B_1\ldots B_{m_2}}_{\bar{I}_1\ldots\bar{I}_{n-m_1},\bar{J}_1\ldots\bar{J}_{n-m_2}}(\beta)=\mathcal{D}_+^{[B_1}\mathcal{G}^{g,A_1\ldots A_{m_1+1},B_2\ldots B_{m_2}]}_{\bar{I}_1\ldots\bar{I}_{n-m_1},\bar{J}_1\ldots\bar{J}_{n-m_2}}(\alpha,\beta)\,,
\end{align}
with the quantity
\begin{align}
&\mathcal{G}^{g,A_1\ldots A_{m_1+1},B_2\ldots B_{m_2}}_{\bar{I}_1\ldots\bar{I}_{n-m_1},\bar{J}_1\ldots\bar{J}_{n-m_2}}(\alpha,\beta):=\frac{1}{(3g+n-4)!}\int_{\mathcal{M}_{(g,n)}}\hspace{-0.7cm}\langle(\mu\cdot G^-)^{3g+n-4}\psi_3(\alpha)\psi_3(\beta)\prod_{a=1}^{m_1+1}\int \bar{\Xi}^{A_a}\cdot\nonumber\\
&\hspace{1cm}\cdot\prod_{b=1}^{n-m_1}\int\bar{\psi}_3\bar{J}_{\bar{I}_b}\prod_{c=2}^{m_2}(\psi_3\Xi^{B_c})\prod_{d=1}^{n-m_2}(J^{++}_{K3}\bar{J}_{\bar{J}_d})\rangle\,.
\end{align}
For further convenience and in order to save writing, we will also introduce a generating functional for $\mathcal{G}^{g,A_1\ldots A_{m_1+1},B_2\ldots B_{m_2}}_{\bar{I}_1\ldots\bar{I}_{n-m_1},\bar{J}_1\ldots\bar{J}_{n-m_2}}(\alpha,\beta)$
\begin{align}
\mathbb{G}^g(\alpha,\beta):=\sum_{n=0}^\infty\sum_{m_1=0}^n\sum_{m_2=0}^n&\frac{\grass ^{\bar{I}_1}\ldots\grass ^{\bar{I}_{n-m_1}}\chi_{A_1}\ldots \chi_{A_{m_1+1}}\eta^{\bar{J}_1}\ldots\eta^{\bar{J}_{n-m_2}}\xi_{B_1}\ldots \xi_{B_{m_2-1}}}{(n-m_1)!(n-m_2)!(m_1+1)!(m_2+1)!}\cdot\nonumber\\
&\cdot\mathcal{G}^{g,A_1\ldots A_{m_1+1},B_1\ldots B_{m_2-1}}_{\bar{I}_1\ldots\bar{I}_{n-m_1},\bar{J}_1\ldots\bar{J}_{n-m_2}}(\alpha,\beta)\,.\label{GeneratingFunctG}
\end{align}
\section{Direct String Derivation of Differential Equations}\label{stringdiffeqs}
\setcounter{equation}{0}
\subsection{String Derivation of Differential Equations}
\begin{figure}[htb]
\begin{center}
\epsfig{file=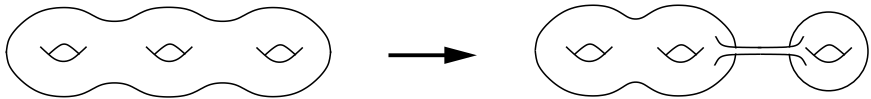, width=11.5cm}
\caption{Pinching of a dividing geodesic of a genus $g=3$ Riemann surface: The shrinking of a non-contractible cycle divides the Riemann surface into two new ones with genus $g=2$ and $g=1$ respectively.}
\label{fig:DegenDivGeo}
\end{center}
\end{figure}

\begin{figure}[htb]
\begin{center}
\epsfig{file=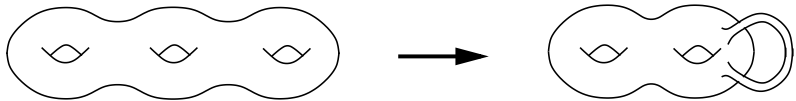, width=10.5cm}
\caption{Pinching of a handle of a genus $g=3$ Riemann surface: The shrinking of a non-contractible cycle results in a Riemann surface of genus $g=2$ with two punctures.}
\label{fig:DegenHandle}
\end{center}
\end{figure}

Our strategy in deriving differential equations with respect to $\N=2$ moduli directly within the setup of topological string amplitudes follows closely \cite{Bershadsky:1993cx}: Moduli derivatives are directly translated into vertex operator insertions within the twisted correlators. In particular we will consider derivatives with respect to vector multiplet and hypermultiplet moduli, for which the relevant insertions take the form
\begin{align}
&\mathcal{D}_{\bar{I}}\longleftrightarrow\int\oint G^+_{T^2}\bar{\psi}_3 \bar{J}_{\bar{I}}\,,&&\text{and}&&\mathcal{D}_i^A\longleftrightarrow \int\oint G^+_{K3,i}\bar{\Xi}^A\,.
\end{align}
We will then further deform the contour integrals (within the topologically twisted setup this is indeed possible on dimensional grounds) thereby trying to manipulate the correlation functions. A generic feature in this type of computations is the appearance of \emph{world-sheet boundary terms}. The latter appear when the above mentioned contour integrals act on the measure of the twisted correlator, where they might give rise to insertions of the total energy momentum tensor $T=T_{T^2}+T_{K3}$ sewed with some Beltrami differential that parametrizes a particular deformation of the genus $g$-world-sheet. The full energy-momentum tensor can in fact be written as a total derivative with respect to this particular modulus of the genus $g$-Riemann surface, thus leading to a boundary contribution in the integral over $\mathcal{M}_{g}$. These boundary contributions can pictorially be quite directly figured as degeneration limits of the genus $g$ Riemann surface and (see e.g. \cite{Antoniadis:1996qg,Cecotti:1992qh,Bershadsky:1993ta,Antoniadis:2006mr,Antoniadis:2007ta}) there are two homologically distinct types corresponding to the pinching of either a dividing geodesic or a handle. Examples for genus $g=3$ are depicted in Figure~\ref{fig:DegenDivGeo} and \ref{fig:DegenHandle} respectively.
Assuming that we are at a generic point in the vector multiplet moduli-space with no charged massless states present there will be no contribution from pinching a handle. It therefore remains to study the pinching of a dividing geodesic where the genus $g$ surface splits into two surfaces of genus $g_s$ and $g-g_s$ respectively\footnote{Note that $g_s$ can also take value zero or $g$ since we have seen that these topological amplitudes start from genus zero.} and to determine which states may propagate on the long and thin tube. In principle there are two different types, namely hyper- and vector multiplet states. We will discuss both in the following.

In our particular case, there is in fact a further type of degeneration besides these purely geometric ones. Indeed, since our world-sheets have in addition an arbitrary number of punctures (and the moduli space is thus rather $\mathcal{M}_{(g,n)}$) we also need to take into account contributions when two of these punctures collide with each other. Such contact terms are notoriously difficult to handle within the string approach. In our differential equations they will manifest themselves as curvature contributions and take into account the fact that the moduli space of the string compactification is in general not flat. We will not be able to say much about these terms in the general case, but we will propose consistent expressions for the case $m_2=0$ as we go along.
\subsection{String Differential Equation for $\tilde{\mathcal{H}}^{g,A_1\ldots A_{m_1},B_1\ldots B_{m_2+1}}_{\bar{I}_1\ldots\bar{I}_{n-m_1},\bar{J}_1\ldots\bar{J}_{n-m_2}}$}\label{DiffTildeH}
This appendix is devoted to the derivation of a differential equation for $\tilde{\mathcal{H}}^{g,A_1\ldots A_{m_1},B_1\ldots B_{m_2+1}}_{\bar{I}_1\ldots\bar{I}_{n-m_1},\bar{J}_1\ldots\bar{J}_{n-m_2}}$ along the lines of section~\ref{Sect:DiffEqTildeH}. The derivation will be divided into first considering the bulk contributions and afterwards the boundary terms, which appear from a particular deformation of the genus $g$ world-sheet.
\subsubsection{Bulk Equation}
We start by explicitly applying the operator (\ref{harop}) to the potential $\tilde{\mathcal{H}}$ given in (\ref{DefTildeH})
\begin{align}
&\epsilon^{ij}\frac{\partial}{\partial\bar{u}^i_+}\mathcal{D}^{[A_{1}}_j\tilde{\mathcal{H}}^{g,A_2\ldots A_{m_1}],B_1\ldots B_{m_2+1}}_{\bar{I}_1\ldots\bar{I}_{n-m_1},\bar{J}_1\ldots\bar{J}_{n-m_2}}=\frac{\epsilon^{ij}}{N}\frac{\partial}{\partial\bar{u}^i_+}\int_{\mathcal{M}_{(g,n+1)}}\hspace{-1cm}\langle\gtmin^{g+m_1+m_2-n}\gkmpin^{2g+2n-m_1-m_2-3}\nonumber\\
&\hspace{0.5cm}\cdot(\mu\cdot J^{--}_{K3})\int\oint G^+_{K3,i}\bar{\Xi}^{[A_{1}}\prod_{a=2}^{m_1}\int \bar{\Xi}^{A_a]} \prod_{b=1}^{m_2+1}(\psi_3\Xi^{B_b})\prod_{c=1}^{n-m_1}\int\bar{\psi}_3\bar{J}_{\bar{I}_c}\prod_{d=1}^{n-m_2}(J^{++}_{K3}\bar{J}_{\bar{J}_d})\rangle\,.
\end{align}
where for convenience we introduce the shorthand notations
\begin{align}
N&:=(g+m_1+m_2-n)!(2g+2n-m_1-m_2-3)!\,,\\
N'&:=(g+m_1+m_2-n)!(2g+2n-m_1-m_2-4)!\,,
\end{align}
in order to save writing. Deforming the contour integral, we find two different terms
\begin{align}
&=\frac{\epsilon^{ij}}{N'}\frac{\partial}{\partial\bar{u}^i_+}\bigg[\bar{u}^k_+\epsilon_{kj}\int_{\mathcal{M}_{(g,n+1)}}\hspace{-0.7cm}\langle\gtmin^{g+m_1+m_2-n}\gkmpin^{2g+2n-m_1-m_2-4}(\mu\cdot J^{--}_{K3})(\mu\cdot T_{K3})\cdot\nonumber\\
&\hspace{1cm}\cdot \prod_{a=1}^{m_1}\int \bar{\Xi}^{A_a}\prod_{b=1}^{m_2+1}(\psi_3\Xi^{B_b})\prod_{c=1}^{n-m_1}\int\bar{\psi}_3\bar{J}_{\bar{I}_c}\prod_{d=1}^{n-m_2}(J^{++}_{K3}\bar{J}_{\bar{J}_d})\rangle\bigg]+\nonumber\\
&+\frac{\epsilon^{ij}}{N}\int_{\mathcal{M}_{(g,n+1)}}\hspace{-0.7cm}\langle\gtmin^{g+m_1+m_2-n}\gkmpin^{2g+2n-m_1-m_2-4}(\mu\cdot G^-_{K3,i})(\mu\cdot G^-_{K3,j})\prod_{a=1}^{m_1}\int \bar{\Xi}^{A_a}\cdot\nonumber\\
&\hspace{1cm}\cdot\prod_{b=1}^{m_2+1}(\psi_3\Xi^{B_b})\prod_{c=1}^{n-m_1}\int\bar{\psi}_3\bar{J}_{\bar{I}_c}\prod_{d=1}^{n-m_2}(J^{++}_{K3}\bar{J}_{\bar{J}_d})\rangle\,.
\end{align}
The second term in fact vanishes due to the anti-symmetrization of the Beltrami differentials and the fermionic nature of the $G^-_{K3,i}$ operators. For the remaining term we can perform the $\bar{u}^i_+$ differential and furthermore use $T=T_{T^2}+T_{K3}$. Denoting the boundary contribution as $\tilde{\mathcal{C}}^{\text{(bdy)},A_1\ldots A_{m_1},B_1\ldots B_{m_2+1}}_{\bar{I}_1\ldots\bar{I}_{n-m_1},\bar{J}_1\ldots\bar{J}_{n-m_2}}$ we obtain
\begin{align}
&=-\frac{2g+2n-m_1-m_2-2}{N'}\int_{\mathcal{M}_{(g,n+1)}}\hspace{-0.7cm}\langle\gtmin^{g+m_1+m_2-n}\gkmpin^{2g+2n-m_1-m_2-4}(\mu\cdot J^{--}_{K3})\cdot\nonumber\\
&\hspace{1.5cm}\cdot (\mu\cdot \oint G^+_{T^2}G^-_{T^2})\prod_{a=1}^{m_1}\int \bar{\Xi}^{A_a}\prod_{b=1}^{m_2+1}(\psi_3\Xi^{B_b})\prod_{c=1}^{n-m_1}\int\bar{\psi}_3\bar{J}_{\bar{I}_c}\prod_{d=1}^{n-m_2}(J^{++}_{K3}\bar{J}_{\bar{J}_d})\rangle+\nonumber\\
&\hspace{0.5cm}+(2g+2n-m_1-m_2-2)\,\tilde{\mathcal{C}}^{\text{(bdy)},A_1\ldots A_{m_1},B_1\ldots B_{m_2+1}}_{\bar{I}_1\ldots\bar{I}_{n-m_1},\bar{J}_1\ldots\bar{J}_{n-m_2}}\,.
\end{align}
Deforming the contour integration, we obtain 
\begin{align}
&=-\frac{2g+2n-m_1-m_2-2}{N'}\int_{\mathcal{M}_{(g,n+1)}}\hspace{-0.7cm}\langle\gtmin^{g+m_1+m_2-n+1}(\mu\cdot J^{--}_{K3})\gkmpin^{2g+2n-m_1-m_2-4}\cdot\nonumber\\
&\hspace{1cm}\cdot \prod_{a=1}^{m_1+1}\int \bar{\Xi}^{A_a}\prod_{b=1}^{m_2+1}(\psi_3\Xi^{B_b})\prod_{c=1}^{n-m_1-1}\int\bar{\psi}_3\bar{J}_{[\bar{I}_c}\int\oint G^+_{T^2}\bar{\psi}_3\bar{J}_{\bar{I}_{n-m_1}} \prod_{d=1}^{n-m_2}(J^{++}_{K3}\bar{J}_{\bar{J}_d})\rangle+\nonumber\\
&\hspace{0.5cm}+(2g+2n-m_1-m_2-2)\,\tilde{\mathcal{C}}^{\text{(bdy)},A_1\ldots A_{m_1+1},B_1\ldots B_{m_2+1}}_{\bar{I}_1\ldots\bar{I}_{n-m_1},\bar{J}_1\ldots\bar{J}_{n-m_2}}\,.
\end{align}
The insertion $\int\oint G^+_{T^2}\bar{\psi}_3\bar{J}_{\bar{I}_{n-m_1}}$ is in fact a derivative with respect to $\mathcal{D}_{\bar{I}_{n-m_1}}$. Pulling it out and comparing with the definition (\ref{DefTildeH}) we find 
\begin{align}
&(2g+2n-m_1-m_2-2)\mathcal{D}_{[\bar{I}_{n-m_1}}\tilde{\mathcal{H}}^{g,A_1\ldots A_{m_1},B_1\ldots B_{m_2+1}}_{\bar{I}_1\ldots\bar{I}_{n-m_1-1}],\bar{J}_1\ldots\bar{J}_{n-m_2}}+\epsilon^{ij}\frac{\partial}{\partial\bar{u}^i_+}\mathcal{D}^{[A_{1}}_j\tilde{\mathcal{H}}^{g,A_2\ldots A_{m_1}],B_1\ldots B_{m_2+1}}_{\bar{I}_1\ldots\bar{I}_{n-m_1},\bar{J}_1\ldots\bar{J}_{n-m_2}}=\nonumber\\
&\hspace{1cm}=(2g+2n-m_1-m_2-2)\,\tilde{\mathcal{C}}^{\text{(bdy)},A_1\ldots A_{m_1},B_1\ldots B_{m_2+1}}_{\bar{I}_1\ldots\bar{I}_{n-m_1},\bar{J}_1\ldots\bar{J}_{n-m_2}}\,.
\end{align}
Notice that the prefactor is in fact the total harmonic charge of $\tilde{\mathcal{H}}^{g,A_1\ldots A_{m_1+1},B_1\ldots B_{m_2+1}}_{\bar{I}_1\ldots\bar{I}_{n-m_1-1},\bar{J}_1\ldots\bar{J}_{n-m_2}}$ shifted by $2$, such that we can summarize this equation by writing
\begin{align}
&(-D_0+2)\mathcal{D}_{[\bar{I}_{n-m_1}}\tilde{\mathcal{H}}^{g,A_1\ldots A_{m_1},B_1\ldots B_{m_2+1}}_{\bar{I}_1\ldots\bar{I}_{n-m_1-1}],\bar{J}_1\ldots\bar{J}_{n-m_2}}+\epsilon^{ij}\frac{\partial}{\partial\bar{u}^i_+}\mathcal{D}^{[A_{1}}_j\tilde{\mathcal{H}}^{g,A_2\ldots A_{m_1}],B_1\ldots B_{m_2+1}}_{\bar{I}_1\ldots\bar{I}_{n-m_1},\bar{J}_1\ldots\bar{J}_{n-m_2}}=\nonumber\\
&\hspace{3cm}=(-D_0+2)\,\tilde{\mathcal{C}}^{\text{(bdy)},A_1\ldots A_{m_1},B_1\ldots B_{m_2+1}}_{\bar{I}_1\ldots\bar{I}_{n-m_1},\bar{J}_1\ldots\bar{J}_{n-m_2}}\,,
\end{align}
where we have explicitly pulled out the factor of $(2g+2n-m_1-m_2-2)$ from the boundary contribution. In the language of the generating functional, this latter factor is converted into the action of the operator $(-D_0+2)$ such that we can write the more compact relation
\begin{align}
\left[(-D_0+2)\grass ^{\bar{I}}\mathcal{D}_{\bar{I}}+\epsilon^{ij}\frac{\partial}{\partial\bar{u}^i_+}\chi_A\mathcal{D}^{A}_j\right]\tilde{\mathbb{H}}^g=(-D_0+2)\tilde{\mathbb{C}}^{\text{bdy}}\,.\label{DiffEqTildeH}
\end{align}
It still remains to determine the boundary contribution $\tilde{\mathbb{C}}^{\text{bdy}}$.
\subsubsection{Boundary Contributions}
The total contribution from the boundary deformations can be divided into two terms
\begin{align}
\tilde{\mathcal{C}}^{\text{(bdy)},A_1\ldots A_{m_1},B_1\ldots B_{m_2+1}}_{\bar{I}_1\ldots\bar{I}_{n-m_1+1},\bar{J}_1\ldots\bar{J}_{n-m_2}}&=\int_{\mathcal{M}_{(g,n+1)}}\hspace{-0.7cm}\langle (\mu\cdot G^-_{T^2})^{g+m_1+m_2-n}(\mu\cdot G^-_{K3,+})^{2g+2n-m_1-m_2-4}(\mu\cdot J^{--}_{K3})(\mu\cdot T)\cdot\nonumber\\
&\hspace{1.5cm}\cdot\prod_{a=1}^{m_1}\int\bar{\Xi}^{A_a}\prod_{b=1}^{m_2+1}(\psi_3\Xi^{B_b})\prod_{c=1}^{n-m_1}\int\bar{\psi}_3\bar{J}_{\bar{I}_c}\prod_{d=1}^{n-m_2}(J^{++}_{K3}\bar{J}_{\bar{J}_d})\rangle=\nonumber\\
&=\tilde{\mathcal{C}}^{\text{(bdy,hyper)},A_1\ldots A_{m_1},B_1\ldots B_{m_2+1}}_{\bar{I}_1\ldots\bar{I}_{n-m_1+1},\bar{J}_1\ldots\bar{J}_{n-m_2}}+\tilde{\mathcal{C}}^{\text{(bdy,vector)},A_1\ldots A_{m_1},B_1\ldots B_{m_2+1}}_{\bar{I}_1\ldots\bar{I}_{n-m_1+1},\bar{J}_1\ldots\bar{J}_{n-m_2}}\label{FullBdyCont}
\end{align}
corresponding to either hypermultiplet or vector multiplet states propagating on the long thin tube when pinching a dividing geodesic. We can calculate both contributions in a straight forward manner.\\[10pt]
\noindent \textbf{Hypermultiplet Propagation}\\
There is not a unique way to distribute the operator insertions on the two Riemann surfaces but rather we have to consider several rather complicated contributions
{\allowdisplaybreaks\begin{align}
&\tilde{\mathcal{C}}^{\text{(bdy,hyper)},A_1\ldots A_{m_1},B_1\ldots B_{m_2+1}}_{\bar{I}_1\ldots\bar{I}_{n-m_1+1},\bar{J}_1\ldots\bar{J}_{n-m_2}}=\sum_{n_s=0}^{n}\sum_{m^s_1=0}^{n_s}\sum_{m^s_2=0}^{n_s}\int_{\mathcal{M}_{(g_s,n_s)}}\hspace{-0.7cm}\langle (\mu\cdot G^-)^{3g_s+n_s-3}\prod_{a=1}^{m^s_1+1}\int \bar{\Xi}^{A_a}\prod_{b=1}^{m_2^s}(\psi_3\Xi^{[B_b})\cdot\nonumber\\
&\hspace{1cm}\cdot \prod_{c=1}^{n_s-m_1^s}\int\bar{\psi}_3\bar{J}_{\bar{I}_c}\prod_{d=1}^{n_s-m_2^s}(J^{++}_{K3}\bar{J}_{\bar{J}_d})\oint G^-\left(\psi_3 \Xi^{B]}\right)\rangle\cdot\Omega_{BC}\cdot\nonumber\\
&\cdot\int_{\mathcal{M}_{(g-g_s,n-n_s+1)}}\hspace{-1.2cm}\langle (\mu\cdot G^-)^{3(g-g_s)+n-n_s-3}\left[(\mu\cdot J^{--}_{K3})\oint (G^-\Xi^C)+(\mu\cdot G^-)\oint (J^{--}_{K3}\Xi^C)\right]\prod_{a=m^s_1+2}^{m_1}\int \bar{\Xi}^{A_a}\cdot\nonumber\\
&\hspace{1cm}\cdot\prod_{b=m_2^s+1}^{m_2+1}(\psi_3\Xi^{B_b})\prod_{c=n_s-m_1^s+1}^{n-m_1+1}\int\bar{\psi}_3\bar{J}_{\bar{I}_c}\prod_{d=n_s-m_2^s+1}^{n-m_2}(J^{++}_{K3}\bar{J}_{\bar{J}_d})\rangle+\nonumber\\
&+\sum_{n_s=0}^{n}\sum_{m^s_1=0}^{n_s}\sum_{m^s_2=0}^{n_s}\int_{\mathcal{M}_{(g_s,n_s)}}\hspace{-0.7cm}\langle (\mu\cdot G^-)^{3g_s+n_s-4}\left[(\mu\cdot J^{--}_{K3})\oint G^-(\psi_3\Xi^B)+(\mu\cdot G^-)\oint J^{--}_{K3}(\psi_3\Xi^B)\right]\cdot\nonumber\\
&\hspace{1cm}\cdot\prod_{a=1}^{m^s_1}\int \bar{\Xi}^{A_a}\prod_{b=1}^{m_2^s}(\psi_3\Xi^{[B_b}) \prod_{c=1}^{n_s-m_1^s}\int\bar{\psi}_3\bar{J}_{\bar{I}_c}\prod_{d=1}^{n_s-m_2^s}(J^{++}_{K3}\bar{J}_{\bar{J}_d})\rangle\cdot\Omega_{BC}\cdot\nonumber\\
&\cdot\int_{\mathcal{M}_{(g-g_s,n-n_s)}}\hspace{-1.2cm}\langle (\mu\cdot G^-)^{3(g-g_s)+n-n_s-2}\prod_{a=m^s_1+1}^{m_1}\int \bar{\Xi}^{A_a}\prod_{b=m_2^s+1}^{m_2+1}(\psi_3\Xi^{B_b})\prod_{c=n_s-m_1^s+1}^{n-m_1+1}\int\bar{\psi}_3\bar{J}_{\bar{I}_c}\cdot\nonumber\\
&\hspace{1cm}\cdot\prod_{d=n_s-m_2^s+1}^{n-m_2}(J^{++}_{K3}\bar{J}_{\bar{J}_d})\oint G^-\Xi^C\rangle\,.
\end{align}}
Fortunately, this expression can be written in terms of topological amplitudes in the following manner
\begin{align}
&\tilde{\mathcal{C}}^{\text{(bdy,hyper)},A_1\ldots A_{m_1},B_1\ldots B_{m_2+1}}_{\bar{I}_1\ldots\bar{I}_{n-m_1+1},\bar{J}_1\ldots\bar{J}_{n-m_2}}=\nonumber\\
&=\sum_{g_s=1}^{g-1}\sum_{n_s=0}^n\sum_{m^s_1=0}^{n_s}\sum_{m^s_2=0}^{n_s}\mathcal{H}^{g_s,A_1\ldots A_{m_1^s+1},B_1\ldots B_{m_2^s}B}_{\bar{I}_1\ldots\bar{I}_{n_s-m_1^s},\bar{J}_1\ldots\bar{J}_{n_s-m_2^s}}\,\Omega_{BC}\,\mathcal{D}_+^C\tilde{\mathcal{H}}^{g-g_s,A_{m_1^s+2}\ldots A_{m_1},B_{m_2^s+1}\ldots B_{m_2+1}}_{\bar{I}_{n_s-m_1^s+1}\ldots\bar{I}_{n-m_1+1},\bar{J}_{n_s-m_2^s+1}\ldots\bar{J}_{n-m_2}}+\nonumber\\
&+\sum_{g_s=1}^{g-1}\sum_{n_s=0}^n\sum_{m^s_1=0}^{n_s}\sum_{m^s_2=0}^{n_s}\tilde{\mathcal{H}}^{g_s,A_1\ldots A_{m_1^s},B_1\ldots B_{m_2^s}B}_{\bar{I}_1\ldots\bar{I}_{n_s-m_1^s},\bar{J}_1\ldots\bar{J}_{n_s-m_2^s}}\,\Omega_{BC}\,\mathcal{D}_+^C\mathcal{H}^{g-g_s,A_{m_1^s+1}\ldots A_{m_1},B_{m_2^s+1}\ldots B_{m_2+1}}_{\bar{I}_{n_s-m_1^s+1}\ldots\bar{I}_{n-m_1+1},\bar{J}_{n_s-m_2^s+1}\ldots\bar{J}_{n-m_2}}+\nonumber\\
&+2\sum_{g_s=1}^{g-1}\sum_{n_s=0}^n\sum_{m^s_1=0}^{n_s}\sum_{m^s_2=0}^{n_s}\mathcal{H}^{g_s,A_1\ldots A_{m_1^s+1},B_1\ldots B_{m_2^s}B}_{\bar{I}_1\ldots\bar{I}_{n_s-m_1^s},\bar{J}_1\ldots\bar{J}_{n_s-m_2^s}}\,\Omega_{BC}\,\mathcal{H}^{g-g_s,A_{m_1^s+2}\ldots A_{m_1}C,B_{m_2^s+1}\ldots B_{m_2+1}}_{\bar{I}_{n_s-m_1^s+1}\ldots\bar{I}_{n-m_1+1},\bar{J}_{n_s-m_2^s+1}\ldots\bar{J}_{n-m_2}}\,.\label{hyperboundary}
\end{align}
\noindent \textbf{Vector Multiplet Propagation}\\
Finally we also have to consider the possibility of having vector multiplet states propagate on the tubes. Also here we obtain several contributions
{\allowdisplaybreaks\begin{align}
&\tilde{\mathcal{C}}^{\text{(bdy,vector)},A_1\ldots A_{m_1},B_1\ldots B_{m_2+1}}_{\bar{I}_1\ldots\bar{I}_{n-m_1+1},\bar{J}_1\ldots\bar{J}_{n-m_2}}=\sum_{n_s=0}^{n}\sum_{m^s_1=0}^{n_s}\sum_{m^s_2=0}^{n_s}\int_{\mathcal{M}_{(g_s,n_s+1)}}\hspace{-0.7cm}\langle (\mu\cdot G^-)^{3g_s+n_s-2}\prod_{a=1}^{m^s_1+1}\int \bar{\Xi}^{A_a}\prod_{b=1}^{m_2^s+1}(\psi_3\Xi^{B_b})\cdot\nonumber\\
&\hspace{1cm}\cdot \prod_{c=1}^{n_s-m_1^s}\int\bar{\psi}_3\bar{J}_{\bar{I}_c}\prod_{d=1}^{n_s-m_2^s}(J^{++}_{K3}\bar{J}_{\bar{J}_d})\oint G^-\left(\psi_3 \bar{J}_L\right)\rangle\cdot G^{L\bar{K}}\cdot\nonumber\\
&\cdot\int_{\mathcal{M}_{(g-g_s,n-n_s)}}\hspace{-1.2cm}\langle (\mu\cdot G^-)^{3(g-g_s)+n-n_s-4}\left[(\mu\cdot J^{--}_{K3})\oint (G^-J^{++}_{K3}\bar{J}_{\bar{K}})+(\mu\cdot G^-)\oint (J^{--}_{K3}J^{++}_{K3}\bar{J}_{\bar{K}})\right]\cdot\nonumber\\
&\hspace{1cm}\cdot\prod_{a=m^s_1+2}^{m_1}\int \bar{\Xi}^{A_a}\prod_{b=m_2^s+2}^{m_2+1}(\psi_3\Xi^{B_b})\prod_{c=n_s-m_1^s+1}^{n-m_1+1}\int\bar{\psi}_3\bar{J}_{\bar{I}_c}\prod_{d=n_s-m_2^s+1}^{n-m_2}(J^{++}_{K3}\bar{J}_{\bar{J}_d})\rangle+\nonumber\\
&+\sum_{n_s=0}^{n}\sum_{m^s_1=0}^{n_s}\sum_{m^s_2=0}^{n_s}\int_{\mathcal{M}_{(g_s,n_s+1)}}\hspace{-0.7cm}\langle (\mu\cdot G^-)^{3g_s+n_s-3}\left[(\mu\cdot J^{--}_{K3})\oint G^-(\psi_3\bar{J}_L)+(\mu\cdot G^-)\oint J^{--}_{K3}(\psi_3\bar{J}_L)\right]\cdot\nonumber\\
&\hspace{1cm}\cdot\prod_{a=1}^{m^s_1}\int \bar{\Xi}^{A_a}\prod_{b=1}^{m_2^s+1}(\psi_3\Xi^{[B_b}) \prod_{c=1}^{n_s-m_1^s}\int\bar{\psi}_3\bar{J}_{\bar{I}_c}\prod_{d=1}^{n_s-m_2^s}(J^{++}_{K3}\bar{J}_{\bar{J}_d})\rangle\cdot G^{L\bar{K}}\cdot\nonumber\\
&\cdot\int_{\mathcal{M}_{(g-g_s,n-n_s)}}\hspace{-1.2cm}\langle (\mu\cdot G^-)^{3(g-g_s)+n-n_s-3}\prod_{a=m^s_1+1}^{m_1}\int \bar{\Xi}^{A_a}\prod_{b=m_2^s+2}^{m_2+1}(\psi_3\Xi^{B_b})\prod_{c=n_s-m_1^s+1}^{n-m_1+1}\int\bar{\psi}_3\bar{J}_{\bar{I}_c}\cdot\nonumber\\
&\hspace{1cm}\cdot\prod_{d=n_s-m_2^s+1}^{n-m_2}(J^{++}_{K3}\bar{J}_{\bar{J}_d})\oint G^-(J^{++}_{K3}\bar{J}_{\bar{K}})\rangle\,.
\end{align}}
that can be fully expressed in terms of topological amplitudes
\begin{align}
&\tilde{\mathcal{C}}^{\text{(bdy,hyper)},A_1\ldots A_{m_1},B_1\ldots B_{m_2+1}}_{\bar{I}_1\ldots\bar{I}_{n-m_1+1},\bar{J}_1\ldots\bar{J}_{n-m_2}}=\nonumber\\
&=\sum_{g_s=1}^{g-1}\sum_{n_s=0}^n\sum_{m^s_1=0}^{n_s}\sum_{m^s_2=0}^{n_s}\partial_L\mathcal{H}^{g_s,A_1\ldots A_{m_1^s+1},B_1\ldots B_{m_2^s}B}_{\bar{I}_1\ldots\bar{I}_{n_s-m_1^s},\bar{J}_1\ldots\bar{J}_{n_s-m_2^s}}\,G^{L\bar{K}}\,\tilde{\mathcal{H}}^{g-g_s,A_{m_1^s+2}\ldots A_{m_1},B_{m_2^s+1}\ldots B_{m_2+1}}_{\bar{I}_{n_s-m_1^s+1}\ldots\bar{I}_{n-m_1+1},\bar{J}_{n_s-m_2^s+1}\ldots\bar{J}_{n-m_2}\bar{K}}+\nonumber\\
&+\sum_{g_s=1}^{g-1}\sum_{n_s=0}^n\sum_{m^s_1=0}^{n_s}\sum_{m^s_2=0}^{n_s}\partial_L\tilde{\mathcal{H}}^{g_s,A_1\ldots A_{m_1^s},B_1\ldots B_{m_2^s}B}_{\bar{I}_1\ldots\bar{I}_{n_s-m_1^s},\bar{J}_1\ldots\bar{J}_{n_s-m_2^s}}\,G^{L\bar{K}}\,\mathcal{H}^{g-g_s,A_{m_1^s+1}\ldots A_{m_1},B_{m_2^s+1}\ldots B_{m_2+1}}_{\bar{I}_{n_s-m_1^s+1}\ldots\bar{I}_{n-m_1+1},\bar{J}_{n_s-m_2^s+1}\ldots\bar{J}_{n-m_2}\bar{K}}\,.\label{vectorboundary}
\end{align}
The final result of the boundary is given by equation (\ref{FullBdyCont}) with the expressions (\ref{hyperboundary}) and (\ref{vectorboundary}), i.e.
{\allowdisplaybreaks
\begin{align}
&\tilde{\mathcal{C}}^{\text{(bdy)},A_1\ldots A_{m_1},B_1\ldots B_{m_2+1}}_{\bar{I}_1\ldots\bar{I}_{n-m_1+1},\bar{J}_1\ldots\bar{J}_{n-m_2}}=\nonumber\\
&=\sum_{g_s=0}^{g}\sum_{n_s=0}^n\sum_{m^s_1=0}^{n_s}\sum_{m^s_2=0}^{n_s}\mathcal{H}^{g_s,A_1\ldots A_{m_1^s+1},B_1\ldots B_{m_2^s}B}_{\bar{I}_1\ldots\bar{I}_{n_s-m_1^s},\bar{J}_1\ldots\bar{J}_{n_s-m_2^s}}\,\Omega_{BC}\,\mathcal{D}_+^C\tilde{\mathcal{H}}^{g-g_s,A_{m_1^s+2}\ldots A_{m_1},B_{m_2^s+1}\ldots B_{m_2+1}}_{\bar{I}_{n_s-m_1^s+1}\ldots\bar{I}_{n-m_1+1},\bar{J}_{n_s-m_2^s+1}\ldots\bar{J}_{n-m_2}}+\nonumber\\
&+\sum_{g_s=0}^{g}\sum_{n_s=0}^n\sum_{m^s_1=0}^{n_s}\sum_{m^s_2=0}^{n_s}\tilde{\mathcal{H}}^{g_s,A_1\ldots A_{m_1^s},B_1\ldots B_{m_2^s}B}_{\bar{I}_1\ldots\bar{I}_{n_s-m_1^s},\bar{J}_1\ldots\bar{J}_{n_s-m_2^s}}\,\Omega_{BC}\,\mathcal{D}_+^C\mathcal{H}^{g-g_s,A_{m_1^s+1}\ldots A_{m_1},B_{m_2^s+1}\ldots B_{m_2+1}}_{\bar{I}_{n_s-m_1^s+1}\ldots\bar{I}_{n-m_1+1},\bar{J}_{n_s-m_2^s+1}\ldots\bar{J}_{n-m_2}}+\nonumber\\
&+2\sum_{g_s=1}^{g-1}\sum_{n_s=0}^n\sum_{m^s_1=0}^{n_s}\sum_{m^s_2=0}^{n_s}\mathcal{H}^{g_s,A_1\ldots A_{m_1^s+1},B_1\ldots B_{m_2^s}B}_{\bar{I}_1\ldots\bar{I}_{n_s-m_1^s},\bar{J}_1\ldots\bar{J}_{n_s-m_2^s}}\,\Omega_{BC}\,\mathcal{H}^{g-g_s,A_{m_1^s+2}\ldots A_{m_1}C,B_{m_2^s+1}\ldots B_{m_2+1}}_{\bar{I}_{n_s-m_1^s+1}\ldots\bar{I}_{n-m_1+1},\bar{J}_{n_s-m_2^s+1}\ldots\bar{J}_{n-m_2}}+\nonumber\\
&+\sum_{g_s=0}^{g}\sum_{n_s=0}^n\sum_{m^s_1=0}^{n_s}\sum_{m^s_2=0}^{n_s}D_L\mathcal{H}^{g_s,A_1\ldots A_{m_1^s+1},B_1\ldots B_{m_2^s}B}_{\bar{I}_1\ldots\bar{I}_{n_s-m_1^s},\bar{J}_1\ldots\bar{J}_{n_s-m_2^s}}\,G^{L\bar{K}}\,\tilde{\mathcal{H}}^{g-g_s,A_{m_1^s+2}\ldots A_{m_1},B_{m_2^s+1}\ldots B_{m_2+1}}_{\bar{I}_{n_s-m_1^s+1}\ldots\bar{I}_{n-m_1+1},\bar{J}_{n_s-m_2^s+1}\ldots\bar{J}_{n-m_2}\bar{K}}+\nonumber\\
&+\sum_{g_s=0}^{g}\sum_{n_s=0}^n\sum_{m^s_1=0}^{n_s}\sum_{m^s_2=0}^{n_s}D_L\tilde{\mathcal{H}}^{g_s,A_1\ldots A_{m_1^s},B_1\ldots B_{m_2^s}B}_{\bar{I}_1\ldots\bar{I}_{n_s-m_1^s},\bar{J}_1\ldots\bar{J}_{n_s-m_2^s}}\,G^{L\bar{K}}\,\mathcal{H}^{g-g_s,A_{m_1^s+1}\ldots A_{m_1},B_{m_2^s+1}\ldots B_{m_2+1}}_{\bar{I}_{n_s-m_1^s+1}\ldots\bar{I}_{n-m_1+1},\bar{J}_{n_s-m_2^s+1}\ldots\bar{J}_{n-m_2}\bar{K}}\,.\label{GenHarmonicityComponent}
\end{align}}
This expression is more compactly written using the generating functional (\ref{GeneratingFunct}) 
\begin{align}
&\tilde{\mathbb{C}}^{\text{(bdy)}}=\sum_{g_s=0}^{g}\left[\left(\mathcal{D}_+^A\tilde{\mathbb{H}}^{g_s}\right)\Omega_{AB}\left(\frac{\partial\mathbb{H}^{g-g_s}}{\partial\xi_B}\right)-\left(\mathcal{D}_+^A\mathbb{H}^{g_s}\right)\Omega_{AB}\left(\frac{\partial\tilde{\mathbb{H}}^{g-g_s}}{\partial\xi_B}\right)+\right.\nonumber\\
&+\left(\frac{\partial\mathbb{H}^{g_s}}{\partial\chi_A}\right)\Omega_{AB}\left(\frac{\partial\mathbb{H}^{g-g_s}}{\partial\xi_B}\right)+\left.\left(\mathcal{D}_I\tilde{\mathbb{H}}^{g_s}\right)G^{I\bar{J}}\left(\frac{\partial \mathbb{H}^{g-g_s}}{\partial\eta^{\bar{J}}}\right)-\left(\mathcal{D}_I\mathbb{H}^{g_s}\right)G^{I\bar{J}}\left(\frac{\partial \tilde{\mathbb{H}}^{g-g_s}}{\partial\eta^{\bar{J}}}\right)\right]\,.\label{CbdyGenFunc}
\end{align}
\subsection{String Differential Equation for $\mathcal{H}^{g,A_1\ldots A_{m_1+1},B_1\ldots B_{m_2+1}}_{\bar{I}_1\ldots\bar{I}_{n-m_1},\bar{J}_1\ldots\bar{J}_{n-m_2}}$}\label{Sect:DiffEqHstring}
\subsubsection{Bulk Equation}
An alternative (and slightly more technical) method to arrive at (\ref{DiffEqH}) is to work directly with the string expression  $\mathcal{H}^{g,A_1\ldots A_{m_1+1},B_1\ldots B_{m_2+1}}_{\bar{I}_1\ldots\bar{I}_{n-m_1},\bar{J}_1\ldots\bar{J}_{n-m_2}}$. To this end we start by taking a derivative with respect to an anti-holomorphic vector multiplet scalar
\begin{align}
&\mathcal{D}_{[\bar{I}_{n-m_1+1}}\mathcal{H}^{g,A_1\ldots A_{m_1+1},B_1\ldots B_{m_2+1}}_{\bar{I}_1\ldots \bar{I}_{n-m_1}],\bar{J}_1\ldots \bar{J}_{n-m_2}}=\frac{1}{(3g-2+n)!}\cdot \int_{\mathcal{M}_{g}}\langle (\mu\cdot G^-)^{3g-2+n}\prod_{a=1}^{m_1+1}\int \bar{\Xi}^{A_a}\nonumber\\
&\hspace{1cm}\cdot\prod_{c=1}^{n-m_1}\int\bar{\psi}_3\bar{J}_{[\bar{I}_c}\oint G^+\bar{\psi}_3\bar{J}_{\bar{I}_{n-m_1+1}]}\prod_{b=1}^{m_2+1}\psi_3 \Xi^{B_b}\prod_{d=1}^{n-m_2} J^{++}_{K3}\bar{J}_{\bar{J}_d}\rangle\,,\nonumber
\end{align}
where we have introduced the operator
\begin{align}
G^+=G^+_{T^2}+G^+_{K3,-}\,.\label{GpDefinition}
\end{align} 
Notice that similar to $G^-$ also this operator is strictly speaking not well defined regarding its harmonic charge. However, just as before the necessity of soaking up all torus fermionic zero modes $\psi_3$ takes care of this ambiguity in a well defined way. Deforming now the $\oint G^+$-contour we obtain
\begin{align}
&\mathcal{D}_{[\bar{I}_{n-m_1+1}}\mathcal{H}^{g,A_1\ldots A_{m_1+1},B_1\ldots B_{m_2+1}}_{\bar{I}_1\ldots \bar{I}_{n-m_1}],\bar{J}_1\ldots \bar{J}_{n-m_2}}=\mathcal{C}^{(\text{bdy}),A_1\ldots A_{m_1+1},B_1\ldots B_{m_2+1}}_{\bar{I}_1\ldots \bar{I}_{n-m_1+1},\bar{J}_1\ldots \bar{J}_{n-m_2}}+\mathcal{C}^{(\text{bulk}),A_1\ldots A_{m_1+1},B_1\ldots B_{m_2+1}}_{\bar{I}_1\ldots \bar{I}_{n-m_1+1},\bar{J}_1\ldots \bar{J}_{n-m_2}}=\nonumber\\
&=(-1)^{n+1}\int_{\mathcal{M}_{g}}\langle (\mu\cdot G^-)^{3g-3+n}(\mu\cdot T)\prod_{a=1}^{m_1+1}\int \bar{\Xi}^{A_a}\prod_{c=1}^{n-m_1+1}\int\bar{\psi}_3\bar{J}_{\bar{I}_c}\prod_{b=1}^{m_2+1}\psi_3 \Xi^{B_b}\prod_{d=1}^{n-m_2}J^{++}_{K3}\bar{J}_{\bar{J}_d}\rangle+\nonumber\\
&+(-1)^{n}\int_{\mathcal{M}_{g}}\hspace{-0.25cm}\langle (\mu\cdot G^-)^{3g-2+n}\oint G^+_{K3,-}\bar{\Xi}^{[A_{1}}\prod_{a=2}^{m_1+1}\int \bar{\Xi}^{A_a]}\hspace{-0.1cm}\prod_{c=1}^{n-m_1+1}\hspace{-0.1cm}\int\bar{\psi}_3\bar{J}_{\bar{I}_c}\prod_{b=1}^{m_2+1}\psi_3 \Xi^{B_b}\prod_{d=1}^{n-m_2} J^{++}_{K3}\bar{J}_{\bar{J}_d}\rangle\nonumber
\end{align}
The first term $\mathcal{C}^{(\text{bdy}),A_1\ldots A_{m_1+1},B_1\ldots B_{m_2+1}}_{\bar{I}_1\ldots \bar{I}_{n-m_1+1},\bar{J}_1\ldots \bar{J}_{n-m_2}}$ corresponds to a boundary contribution since it contains an insertion of the full energy-momentum tensor. The term $\mathcal{C}^{(\text{bulk}),A_1\ldots A_{m_1+1},B_1\ldots B_{m_2+1}}_{\bar{I}_1\ldots \bar{I}_{n-m_1+1},\bar{J}_1\ldots \bar{J}_{n-m_2}}$ on the other hand is a bulk contribution and is in fact just a vector multiplet derivative of $\mathcal{H}^{g,A_1\ldots A_{m_1},B_1\ldots B_{m_2+1}}_{\bar{I}_1\ldots \bar{I}_{n-m_1+1},\bar{J}_1\ldots \bar{J}_{n-m_2}}$. Thus we can write
\begin{align}\label{neweq}
&\mathcal{D}_{[\bar{I}_{n-m_1+1}}\mathcal{H}^{g,A_1\ldots A_{m_1+1},B_1\ldots B_{m_2+1}}_{\bar{I}_1\ldots \bar{I}_{n-m_1}],\bar{J}_1\ldots \bar{J}_{n-m_2}}+(-1)^{n+1}\mathcal{D}_-^{[A_{1}}\mathcal{H}^{g,A_2\ldots A_{m_1+1}],B_1\ldots B_{m_2+1}}_{\bar{I}_1\ldots \bar{I}_{n-m_1+1},\bar{J}_1\ldots \bar{J}_{n-m_2}}\nonumber\\
&=(-1)^{n+1}\int_{\mathcal{M}_{g}}\langle (\mu\cdot G^-)^{3g-3+n}(\mu\cdot T)\prod_{a=1}^{m_1+1}\int \bar{\Xi}^{A_a}\prod_{c=1}^{n-m_1+1}\int\bar{\psi}_3\bar{J}_{\bar{I}_c}\prod_{b=1}^{m_2+1}\psi_3 \Xi^{B_b}\prod_{d=1}^{n-m_2}J^{++}_{K3}\bar{J}_{\bar{J}_d}\rangle\nonumber\\
&=\mathcal{C}^{(\text{bdy}),A_1\ldots A_{m_1+1},B_1\ldots B_{m_2+1}}_{\bar{I}_1\ldots \bar{I}_{n-m_1+1},\bar{J}_1\ldots \bar{J}_{n-m_2}}\,.
\end{align}
\subsubsection{Boundary Contribution}
We can directly discuss the contribution of vector- and hypermultiplets separately.\\[10pt]
\noindent \textbf{Hypermultiplet Propagation}\\
Assuming that the two surfaces have genus $g_s$ and $g-g_s$ respectively we get the contribution
{\allowdisplaybreaks
\begin{align}
&\mathcal{C}^{(\text{bdy,hyper}),A_1\ldots A_{m_1+1},B_1\ldots B_{m_2+1}}_{\bar{I}_1\ldots \bar{I}_{n-m_1},\bar{J}_1\ldots \bar{J}_{n-m_2}}=\nonumber\\
&=\sum_{n_s=0}^n\sum_{m^s_1=0}^{n_s}\sum_{m^s_2=0}^{n_s}\int_{\mathcal{M}_{g_s}}\langle (\mu\cdot G^-)^{3g_s-3+n_s}\prod_{a=1}^{m^s_1+1}\int \bar{\Xi}^{A_a}\prod_{c=1}^{n_s-m_1^s}\int\bar{\psi}_3\bar{J}_{\bar{I}_c}\prod_{b=1}^{m_2^s}\psi_3 \Xi^{[B_b}\prod_{d=1}^{n_s-m_2^s} J^{++}_{K3}\bar{J}_{\bar{J}_d}\cdot\nonumber\\
&\hspace{0.5cm}\cdot \oint G^-\left(\psi_3 \Xi^{B]}\right)\rangle\,\Omega_{BC}\,\int_{\mathcal{M}_{g-g_s}}\langle (\mu\cdot G^-)^{3(g-g_s)-2+(n-n_s)}\prod_{a=m^s_1+2}^{m_1+1}\int \bar{\Xi}^{A_a}\prod_{c=n_s-m_1^s+1}^{n-m_1+1}\int\bar{\psi}_3\bar{J}_{\bar{I}_c}\cdot\nonumber\\
&\hspace{0.5cm}\cdot \prod_{b=m_2^s+1}^{m_2+1}\psi_3 \Xi^{B_b} \prod_{d=n_s-m_2^s+1}^{n-m_2} J^{++}_{K3}\bar{J}_{\bar{J}_d}\oint G^-\Xi^C\rangle\,.
\end{align}}
Here the operator $\psi_3 \Xi^B$ on genus $g_s$ and the operator $\Xi^C$ on genus $g-g_s$ surfaces appear at the nodes respectively and $\Omega_{BC}$ is the propagator on the tube. These operators are dimension zero and the net charge is $+3$ as it should be for the sphere. In the above expression $G^-$ contours around these operators are coming from the Beltrami differentials associated with the punctures at the nodes.
Note however that when $g_s$ or $g-g_s$ is one, then the operator at the node on the genus 1 surface is unintegrated and the corresponding $G^-$ is folded with the Beltrami differential dual to the torus modulus. Similarly when one of the surfaces is of genus zero then three of the dimension zero operators on the sphere are unintegrated. As a result 
the counting of $G^-$ in all cases gives the correct values.

Reinterpreting this expression making use of the anti-symmetrization of the $B$-indices we obtain the expression
\begin{align}
&\mathcal{C}^{(\text{bdy,hyper}),A_1\ldots A_{m_1+1},B_1\ldots B_{m_2+1}}_{\bar{I}_1\ldots \bar{I}_{n-m_1},\bar{J}_1\ldots \bar{J}_{n-m_2}}=\nonumber\\
&\hspace{0.5cm}=\sum_{n_s=0}^n\sum_{m_s^1=0}^{n_s}\sum_{m_s^2=0}^{n_s}\mathcal{H}^{g_s,A_1\ldots A_{m_1^s+1},B_1\ldots B_{m_2^s}B}_{\bar{I}_1\ldots \bar{I}_{n_s-m_1^s},\bar{J}_1\ldots\bar{J}_{n_s-m_2^s}}\Omega_{BC}\mathcal{D}_+^ C\mathcal{H}^{g-g_s,A_{m_1^s+2}\ldots A_{m_1+1},B_{m_2^s+1}\ldots B_{m_2+1}}_{\bar{I}_{n_s-m_1^s+1}\ldots\bar{I}_{n-m_1+1},\bar{J}_{n_s-m_2^s+1}\ldots \bar{J}_{n-m_2}}\,.
\end{align}
\noindent
\textbf{Vector Multiplet Propagation}\\
In the case $m_1\neq 0\neq m_2$ we will also obtain vector multiplet contributions for the pinching of Riemann surfaces of generic genus. 
{\allowdisplaybreaks\begin{align}
&\mathcal{C}^{(\text{bdy,vector}),A_1\ldots A_{m_1+1},B_1\ldots B_{m_2+1}}_{\bar{I}_1\ldots \bar{I}_{n-m_1},\bar{J}_1\ldots \bar{J}_{n-m_2}}=\nonumber\\
&=\sum_{n_s=0}^n\sum_{m^s_1=0}^{n_s}\sum_{m^s_2=0}^{n_s}\int_{\mathcal{M}_{g_s}}\langle (\mu\cdot G^-)^{3g_s-2+n_s}\prod_{a=1}^{m^s_1+1}\int \bar{\Xi}^{A_a}\prod_{b=1}^{m_2^s+1}\psi_3 \Xi^{B_b}\prod_{c=1}^{n_s-m_1^s}\int\bar{\psi}_3\bar{J}_{\bar{I}_c}\prod_{d=1}^{n_s-m_2^s}J^{++}_{K3}\bar{J}_{\bar{J}_d}\cdot\nonumber\\
&\hspace{0.5cm}\cdot \oint G^-\left(\psi_3 \bar{J}_L\right)\rangle\,G^{L\bar{K}}\,\int_{\mathcal{M}_{g-g_s}}\langle (\mu\cdot G^-)^{3(g-g_s)-3+(n-n_s)}\prod_{a=m^s_1+2}^{m_1+1}\int \bar{\Xi}^{A_a}\prod_{b=m_2^s+2}^{m_2+1}\psi_3 \Xi^{B_b}\cdot\nonumber\\
&\hspace{0.5cm}\cdot \prod_{c=n_s-m_1^s+1}^{n-m_1+1}\int\bar{\psi}_3\bar{J}_{\bar{I}_c} \prod_{d=n_s-m_2^s+1}^{n-m_2} J^{++}_{K3}\bar{J}_{\bar{J}_d}\oint G^-\left(J^{++}_{K3}\bar{J}_{\bar{K}}\right)\rangle\,.
\end{align}}
Notice that for $m_2=0$ there will only be contributions to this expression if one of the two surfaces is genus $1$ or $0$, otherwise there will not be enough $\psi_3$-insertions on both Riemann surfaces to soak up all zero modes (see also section~\ref{Sect:DiffEqHm20}). Rewriting this expression using once more the anti-symmetrization of the $B$-indices, we obtain the following relation
\begin{align}
&\mathcal{C}^{(1,\text{vector}),A_1\ldots A_{m_1+1},B_1\ldots B_{m_2+1}}_{\bar{I}_1\ldots \bar{I}_{n-m_1},\bar{J}_1\ldots \bar{J}_{n-m_2}}=\nonumber\\
&\hspace{1.5cm}=\mathcal{D}_L\mathcal{H}^{g_s,A_1\ldots A_{m_1^s+1},B_1\ldots B_{m_2^s+1}}_{\bar{I}_1\ldots \bar{I}_{n_s-m_1^s},\bar{J}_1\ldots\bar{J}_{n_s-m_2^s}}G^{L\bar{K}}\mathcal{H}^{g-g_s,A_{m_1^s+2}\ldots A_{m_1+1},B_{m_2^s+2}\ldots B_{m_2+1}}_{\bar{I}_{n_s-m_1^s+1}\ldots\bar{I}_{n-m_1+1},\bar{J}_{n_s-m_2^s+1}\ldots \bar{J}_{n-m_2}\bar{K}}\,.
\end{align}
Taking all results together we find for the equation (\ref{neweq})
\begin{align}
&\mathcal{D}_{[\bar{I}_{n-m_1+1}}\mathcal{H}^{g,A_1\ldots A_{m_1+1},B_1\ldots B_{m_2+1}}_{\bar{I}_1\ldots \bar{I}_{n-m_1}],\bar{J}_1\ldots \bar{J}_{n-m_2}}+(-1)^{n-m_1+1}\mathcal{D}_-^{[A_{m_1+1}}\mathcal{H}^{g,A_1\ldots A_{m_1}],B_1\ldots B_{m_2+1}}_{\bar{I}_{1}\ldots \bar{I}_{n-m_1+1,\bar{J}_1\ldots \bar{J}_{n-m_2}}}=\nonumber\\
&\hspace{0.3cm}=\sum_{g_s=0}^{g}\sum_{n_s=0}^n\sum_{m_1^s=0}^{n_s}\sum_{m_2^s=0}^{n_s}\mathcal{H}^{g_s,A_1\ldots A_{m_1^s+1},B_1\ldots B_{m_2^s}B}_{\bar{I}_1\ldots \bar{I}_{n_s-m_1^s},\bar{J}_1\ldots\bar{J}_{n_s-m_2^s}}\Omega_{BC}\mathcal{D}_+^ C\mathcal{H}^{g-g_s,A_{m_1^s+2}\ldots A_{m_1+1},B_{m_2^s+1}\ldots B_{m_2+1}}_{\bar{I}_{n_s-m_1^s+1}\ldots\bar{I}_{n-m_1+1},\bar{J}_{n_s-m_2^s+1}\ldots \bar{J}_{n-m_2}}+\nonumber\\
&\hspace{0.3cm}+\sum_{g_s=0}^{g}\sum_{n_s=0}^n\sum_{m_1^s=0}^{n_s}\sum_{m_2^s=0}^{n_s}\mathcal{D}_L\mathcal{H}^{g_s,A_1\ldots A_{m_1^s+1},B_1\ldots B_{m_2^s+1}}_{\bar{I}_1\ldots \bar{I}_{n_s-m_1^s},\bar{J}_1\ldots\bar{J}_{n_s-m_2^s}}G^{L\bar{K}}\mathcal{H}^{g-g_s,A_{m_1^s+2}\ldots A_{m_1+1},B_{m_2^s+2}\ldots B_{m_2+1}}_{\bar{I}_{n_s-m_1^s+1}\ldots\bar{I}_{n-m_1+1},\bar{J}_{n_s-m_2^s+1}\ldots \bar{J}_{n-m_2}\bar{K}}\,.\label{DiffEqFin}
\end{align}
Since this equation is very involved due to the many indices of the various expressions, we resort to reformulating it in terms of the generating functional (\ref{GeneratingFunct})
\begin{align}
\left(\grass ^{\bar{I}}\mathcal{D}_{\bar{I}}+\chi_A\mathcal{D}_-^A\right)\mathbb{H}^g=\sum_{g_s=0}^{g}\left[\left(\frac{\partial\mathbb{H}^{g_s}}{\partial\xi_A}\right)\Omega_{AB}\left(\mathcal{D}_+^B\mathbb{H}^{g-g_s}\right)+\left(\mathcal{D}_I\mathbb{H}^{g_s}\right)G^{I\bar{J}}\left(\frac{\partial \mathbb{H}^{g-g_s}}{\partial\eta^{\bar{J}}}\right)\right]\,.\label{exteriorderivative}
\end{align}
Up to possible contact terms (which are incarnations of the curvature
contributions), which we did not consider in the computation so far,
this precisely agrees with (\ref{DiffEqH}). The left hand side of this
equation can be attributed as a bulk equation, while the right hand
side is an additional boundary contribution. Since in the present case
only boundaries corresponding to degeneration of the genus $g$ surface
along dividing geodesics contribute, these anomalous terms should be
due to reducible graphs involving elimination of auxiliary fields.

\subsection{String Differential Equation for $\hat{\tilde{\mathcal{H}}}^{g,A_1\ldots A_m}_{\bar{I}_1\ldots\bar{I}_{n-m},\bar{J}_1\ldots\bar{J}_n}$}\label{Sect:StringDerhattilde}
We start by considering a differentiation of (\ref{M20RedCase}) with respect to one of the vector multiplet moduli
\begin{align}
&\mathcal{D}_{[\bar{I}_{m-n+1}}\ftopmod{g}{m}{n-m}{n}=\frac{1}{(3g+n-4)!}\int_{\mathcal{M}_{(g,n)}}\hspace{-0.5cm}\langle(\mu\cdot G^-)^{3g+n-4}(\mu\cdot J^{--}_{K3})\psi_3(\alpha)\cdot\nonumber\\
&\hspace{1cm}\cdot\prod_{a=1}^m\int\bar{\Xi}^{A_a}\prod_{b=1}^{n-m}\int \bar{\psi}_3\bar{J}_{[\bar{I}_{b}}\prod_{c=1}^n(J^{++}_{K3}\bar{J}_{\bar{J}_c})\oint G^+\bar{\psi}_3\bar{J}_{\bar{I}_{m-n+1}]}\rangle\,.\nonumber
\end{align}
Deforming the contour integral of the $G^+$ operator we obtain the following three contributions
\begin{align}
&\mathcal{D}_{[\bar{I}_{m-n+1}}\ftopmod{g}{m}{n-m}{n}=\hat{\tilde{\mathcal{C}}}^{\text{(bdy)},A_1\ldots A_{m}}_{\bar{I}_1\ldots\bar{I}_{n-m+1},\bar{J}_1\ldots\bar{J}_{n}}+\hat{\tilde{\mathcal{C}}}^{\text{(bulk)},A_1\ldots A_{m}}_{\bar{I}_1\ldots\bar{I}_{n-m+1},\bar{J}_1\ldots\bar{J}_{n}}=\nonumber\\
&=(-1)^n \int_{\mathcal{M}_{(g,n)}}\hspace{-0.5cm}\langle(\mu\cdot G^-)^{3g+n-5}(\mu\cdot T)(\mu\cdot J^{--}_{K3})\psi_3(\alpha)\prod_{a=1}^m\int\bar{\Xi}^{A_a}\prod_{b=1}^{n-m+1}\int \bar{\psi}_3\bar{J}_{\bar{I}_{b}}\prod_{c=1}^n(J^{++}_{K3}\bar{J}_{\bar{J}_c})\rangle\nonumber\\
&+(-1)^n\int_{\mathcal{M}_{(g,n)}}\hspace{-0.5cm}\langle(\mu\cdot G^-)^{3g+n-4}(\mu\cdot G^{-}_{K3,-})\psi_3(\alpha)\prod_{a=1}^m\int\bar{\Xi}^{A_a}\prod_{b=1}^{n-m+1}\int \bar{\psi}_3\bar{J}_{\bar{I}_{b}}\prod_{c=1}^n(J^{++}_{K3}\bar{J}_{\bar{J}_c})\rangle\nonumber\\
&+(-1)^{n}\int_{\mathcal{M}_{(g,n)}}\hspace{-0.7cm}\langle(\mu\cdot G^-)^{3g+n-4}(\mu\cdot J^{--}_{K3})\psi_3(\alpha)\oint G^+_{K3,-}\bar{\Xi}^{[A_1}\prod_{a=2}^{m}\int\bar{\Xi}^{A_a]}\prod_{b=1}^{n-m+1}\!\!\int \bar{\psi}_3\bar{J}_{\bar{I}_{b}}\prod_{c=1}^nJ^{++}_{K3}\bar{J}_{\bar{J}_c}\rangle\label{HarmonicityRelSplittingred}
\end{align}
The first line $\hat{\tilde{\mathcal{C}}}^{\text{(bdy)},A_1\ldots A_{m}}_{\bar{I}_1\ldots\bar{I}_{n-m+1},\bar{J}_1\ldots\bar{J}_{n}}$ is a boundary contribution due to the insertion of the energy momentum tensor sewed with one of the Beltrami differentials. Following the same lines as before, we can immediately state the result
\begin{align}
\hat{\tilde{\mathcal{C}}}^{\text{(bdy)},A_1\ldots A_{m}}_{\bar{I}_1\ldots\bar{I}_{n-m+1},\bar{J}_1\ldots\bar{J}_{n}}=\hat{\tilde{\mathcal{C}}}^{\text{(bdy,hyper)},A_1\ldots A_{m}}_{\bar{I}_1\ldots\bar{I}_{n-m+1},\bar{J}_1\ldots\bar{J}_{n}}+\hat{\tilde{\mathcal{C}}}^{\text{(bdy,vector)},A_1\ldots A_{m}}_{\bar{I}_1\ldots\bar{I}_{n-m+1},\bar{J}_1\ldots\bar{J}_{n}}\,,\label{HatTildeBDY}
\end{align}
with the explicit expressions
\begin{align}
&\hat{\tilde{\mathcal{C}}}^{\text{(bdy,hyper)},A_1\ldots A_{m}}_{\bar{I}_1\ldots\bar{I}_{n-m+1},\bar{J}_1\ldots\bar{J}_{n}}=\sum_{g_1=0}^{g}\sum_{n_s=0}^n\sum_{m_s=0}^{n_s}\mathcal{D}^{B}_+\hat{\tilde{\mathcal{H}}}^{g,A_1\ldots A_{m_s}}_{\bar{I}_1\ldots\bar{I}_{n_s-m_s},\bar{J}_1\ldots\bar{J}_{n_s}}\,\Omega_{BC}\,\mathcal{D}^{C}_+\hat{\mathcal{H}}^{g,A_{m_s+1}\ldots A_{m}}_{\bar{I}_{n_s-m_s+1}\ldots\bar{I}_{n-m+1},\bar{J}_{n_s+1}\ldots\bar{J}_{n}}\,,\nonumber\\
&\hat{\tilde{\mathcal{C}}}^{\text{(bdy,vector)},A_1\ldots A_{m}}_{\bar{I}_1\ldots\bar{I}_{n-m+1},\bar{J}_1\ldots\bar{J}_{n}}=0\,.\nonumber
\end{align}
Here the propagation of the vector multiplet states is in fact cancelled against the contribution stemming from the exchange of a state $\rho$ of charge $+3$ and the unit operator. As already remarked in footnote \footnotemark[6], this subtle cancellation was overlooked in the equations derived in \cite{Antoniadis:2009nv}, but is strictly necessary for the integrability of the final differential equation. For later convenience, we introduce a generating functional for the quantity (\ref{HatTildeBDY})                                                                                                                                                                                                                                                                                                                         
\begin{align}
\hat{\tilde{\mathbb{C}}}^g_{\text{bdy}}:=\sum_{g_s=1}^{g-1}\mathcal{D}_+^B\hat{\tilde{\mathbb{H}}}^g_{\text{red}}\,\Omega_{BC}\,\mathcal{D}^C_+\hat{\mathbb{H}}^g_{\text{red}}\,.\label{Bdytildehat}
\end{align}
In the remaining two lines of (\ref{HarmonicityRelSplittingred}) which we called $\hat{\tilde{\mathcal{C}}}^{\text{(bulk)},A_1\ldots A_{m}}_{\bar{I}_1\ldots\bar{I}_{n-m+1},\bar{J}_1\ldots\bar{J}_{n}}$ we readily recognize again a bulk contribution, which we may rewrite using the harmonicity operator (\ref{harop})
\begin{align}
&\hat{\tilde{\mathcal{C}}}^{\text{(bulk)},A_1\ldots A_{m}}_{\bar{I}_1\ldots\bar{I}_{n-m+1},\bar{J}_1\ldots\bar{J}_{n}}=(-1)^{n-m_1}\nabla^{[A_{m}}_-\hat{\tilde{\mathcal{H}}}^{g,A_1\ldots A_{m-1}]}_{\bar{I}_1\ldots \bar{I}_{n-m+1},\bar{J}\ldots \bar{J}_{n}}\,,
\end{align}

\end{document}